\documentclass{lmcs}
\pdfoutput=1

\usepackage{lastpage}
\lmcsdoi{19}{4}{12}
\lmcsheading{}{\pageref{LastPage}}{}{}%
{Jun.~09,~2022}{Nov.~22,~2023}{}

\usepackage{todonotes}

\usepackage{hyperref}

\usepackage[utf8]{inputenc}
\usepackage{amsmath}
\usepackage{amssymb}
\usepackage{mathrsfs}
\usepackage{graphicx}
\usepackage{mathtools}
\usepackage{verbatim}
\usepackage{adjustbox}
\usepackage{graphicx}
\usepackage{bm}
\usepackage{tikz}
\usetikzlibrary{automata}
\usetikzlibrary{decorations.pathmorphing,arrows,positioning}
\usetikzlibrary{calc}
\usepackage{tikzscale}
\usetikzlibrary{shapes.geometric}
\usetikzlibrary{arrows,automata}
\usetikzlibrary{matrix}
\usetikzlibrary{decorations.markings}
\usetikzlibrary{decorations.shapes}
\usepackage{stmaryrd}
\usepackage{microtype}
\usepackage{bbm}
\usepackage{braket}
\usepackage{array}
\usepackage[normalem]{ulem} 

\usepackage{extarrows} 

\newcommand{\llp}{\llparenthesis}
\newcommand{\rrp}{\rrparenthesis}
\newcommand{\derives}{\overset{*}{\Longrightarrow}}
\newcommand{\avv}{\curvearrowright}
\newcommand{\powerset}{\raisebox{.15\baselineskip}{\large\ensuremath{\wp}}}
\newcommand{\dd}[1]{#1^{\shortdownarrow}}
\newcommand{\uu}[1]{#1^{\shortuparrow}}

\DeclareMathOperator{\hole}{hole}

\newcommand{\boldtrans}[1]{\stackrel{#1}{\pmb\longrightarrow}}
\newcommand{\boldtranss}[2]{\underset{#1}{\overset{#2}{\pmb\longrightarrow}}}

\DeclareMathOperator{\treec}{TreeC}

\newcommand{\g}[1]{\textbf{\textit{#1}}}

\newcommand{\ee}{\g{e}}

\begin{document}

\title[Aperiodic = Star-free = FO-definable for OP Languages]{Aperiodicity, Star-freeness, and First-order Logic Definability of Operator Precedence Languages}
\thanks{The paper \cite{MPC20} presents a preliminary version of the results of Sections~\ref{subsec:OPE} to \ref{NC-closure}.}

\author[D. Mandrioli]{Dino Mandrioli\lmcsorcid{0000-0002-0945-5947}}[a]
\address{DEIB, Politecnico di Milano, Italy}
\email{dino.mandrioli@polimi.it}

\author[M. Pradella]{Matteo Pradella\lmcsorcid{0000-0003-3039-1084}}[b]
\address{DEIB, Politecnico di Milano, Italy and
IEIIT, Consiglio Nazionale delle Ricerche}
\email{matteo.pradella@polimi.it, stefano.crespireghizzi@polimi.it}

\author[S. Crespi Reghizzi]{Stefano {Crespi Reghizzi}\lmcsorcid{0000-0001-5061-7402}}[b]

\begin{abstract}
A classic result in formal language theory is the equivalence among non-counting, or aperiodic, regular languages, and languages defined through star-free regular expressions, or first-order logic.
Past attempts to extend this result beyond the realm of regular languages have met with difficulties:
for instance it is known  that  star-free tree languages may violate the
non-counting property and there are aperiodic tree languages that cannot be defined through first-order logic.

We extend such classic equivalence results to a significant family of deterministic context-free languages,
the operator-precedence languages (OPL), which strictly includes
the widely investigated visibly pushdown, alias input-driven, family and other structured context-free languages.
The OP model originated in the '60s for  
defining programming languages and is still used by high performance compilers;
its rich algebraic properties have been investigated initially in connection with grammar learning  and
recently completed with further closure properties and with monadic second order logic definition.

We introduce an extension of regular expressions, the OP-expressions (OPE) which define the OPLs and,  
under the star-free hypothesis, define first-order definable and non-counting OPLs.
Then, we prove, through a fairly articulated grammar transformation, that aperiodic OPLs are first-order definable.
Thus, the classic equivalence of star-freeness, aperiodicity, and first-order definability is established for the large and powerful class of OPLs. 

We argue that the same approach can be exploited to obtain analogous results for visibly pushdown languages too.
\end{abstract}



\maketitle

\section{Introduction} 

From a long time much research effort in the field of
formal language theory has been devoted to extend as much as possible the nice
algebraic and logic properties of regular languages to larger families of
languages, typically the context-free (CF) ones or subfamilies thereof. Regular
languages in fact are closed w.r.t.\ all basic algebraic operations and are
characterized also in terms of classic monadic second-order (MSO) logic (with the ordering relation between character positions)
\cite{bib:Buchi1960a,Elg61,Tra61}, but not so for general CF languages. 

On the other hand, some important algebraic and logic properties of regular languages are preserved by certain  subfamilies of the CF languages, that may be referred to as \emph{structured
CF languages} because the syntax structure  is immediately visible in their sentences.  Two first and practically equivalent examples of such
languages are \emph{parenthesis languages} and \emph{tree languages} introduced
respectively by McNaughton \cite{McNaughton67} and Thatcher \cite{Tha67}. More
recently, \emph{visibly pushdown languages (VPL)} \cite{jacm/AlurM09},
originally introduced as  \emph{input-driven languages (IDL)} \cite{Input-driven},
\emph{height-deterministic} \cite{conf/mfcs/NowotkaS07} and \emph{synchronized languages}  \cite{conf/dlt/Caucal06} have also
been shown to share many important properties of regular languages.
In particular tree languages and VPLs are closed w.r.t.\ Boolean operations,
concatenation, Kleene $^\ast$ and are characterized in terms of some MSO logic,
although  such operations and the adopted logic language are not the same in the two cases. For a complete analysis of
structured languages and how they extend algebraic and logic properties of
regular languages, see~\cite{DBLP:journals/csr/MandrioliP18}.

In this paper we study for structured CF languages three important language features, namely the \emph{non-counting} (NC) or \emph{aperiodicity},\footnote{The two terms are synonyms in the literature, so we will use them interchangeably.}
the \emph{star-freeness} (SF), and the \emph{first-order} (FO) logic definability properties, which for regular languages are known to be equivalent~\cite{McNaughtPap71}. 

Intuitively, a language has the aperiodicity property
if the recognizing device ---a finite state automaton in the case of regular languages--- cannot
separate two strings that 
only differ by the count, modulo an integer greater than 1, of the occurrences of some substring.
Linguists and computer scientists alike have  observed that human languages, both natural and artificial, do not rely on modulo counting.
For programming languages the early and fairly obvious observation that they do not include syntactic constructs based on modulo counting motivated the definition of non-counting context-free grammar~\cite{CreGuiMan78}, and that of aperiodic tree languages \cite{DBLP:conf/caap/Thomas84}.  
The theory of Linguistic Universals~\cite{Chomsky} postulates that all human languages have some common features that are necessary for their acquisition and use. 
The list of such features has evolved over time and is not agreed upon by everybody. Some feature lists included the fact that syntactic categories, hence grammaticality of a sentence, are not based on modulo arithmetic. 
A possible reason for that is that in noisy linguistic communication, the interpretation of the message would be very error prone.

SF regular languages are definable through a star-free regular expression (RE), i,e, an expression composed exclusively by means of Boolean operations and concatenation. FO logic defined regular languages are characterized by the first-order (FO) restriction of MSO logic.

The above properties, together with other equivalent ones which are not the object of the present investigation \cite{McNaughtPap71}, have ignited various important practical applications in the realm of regular languages.  FO
definition, in particular, has a tremendous impact on the success of
model-checking algorithms, thanks to the first-order completeness of linear
temporal logic\footnote{This result is due to H.W. Kamp. From his thesis
several simplified proofs have been derived, e.g.,~\cite{DBLP:journals/corr/Rabinovich14}.}: most model-checkers of practical usefulness exploit NC languages.

Moving from regular languages to suitable families of structured CF languages is certainly a well motivated goal:  the aperiodicity property, in fact,
is perhaps even more important for CF languages than for regular
ones: whereas various hardware devices, e.g., count modulo some natural number,
it is quite unlikely that a programming, a data description, or a natural
language exhibits counting features such as forbidding an even number of nested
loops or recursive procedure calls.  We could claim that most if not all of
CF languages of practical interest have an aperiodic structure.


Non-counting parenthesis languages were first introduced in
\cite{CreGuiMan78}. 
Then, an equivalent definition of aperiodicity in terms of tree languages
was given in \cite{DBLP:conf/caap/Thomas84}. 
It was immediately clear, however, that the above properties holding for regular word languages do not extend naturally to regular tree languages: 
in \cite{DBLP:conf/caap/Thomas84} itself it is shown that SF regular expressions for tree languages may define even counting languages; 
this is due to the fact that string concatenation is replaced by the append operation in tree languages. 
The same paper shows further intricacies in the investigation of algebraic and logic characterization of tree languages.
Subsequent studies (e.g.,
\cite{DBLP:journals/ita/Heuter91,DBLP:conf/cai/EsikI07,DBLP:journals/jolli/Langholm06,DBLP:conf/tapsoft/Potthoff95,DBLP:journals/tcs/Potthoff94})
provided partial results by investigating algebraic and logic properties of various subclasses of tree languages. We mention in particular another negative result, i.e.,
the existence of aperiodic tree languages that are not FO-definable \cite{DBLP:journals/ita/Heuter91,DBLP:conf/tapsoft/Potthoff95}.
To summarize, we quote Heuter: ``The equivalence of the notions {\em first-order, star-free} and {\em aperiodic} for regular word languages completely 
fails in the corresponding case of tree languages.'' 

In contrast, here we show that the three equivalent characterizations holding for NC
regular languages can be extended to the  family of \emph{operator precedence languages} (OPLs). It is worthwhile to outline their history and their practical and theoretical development.

Invented by R. Floyd \cite{Floyd1963} to support fast deterministic parsing,  operator precedence grammars (OPG) are still used within modern compilers to parse expressions with operators ranked by  priority.
The syntax tree  of a sentence is determined by  three binary precedence relations over the terminal alphabet that are easily pre-computed from the grammar productions. 
We classify OPLs as ``structured but non-visible'' languages since their
structure is implicitly assigned by such precedence relations. For readers unacquainted with
 OPLs, we provide a preliminary example:  the arithmetic sentence  $a + b * c$ does not make manifest the natural structure
$(a + (b * c))$,  but the latter is implied by the fact that the plus operator yields precedence to the times.

Early theoretical investigation \cite{Crespi-ReghizziMM1978}, originally motivated by grammar inference goals, realized that, 
thanks to the  tree structure assigned
to  strings by the precedence relations, many closure
properties  of regular languages and other structured CF ones hold for OPLs too;
this despite the fact that, unlike other better known
structured languages, OPLs  need a simple parsing process to make
their syntax trees explicit.  This fact accounts for the  wider generative capacity that makes OPLs suitable to define  programming and data description languages.

After a long intermission, theoretical research \cite{Crespi-ReghizziM12} proved further algebraic properties of OPLs, thus moving some  steps ahead from regular to
structured CF languages. At the same time, it was found 
  that the VPLs are a particular case of the OPLs characterized by the precedence relations encoded in the 3-partition of their alphabet;
OPLs considerably generalize VPLs while  retaining their closure properties.
Then in \cite{LonatiEtAl2015} the  \emph{Operator
Precedence automata (OPA)} recognizing OPLs were introduced to formalize the efficient parallel parsing algorithm implemented in \cite{BarenghiEtAl2015}.
In the same paper an MSO logic characterization of OPLs
that  naturally extends  the classic one for regular languages was also produced.
Recently, yet another characterization of regular languages has been extended to OPLs, namely, in terms of a congruence such that a language is an OPL iff the equivalence classes of the congruence are finite \cite{Henzinger23}.

Thus, OPLs' potential for 
 practical applications is broader than other structured CF languages:  the following example hints at applications  for  automatic proof of systems properties.  OPLs  with their corresponding MSO logic
may be used  to specify and prove properties of software systems where the typical LIFO
policy of procedure calls and returns can be broken by unexpected events such
as interrupts or exceptions
\cite{LonatiEtAl2015,DBLP:journals/csr/MandrioliP18}, a feature that is not available in VPLs and their MSO logic \cite{AF16}.  

In summary, to the best
of our knowledge, OPLs are currently the largest language family that retains the main
closure and decidability properties of regular languages, including a logical
characterization naturally extending the classic one.

We recently realized that a NC subclass of OPLs introduced long ago in the course of grammar-inference studies \cite{DBLP:journals/cacm/Crespi-ReghizziML73,DBLP:journals/ipl/Crespi-ReghizziM78} is  FO logic definable
\cite{LMPP15}. This led us to the present successful search for 
equivalent characterizations  of aperiodic, star-free and FO definable OPLs. 
Our approach is based on two key ideas:

\begin{enumerate}
    \item 
   Since the traditional attempt at extending NC regular language properties to tree languages 
   failed and produced only partial results,
   we went back to string languages. Accordingly, we  use the operation of string concatenation and not the append operation of tree languages. 
  \item
    We kept using the  MSO logic of our past work \cite{LonatiEtAl2015,LMPP15}, which  had been inspired by previous work on CF string languages \cite{Lautemann94} and on VPLs \cite{jacm/AlurM09}. Such logics too are defined on strings rather than on trees as a natural extension of the traditional one for regular languages. We examined its restriction to the FO case. 
\end{enumerate}

The main results of this paper are:
\begin{itemize}
 \item 
    The introduction (in Section~\ref{subsec:OPE}) of \emph{operator precedence expressions (OPE)} which extend
    regular expressions: they  add to the classical operations a new one, called \emph{fence},  that imposes a matching between
    two (hidden) parentheses: we show that OPEs define the OPL family.
      \item
   The proof (in Section~\ref{sec:SF-FO}) that the  OPLs defined by star-free OPEs coincide with the ones defined 
    by FO formulas, and (in Section~\ref{NC-closure}) the proof that they have  the aperiodicity property.
    \item
    Finally, (in Section~\ref{sec:FO}) the proof that every NC OPL can be defined
    by means of an FO formula. The proof, articulated in several lemmas,
     exploits a
    \emph{regular language control theorem} (in Section~\ref{sec:contrMSO}) which, informally,
    ``splits'' the logic formulas defining an OPL into a part  describing its  tree-like structure and another part that imposes a regular control  on the strings derived from the grammar's nonterminal symbols.
   After a series of nontrivial transformations of finite automata, we
    obtain the result that the control language can be made NC if the original OPL is in
    turn NC.
    Thanks to the fact that both parts of the logic formulas can be defined in FO logic, we obtain the language family identities:
   \begin{center}   
   \emph{OPLs = OPE-languages = MSO-languages} 
  
    \emph{NC-OPLs = SF-OPE-languages = FO-languages} 
    \end{center}
    which extend the classic equivalences for regular languages and could be transposed to VPLs, by following a similar path.
\end{itemize}
Section~\ref{sec:preliminaires} provides the necessary terminology and background on OPLs, aperiodicity, parenthesis languages, MSO and FO logic  characterization.
The conclusion mentions new application-oriented developments rooted in the present results, consisting of a suitable, FO-complete, temporal logic and a model-checker to prove properties of aperiodic OPLs. New directions for future research are also suggested.

\section{Preliminaries}\label{sec:preliminaires}

We assume some familiarity with the classical literature on formal language and automata theory, e.g., \cite{Salomaa73,Harrison78}.
Here, we  just list and explain our notations for the basic concepts we use from this theory. 
The terminal alphabet is usually denoted by $\Sigma$, and the empty string is $\varepsilon$.
For a string, or set, $x$, $|x|$ denotes the length, or the cardinality, of $x$.  
The character $\#$, not present in the terminal alphabet, is used as string \textit{delimiter}, and we define the alphabet  $\Sigma_\# = \Sigma \cup \{\# \}$.
Other special symbols augmenting $\Sigma$ will be introduced in the following.


\subsection{Regular languages: automata, regular expressions, logic}\label{subsec:FA}

\paragraph*{Finite Automata}
A \emph{finite automaton} (FA) $\mathcal{A}$ is defined  by a  5-tuple
 $( Q,\Sigma, \delta, I, F)$  where
$Q$ is the set of states, $\delta$ the \emph{state-transition relation} (or its \emph{graph} denoted by $\longrightarrow$),  $\delta\subseteq Q \times \Sigma \times Q$;
$I$ and $F$ are the nonempty subsets of $Q$ respectively comprising the  initial  and  final states. 
If the tuple $(q,a,q')$ is  in the relation $\delta$, the edge $q\xlongrightarrow {a} q'$ is in the graph. 
The transitive closure of the relation is defined as usual. 
Thus, for a string $x \in \Sigma^*$ such that there is a path from state $q$ to $q'$ labeled with $x$, the notation $q\xlongrightarrow  {x} q'$ is equivalent to    
 $(q,x,q') \in \delta^*$;
if $q \in I$ and $q'\in F$, then the string  $x$ is {\em accepted}\/ by $\mathcal{A}$.
The language of the accepted strings  is denoted  by $L(\mathcal{A})$; it is called a \emph{regular language}.

In this paper we make use of two well-known extensions of the previous FA definition, both not impacting on the language family recognized.
In the first extension, we permit an edge label to be the empty string; such an edge is called a {\em spontaneous}\/ transition or step. 
In the second one, an edge label may be a string in $\Sigma^+$. These two classical extensions are formalized by letting $\pmb\delta\subseteq Q \times \Sigma^*  \times Q$, where for clarity, the extended transition relation is written in boldface.
An edge $(q, x, q') \in \pmb\delta$  is called a {\em  macro-transition}\/ or \emph{macro-step}
and is denoted by $q \boldtranss{\pmb\delta}{x} q'$. Whenever there will be no risk on ambiguity we will omit the label $\pmb\delta$ in the edge. 

\paragraph*{Regular expressions and star-free languages}
A \emph{regular expression} (RE) over an alphabet $\Sigma$ is a well-formed formula made with the characters of $\Sigma$, $\emptyset$, $\varepsilon$, the Boolean  operators $\cup, \neg, \cap$, the concatenation  `$\cdot$', and the Kleene star operator `$^*$'. We may also use the  operator `$^+$'. When neither `$^*$' nor `$^+$' are used, the RE is called \emph{star-free} (SF).
An RE $E$ defines a language over $\Sigma$, denoted by $L(E)$.

\paragraph*{Monadic second and first order logics to define languages \cite{bib:Thomas1990a}} A \emph{monadic second order (MSO) logic} on an alphabet $\Sigma$ is a well-formed formula made with  first and second order variables interpreted, respectively, as string positions and sets of string positions, monadic predicates on string positions biunivocally associated to $\Sigma$ elements, an ordering relation, and the usual logical connectors and quantifiers. When the logic is restricted to first-order variables only, it is named an FO-logic.

\paragraph*{Non-counting or aperiodic regular languages}
A regular language $L$ over $\Sigma$ is called \emph{non-counting} (NC) or \emph{aperiodic} if there exists an integer $n\geq 1$ such that for all $x,y,z \in \Sigma^*$, $x y^n z\in L$ iff   $x y^{n+m} z\in L$,
$\forall m \ge 0$. 

\begin{prop}\label{propo:RE}
Finite automata, regular expressions and MSO logic define  the family of \emph{regular} (or rational)  languages (REG) \cite{bib:Buchi1960a,Elg61,Tra61}.   The family of aperiodic regular languages coincides with the families of languages defined by star-free REs and by FO-logic \cite{McNaughtPap71}.
\end{prop}

\subsection{Grammars}

\begin{defi}[Grammar and language]\label{def:grammar}
A (CF) \emph{grammar} is a tuple $G=(\Sigma, V_N,  P, S)$ where $\Sigma$ and $V_N$, with $ \Sigma \cap V_N = \emptyset$, are resp.\ the terminal and the nonterminal alphabets, the total alphabet is $V = \Sigma \cup V_N$, $P\subseteq V_N \times V^*$ is the rule (or production) set,  and $S \subseteq V_N$, $S \neq \emptyset$,  is the axiom set.
For a generic rule, denoted as $A \to \alpha$,  where $A$ 
and $\alpha$ are resp.\ called the left/right hand sides (lhs / rhs), the following forms  are relevant:
\begin{center}
\begin{tabular}{l@{\;:\; }p{10cm}}
axiomatic & $A \in S$
\\
terminal & $\alpha \in \Sigma^+$
\\
empty & $\alpha =\varepsilon$
\\
renaming & $\alpha  \in V_N$
\\ 
linear & $\alpha\in \Sigma^* V_N \Sigma^* \cup \Sigma^* $\\
operator & $\alpha \not\in V^* V_N V_N V^*$, i.e., at least one terminal is interposed between any two nonterminals occurring in $\alpha$
\\

parenthesized & $\alpha = \llp \beta \rrp$ where $\beta \in V^*$, and $\llp$, $\rrp$ are new terminals.
\end{tabular}
\end{center}

\par
A grammar is called \emph{backward deterministic} or a BD-grammar  (or \emph{invertible}) if $(B \to \alpha, C \to \alpha \in P)$ implies $B=C$. 
\par
If all rules of a grammar  are in operator (respectively, linear) form, the grammar is called an \emph{operator grammar} or O-grammar (respectively,  \emph{linear grammar}) .
\par
A  grammar
$G_p =\left(\Sigma \cup \{\llp ,\rrp\},V_N, P_p, S \right)$
is a \emph{parenthesis} grammar (Par-grammar) if the rhs of every rule is parenthesized. 
$G_p$ is called
 the \emph{parenthesized version} of $G$,  if $P_p$  consists
of all rules  $A\to \llp \alpha \rrp$ such that $A \to \alpha$ is in $P$.

For brevity, we assume the reader is familiar with the usual definition of \emph{derivation} denoted by the symbols $\xLongrightarrow[G]{}$ (immediate derivation), $\xLongrightarrow[G]{\ast}$ (reflexive and transitive closure of $\xLongrightarrow[G]{}$),
$\xLongrightarrow[G]{+}$ (transitive closure of $\xLongrightarrow[G]{}$),
$\xLongrightarrow[G]{m}$ (derivation in $m$ steps);
 the subscript $G$ will be omitted whenever clear from the context.
 
 We also suppose that the reader is familiar with the notion of \emph{syntax tree} and that a parenthesized string is an equivalent way to represent a syntax tree of a CF grammar where internal nodes are unlabeled.
 As usual, the {\em frontier} of a syntax tree is the ordered left to right sequence of the leaves of the tree.
\par
The \emph{language} defined by a grammar starting from a nonterminal  $A$ is 
\[
L_G (A) = \left\{w  \mid w \in \Sigma^*, A \xLongrightarrow[G]{\ast} w \right\}.
\]
We call $w$ a \emph{sentence} if $A \in S$. 
The union of  $L_G(A)$ for all $A\in S$ is the language $L(G)$ defined by $G$. 
The language generated by a Par-grammar is called a \emph{parenthesis language}, and its sentences are well-parenthesized strings.

Two grammars defining the same language are \emph{equivalent}. Two grammars such that their parenthesized versions are equivalent, are \emph{structurally equivalent}.
\end{defi}

\noindent {\em Notation:} In the following, \emph{unless otherwise explicitly stated}, lowercase letters at the beginning of the alphabet will denote terminal symbols, lowercase letters at the end of the alphabet will denote strings of terminals, 
Greek letters at the beginning of the alphabet will denote strings in $V^*$. Capital letters will be used for nonterminal symbols.

Any grammar can be effectively transformed into an equivalent BD-grammar,
and also into an O-grammar~\cite{DBLP:reference/hfl/AutebertBB97,Harrison78} without renaming rules and without empty rules but possibly a single rule whose lhs is an axiom not otherwise occurring in any other production.
\emph{From now on, w.l.o.g., we exclusively deal with O-grammars without renaming and empty rules, with the only exception that, if $\varepsilon$ is part of the language, there is a unique empty rule whose lhs is an axiom that does not appear in the rhs of any production.}

\begin{defi}[Backward deterministic reduced  grammar \cite{McNaughton67,Salomaa73}] \label{def:bdrG}
A \emph{context} over an alphabet $\Sigma$ is a string in
$\Sigma^*\{-\}\Sigma^*$ , where the character `$-$' $\notin \Sigma$
is called a blank. 
We denote by
$\alpha[x]$ the context $\alpha$ with its blank replaced by the string $x$.
 Two nonterminals
$B$ and $C$ of a  grammar $G$ are termed \emph{equivalent} if, for every context $\alpha$,
$\alpha[B]$ is derivable from an axiom of $G$ iff so is $\alpha[C]$ (not necessarily from the same axiom).
\par 
A nonterminal $A$ is \emph{useless} if
there is no context $\alpha$ such that $\alpha[A]$ is derivable from  an axiom or $A$ generates no terminal string.
A terminal $a$ is useless if it does not appear in any sentence of $L(G)$. 
\par
A grammar is \emph{clean} if it has no useless nonterminals  and terminals.
A grammar is \emph{reduced}  if it is clean and  no two nonterminals are equivalent.
\par
A BDR-grammar is both backward deterministic and reduced. 
\end{defi}
\par
From \cite{McNaughton67}, every parenthesis language is generated by a unique, up to an isomorphism of its nonterminal alphabet, Par-grammar that is BDR.


\subsubsection{Operator precedence grammars}
We define the  operator precedence grammars (OPGs) following primarily \cite{DBLP:journals/csr/MandrioliP18}. 

Intuitively, operator precedence grammars are O-grammars whose parsing is driven by three \emph{precedence relations}, called \emph{equal}, \emph{yield} and \emph{take}, included in $\Sigma_\# \times \Sigma_\#$.
They are defined in such a way that two consecutive terminals of a grammar's rhs
---ignoring possible nonterminals in between---
are in the equal relation, while the two extreme ones ---again, whether or not preceded or followed by a nonterminal--- 
are preceded by a yield and followed by a take relation, respectively; 
in this way a complete rhs of a grammar rule is identified and can be \emph{reduced} to a corresponding lhs by a typical bottom-up parsing.
More precisely, the three relations are defined as follows. Subsequently we show how they can drive the bottom-up parsing of sentences.

%
%

\begin{defiC}[\cite{Floyd1963}]\label{def:OP-gramm}\label{def:OPG}
Let $G=(\Sigma, V_N, P, S)$ be an O-grammar.
Let $a,b$ denote elements in $\Sigma$,  $A, B$ in $V_N$, $C$ either an element of $V_N$ or the empty string $\varepsilon$, and $\alpha, \beta$ range over  $V^*$.
The \textit{left and right terminal sets} of terminals associated to nonterminals are respectively:
\[
  \mathcal{L}_G(A)  = \left\{a \in \Sigma \mid \exists C: A \xLongrightarrow[G]{\ast} C a \alpha
  \right\}
  \text{ and }
  \ \mathcal{R}_G(A)  =  \left\{a \in \Sigma \mid \exists C: A \xLongrightarrow[G]{\ast} \alpha a C \right\}.
\]
(The grammar name will be omitted unless necessary to prevent confusion.) 


The \emph{operator precedence relations} (OPRs) are defined over $\Sigma_\# \times \Sigma_\# $ 
as follows:
\begin{itemize}
\item equal in precedence:
  $
    a\doteq b  \iff 
    \exists A\to\alpha aCb\beta \in P \ \ 
    $
\item takes precedence:
$  a\gtrdot b  \iff \exists 
  A\to\alpha B b\beta \in P,
  a\in \mathcal{R}(B); 
  \\
  a\gtrdot \#  \iff  a\in \mathcal{R}(B), B \in S$
\item yields precedence:
  $
  a\lessdot b \iff  \exists  
  A\to\alpha aB\beta \in P, 
	b\in \mathcal{L}(B); \\
  \# \lessdot b \iff b\in \mathcal{L}(B), B \in S.
$
\end{itemize}
The OPRs can be collected  into a  $|\Sigma_\#| \times  |\Sigma_\#|$ array, called the \emph{operator precedence matrix} of the grammar,  $OPM(G)$: 
for each (ordered) pair $(a,b)\in \Sigma_\# \times  \Sigma_\#$,  $OPM_{a,b}(G)$ contains the OP relations holding between $a$ and $b$. 
\end{defiC}
\par
More formally, consider   a square matrix: 
\begin{equation}\label{eq:OPmatrix}
M=\left\{M_{a,b}\subseteq \left\{\doteq, \lessdot, \gtrdot \right\} \,\mid \, a, b \in \Sigma_\# \right\}
\end{equation}
 Such a matrix  is  called  \emph{conflict-free} iff $\forall a,b \in \Sigma_\#$, $0 \leq |M_{a,b}|\leq 1$. A conflict-free matrix  is called \emph{total} iff $\forall a,b \in \Sigma_\#$, $|M_{a,b}|=1$. By convention, if $M_{\#,\#}$ is not empty, $M_{\#,\#} = \{\doteq\}$.
A matrix  is $\dot=$-\emph{acyclic} if the transitive closure of the $\dot=$ relation over $\Sigma \times \Sigma$ is irreflexive.

We extend the set inclusion relations and the Boolean operations in the obvious cell by cell way, to any two matrices having the same terminal alphabet.
Two matrices are \emph{compatible} iff their union is conflict-free.

\begin{defi}[Operator precedence grammar]\label{def:OPgrammar}
A grammar $G$ is an \emph{operator precedence} (or Floyd's)  grammar, for short an OPG, iff the matrix  $OPM(G)$ is conflict-free, i.e.\ the three OP relations are disjoint.  
An OPG is $\dot=$-\emph{acyclic} if $OPM(G)$ is so. 
An \emph{operator precedence language (OPL)} is a language generated by an OPG.
\end{defi}

Figure \ref{fig:OPM1} (left) displays an  OPG, $G_{AE}$, which generates simple, unparenthesized arithmetic expressions and its OPM (center). The left and right terminal sets
of $G_{AE}$'s nonterminals $E$, $T$ and $F$ are, respectively:
$\mathcal{L}(E)  = \{+, *,e\}$,
$\mathcal{L}(T)  = \{*,e\}$, 
$\mathcal{L}(F)  = \{e\}$,
$\mathcal{R}(E) =  \{+, *,e\}$,
$\mathcal{R}(T) =  \{*,e\}$, and 
$\mathcal{R}(F) =  \{e\}$. 
  
\noindent\emph{Remarks.} If the relation $\dot=$ is acyclic, then the length of the rhs of any rule of $G$ is bounded by the length of the longest $\dot=$-chain in $OPM(G)$.

\begin{figure}
\begin{tabular}{m{0.4\textwidth}m{0.3\textwidth}m{0.2\textwidth}}
$
 	\begin{array}{ll}
	G_{AE}: &	S = \{E ,T ,F\} \\
 	&	E \to  E + T \mid T * F \mid e  \\  	
 	&	T \to  T * F \mid e      \\
 	&	F \to  e  \\
 	\end{array}
$ &
$
\begin{array}{c|cccc}
    &+ & *  & e  & \#\\
\hline
+ & \gtrdot & \lessdot & \lessdot & \gtrdot\\
* & \gtrdot & \gtrdot  & \lessdot & \gtrdot\\
e & \gtrdot & \gtrdot  &         & \gtrdot \\
\# & \lessdot & \lessdot &\lessdot \\
\end{array}
$ &
\centering\begin{tikzpicture}[scale=0.4]
\node{$N$}
child{ node{$\#$} }
child{
node{$N$}
child{ node{$N$} 
    child{ node{$N$} child { node{$e$} } }
    child{ node{$+$} }
    child{ node{$N$} 
        child{ node{$N$} child { node{$e$} }}
        child{ node{$*$} }
        child{ node{$N$} child { node{$e$} }}
    }
}
child{ node{$+$} }
child{ node{$N$} child { node{$e$} } }
}
child{ node{$\#$} };
\end{tikzpicture}
\end{tabular}
\caption{$G_{AE}$ (left), its OPM (center), and the syntax tree of $e + e * e + e$ according to the OPM (right).}
\label{fig:OPM1}
\label{fig:ST}
\end{figure}

Unlike the  arithmetic relations having similar typography, the OP relations do not enjoy any of the transitive, symmetric, reflexive properties. We kept the original Floyd's notation but we urge the reader not to be confused by the similarity of the two notations.

It is known that the family of OPLs is strictly included within the deterministic and reverse-deterministic CF family, i.e., the languages that can be deterministically parsed both from left to right and from right to left. 

The key feature of OPLs is that a conflict-free OPM $M$ defines a universe of \emph{strings compatible with $M$} and associates to each of them a unique \emph{syntax tree} whose internal nodes are unlabeled and whose leaves are elements of $\Sigma$, or, equivalently, a unique parenthesization. We illustrate such a feature through a simple example and refer the reader to previous literature  for a thorough description of OP parsing \cite{GruneJacobs:08, DBLP:journals/csr/MandrioliP18}.

\begin{exa}\label{ex:OPG:GAE_1}
Consider the $OPM(G_{AE})$ of Figure \ref{fig:OPM1} and the string $e + e * e + e$. Display all precedence relations holding between consecutive terminal characters, \emph{including the relations with the delimiters \#} as shown below:
\[ 
\# \lessdot e \gtrdot + \lessdot e \gtrdot * \lessdot  e \gtrdot  + \lessdot e \gtrdot  \#
\]
each pair $ \lessdot , \gtrdot $ (with no further $ \lessdot , \gtrdot $ in between) includes a \emph{possible} rhs of a production of \emph{any OPG} sharing the OPM with $G_{AE}$, not necessarily a $G_{AE}$ rhs. 
Thus, as it happens in typical bottom-up parsing, we replace each string included within the pair $ \lessdot , \gtrdot $ with a \emph{dummy nonterminal $N$}; this is because nonterminals are irrelevant for OPMs. 
The result is the string $\# N + N * N + N \# $. Next, we compute again the precedence relation between consecutive terminal characters by \emph{ignoring nonterminals}: the result is 
$ \# \lessdot N + \lessdot N * N \gtrdot  +  N \gtrdot  \# $.

This time, there is only one pair  $ \lessdot , \gtrdot $ including a potential rhs determined by the OPM (the fact that the external $ \lessdot $ and  $ \gtrdot $ ``look matched" is coincidental as it can be easily verified by repeating the previous procedure with the string $e + e * e + e + e$).
Again, we replace the pattern $N * N$, with the dummy nonterminal $N$; notice that there is no doubt about associating the two $N$ to the $*$ rather than to one of the adjacent $+$ symbols: 
if we replaced, say, just the $*$ with an $N$ we would obtain the string
$N + NNN + N$ which cannot be derived by an O-grammar.
By recomputing the precedence relations we obtain the string $\# \lessdot N + N \gtrdot + N \gtrdot \# $.
Finally, by applying twice the replacing of $N + N$ by $N$ we obtain $ \# N\# $. 
The result of the whole bottom-up reduction procedure is synthetically represented by the \emph{syntax tree} of Figure \ref{fig:ST} (right) which shows the precedence of the multiplication operation over the additive one in traditional arithmetics.

Notice that the tree of Figure \ref{fig:ST} has been obtained by using exclusively the OPM, not the grammar $G_{AE}$ although the string $e + e * e + e \in L(G_{AE})$
\footnote{As a side remark, the above procedure that led to the syntax tree of Figure \ref{fig:ST} could be easily adapted to become an algorithm that produces a new syntax tree whose internal nodes are labeled by $G_{AE}$'s nonterminals.
Such an algorithm could be made deterministic by transforming $G_{AE}$ into an equivalend BD grammar (sharing the same OPM). This aspect, however, belongs to the realm of efficient parsing which is not a major concern in this paper.}.
There is an obvious one-to-one correspondence between the trees whose internal nodes are unlabeled or labeled by a unique character, and well-parenthesized strings on the enriched alphabet $\Sigma \cup \{\llp, \rrp\})$; e.g., the parenthesized string corresponding to the tree of Figure \ref{fig:ST} is $\llp \llp  \llp e \rrp + \llp \llp e \rrp * \llp e \rrp \rrp \rrp  + \llp e \rrp \rrp$.

Obviously, all sentences of $L(G_{AE})$ can be given a syntax tree by $OPM(G_{AE})$, but there are also strings in $\Sigma^*$ that can be parsed according to the same OPM but are not in $L(G_{AE})$. E.g., the string $+++$ is parsed according to the $OPM(G_{AE})$ as the parenthesis string $\llp \llp  \llp + \rrp + \rrp + \rrp $. Notice also that, in general, not every string in $\Sigma^*$ is assigned a syntax tree ---or parenthesized string--- by an OPM; e.g., in the case of $OPM(G_{AE})$ the parsing procedure applied to $ee$ is immediately blocked since there is no precedence relation between $e$ and itself.
\end{exa}

The following definition synthesizes the concepts introduced by Example \ref{ex:OPG:GAE_1}.

\begin{defi}[OP-alphabet and Maxlanguage]\label{def:OP structures}\hfill
\begin{itemize}
\item
A string in $\Sigma^*$ is \emph{compatible} with an OPM $M$ iff the procedure described in Example~\ref{ex:OPG:GAE_1} terminates by producing the pattern $\#N\#$. The set of all strings compatible with an OPM $M$ is called the \emph{maxlanguage} or the \emph{universe} of $M$ and is simply denoted as $L(M)$.
\item
 Let $M$ be a conflict-free OPM over $\Sigma_\# \times \Sigma_\#$. We use the same identifier $M$ to denote the ---partial--- function $M: \Sigma^* \to (\Sigma \cup \{\llp, \rrp\})^*$ that assigns to strings in $\Sigma^*$ their unique well-parentesization as informally illustrated in Example~\ref{ex:OPG:GAE_1}.
 \item 
 The pair $(\Sigma, M)$ where $M$ is a conflict-free OPM over $\Sigma_\# \times \Sigma_\#$, is called an \emph{OP-alphabet}. We introduce the concept of OP-alphabet as a pair to emphasize that it defines a universe of strings on the alphabet $\Sigma$ ---not necessarily covering the whole $\Sigma^*$--- and implicitly assigns them a structure univocally determined by the OPM, or, equivalently, by the function $M$.
 \item 
 Let $(\Sigma, M)$ be an OP-alphabet. The class  of \emph{$(\Sigma, M)$-compatible} OPGs and OPLs are: 
\[
\mathscr{G}_M = \{G  \mid G \text{ is an OPG and } OPM(G) \subseteq M \},
\quad
 \mathscr{L}_M =  \{ L(G) \mid G \in \mathscr{G}_M\}. 
\]
\end{itemize}
\end{defi}

\par
Various formal properties of OPGs and OPLs are documented in the literature, chiefly in~\cite{Crespi-ReghizziMM1978,Crespi-ReghizziM12,DBLP:journals/csr/MandrioliP18}. 
In particular, in \cite{Crespi-ReghizziM12} it is proved that Visibly Pushdown Languages are strictly included in OPLs.
In VPLs the input alphabet is partitioned into three disjoint sets, 
namely {\em call} ($\Sigma_c$), {\em return} ($\Sigma_r$), and {\em internals} ($\Sigma_i$), 
where \emph{call} and \emph{return} play the role of open and closed parentheses. 
Intuitively, the string structure determined by these alphabets can be represented through an OP matrix in the following way: 
$a \lessdot b$, for any $a \in \Sigma_c$, $b \in \Sigma_c \cup \Sigma_i$; 
$a \doteq b$, for any $a \in \Sigma_c$, $b \in \Sigma_r$; 
$a \gtrdot b$, for all the other cases.

For convenience, we just recall and collect the 
OPL properties that are relevant for this article in the next proposition.

\begin{prop}[Algebraic properties of OPGs and OPLs]\label{propo:OPGandOPL}\label{PropOPL}\hfill
\begin{enumerate}
\item 

If an OPM $M$ is total, then the corresponding homonymous function, defined in the second bullet of Definition~\ref{def:OP structures}, is total as well, i.e., $L(M)= \Sigma^*$.


\item
Let $(\Sigma, M)$ be an OP-alphabet where $M$ is $\dot=$-acyclic. The class $\mathscr{G}_M$ contains an OPG, called the \emph{maxgrammar} of $M$, denoted by $G_{max, M}$, which generates the maxlanguage $L(M)$. 
For all grammars $G \in \mathscr{G}_M$, the inclusions  $L(G) \subseteq L(M)$ and $L(G_p) \subseteq L(G_{p,max, M}) = L_p(M)$ hold, where $G_p$ and $G_{p,max, M}$ are the parenthesized versions of $G$ and $G_{max, M}$, and $L_p(M)$ is the parenthesized version of $L(M)$.

\item The closure properties of the family $\mathscr{L}_M$ of $(\Sigma, M)$-compatible OPLs defined by a total OPM are the following: 
\begin{itemize}
\item $\mathscr{L}_M$ is closed under union, intersection and set-difference, therefore also under complement. 
\item $\mathscr{L}_M$ is closed under concatenation.
\item if matrix $M$ is $\dot=$-acyclic, $\mathscr{L}_M$ is closed under Kleene star.
\end{itemize}
\end{enumerate}
\end{prop}

\noindent \emph{Remark}. Thanks to the fact that a conflict-free OPM assigns to each string at most one parenthesization 
---and exactly one if the OPM is total--- 
the above closure properties of OPLs w.r.t.\ Boolean operations automatically extend to their parenthesized versions\footnote{The same does not apply to the case of concatenation.}.
In particular, any total, conflict-free, $\dot=$-acyclic OPM defines a \emph{universal parenthesized language} 
$L_{pU}$ 
such that its image under the homomorphism that erases parentheses is $\Sigma^*$ and
the result of applying Boolean operations to the parenthesized versions of some OPLs 
is the same as the result of parenthesizing the result of applying the same operations to the unparenthesized languages.

\emph{In the following we will assume that an OPM is $\dot=$-acyclic unless we explicitly point out the opposite.} 
Such a hypothesis is stated for simplicity despite the fact that, rigorously speaking, it affects the expressive power of OPLs
\footnote{An example language that cannot be generated with an $\doteq$-acyclic OPM is the following: 
	$L = \{a^n {(b c)}^n \mid n\geq 0\} \cup \{b^n {(c a)}^n \mid n\geq 0\} \cup \{c^n (a b)^n \mid n\geq 0\} $ since it requires the relations $a \doteq b, b \doteq c, c \doteq a $ \cite{HOP20}.}
 : it guarantees the closure w.r.t.\ Kleene star and therefore the possibility of generating $\Sigma^*$; 
this limitation however, is not necessary if we define OPLs by means of automata or MSO logic~\cite{LonatiEtAl2015}; in the case of OPGs a $\dot=$-cyclic OPM could require rhs of unbounded length; thus, the assumption could be avoided
by adopting OPGs extended by the possibility of including regular expressions in production rhs \cite{HOP20}, which however would require a much heavier notation.

\subsection{Logic characterization of operator precedence languages}\label{sec:logic}

In \cite{LonatiEtAl2015} the traditional monadic second order logic (MSO) characterization of regular languages by B\"uchi, Elgot, and Trakhtenbrot
\cite{bib:Buchi1960a,Elg61,Tra61} is extended to the case of OPLs\@.
Historically, a first attempt to extend the MSO logic for regular languages to deal with the typical tree structure of CF languages was proposed in \cite{Lautemann94} and then resumed by \cite{jacm/AlurM09}. 
In essence, the approach consists in adding to the normal syntax of the original logic a new binary relation symbol, named \emph{matching relation}, which joins the positions of two characters that somewhat extend the use of parentheses of \cite{McNaughton67}; 
e.g., in VPLs the matching relation pairs a \emph{call} with a \emph{return} according to the traditional LIFO policy of pushdown automata.

Such a matching relation, however, is typically one-to-one ---with an exception of minor relevance--- but cannot be extended to languages whose structure is not made immediately visible by explicit parentheses.
Thus, in \cite{LonatiEtAl2015} we introduced a new binary relation between string positions which, instead of joining the extreme positions of subtrees of the syntax trees, 
joins their contexts, i.e., the positions of the terminal characters immediately at the left and at the right of every subtree, i.e.,
respectively, of the character that yields precedence to the subtree's leftmost leaf, and of the one over which the subtree's rightmost leaf takes precedence. 
The new relation is denoted by the symbol $\avv$ and we write $\bm {x} \avv \bm {y}$ to state that it holds  between position $\bm {x}$ and position $\bm {y}$.


\begin{figure} 
\centering
\begin{tikzpicture}[flush/.style={double, >=stealth, thin, rounded corners}]
\matrix (m) [matrix of nodes]
 {\# & $e$ & $+$ & $e$ & $*$ & $e$ & $+$ & $e$  & \# \\
    0 & 1 & 2 & 3 & 4 & 5 & 6 & 7 & 8  \\
 };

\draw[->] (m-1-1)  to [out=60, in=120] (m-1-3);
\draw[->] (m-1-3)  to [out=60, in=120] (m-1-5);
\draw[->] (m-1-5)  to [out=60, in=120] (m-1-7);
\draw[->] (m-1-7)  to [out=60, in=120] (m-1-9);
\draw[->] (m-1-3)  to [out=60, in=120] (m-1-7);
\draw[->] (m-1-1)  to [out=60, in=120] (m-1-7);
\draw[->] (m-1-1)  to [out=60, in=120] (m-1-9);

\end{tikzpicture}
\caption{The string $e + e * e + e$, with relation $\avv$.}\label{fig:log:avv}
\end{figure}

Unlike the similar but simpler matching relation adopted in \cite{Lautemann94} and \cite{jacm/AlurM09}, the $\avv$ relation is not one-to-one.
For instance, Figure~\ref{fig:log:avv} displays the $\avv$ relation holding for the sentence $e + e * e + e $ generated by grammar $G_{AE}$:  we have
$0  \avv  2$, $2  \avv  4$, $4 \avv  6$, $6  \avv  8$, $2  \avv  6$, $0  \avv  6$, and $0 \avv  8$.
Such pairs correspond to
contexts where a reduce operation is executed  during the left-to-right, bottom-up parsing of the string (they are listed according to their execution order).
By comparing Figure~\ref{fig:log:avv} with Figure \ref{fig:OPM1} it is immediate to realize that every $\avv$ "embraces" a subtree of the syntax tree of the string $e+e*e+e$.

Formally, we define a countable infinite set of
first-order variables $\g{x}, \g{y}, \dots$ and a countable infinite set of
monadic second-order (set) variables $\g{X},\g{Y}, \dots$. We
adopt the convention to denote first and second-order variables in boldface
font.

\begin{defi}[Monadic Second-Order Logic for OPLs]
    Let $(\Sigma,M)$ be an OP-alphabet, $\mathcal{V}_1$ a set of first-order variables, and
$\mathcal{V}_2$ a set of second-order (or set) variables.
	The MSO$_{(\Sigma,M)}$ (\emph{monadic second-order logic} over $(\Sigma,
	M)$) is defined by the following syntax (the OP-alphabet will be omitted unless necessary to prevent confusion):
\[
	\varphi :=
		c(\g{x}) \mid
		\g{x} \in \g{X} \mid
		\g{x} < \g{y} \mid
		\g{x} \avv \g{y} \mid
		\neg \varphi \mid
		\varphi \lor \varphi \mid
		\exists \g{x}.\varphi \mid
		\exists \g{X}.\varphi
\]
where $c \in \Sigma_\#$, $\g{x}, \g{y} \in \mathcal{V}_1$, and $\g{X} \in \mathcal{V}_2$.\footnote{This is the usual MSO over strings, augmented with the $\avv$ predicate.}

A MSO formula is interpreted over a $(\Sigma, M)$ string $w$ compatible with $M$, 
with respect to assignments $\nu_1: \mathcal{V}_1 \to \{0, 1, \ldots, |w|+1\}$ and
$\nu_2: \mathcal{V}_2 \to \powerset(\{0, 1, \ldots, |w|+1\})$, in this way:

\begin{itemize}
    \item $\# w \#, M, \nu_1, \nu_2 \models c(\g{x})$ iff $\# w \# = w_1 c w_2$ and $|w_1| = \nu_1(\g{x})$.
    \item $\# w \#, M, \nu_1, \nu_2 \models \g{x} \in \g{X}$ iff $\nu_1(\g{x}) \in \nu_2(\g{X})$.
    \item $\# w \#, M, \nu_1, \nu_2 \models \g{x} < \g{y}$ iff $\nu_1(\g{x}) < \nu_1(\g{y})$.

    \item $\# w \#, M, \nu_1, \nu_2 \models \g{x} \avv \g{y}$ iff $\# w \# = w_1
      a w_2 b w_3$, $|w_1| = \nu_1(\g{x})$, $|w_1 a w_2| = \nu_1(\g{y})$, and
      $w_2$ is 
      the frontier of a subtree of the syntax tree of $w$, i.e., $w_2$ is well parenthesized within $M(w)$.
    \item $\# w \#, M, \nu_1, \nu_2 \models \neg\varphi$ iff $\# w \#, M, \nu_1, \nu_2 \not\models \varphi$.
    \item $\# w \#, M, \nu_1, \nu_2 \models \varphi_1 \lor \varphi_2$ iff $\# w \#, M, \nu_1, \nu_2 \models \varphi_1$ or  $\# w \#, M, \nu_1, \nu_2 \models \varphi_2$.
    \item $\# w \#, M, \nu_1, \nu_2 \models \exists \g{x} \varphi$ iff $\# w \#, M, \nu'_1, \nu_2 \models \varphi$, for some $\nu'_1$ with $\nu'_1(\g{y}) = \nu_1(\g{y})$ for all $\g{y} \in \mathcal{V}_1 - \{\g{x}\}$.
    \item $\# w \#, M, \nu_1, \nu_2 \models \exists \g{X} \varphi$ iff $\# w \#, M, \nu_1, \nu'_2 \models \varphi$, for some $\nu'_2$ with $\nu'_2(\g{Y}) = \nu_2(\g{Y})$ for all $\g{Y} \in \mathcal{V}_2 - \{\g{X}\}$.

\end{itemize}

To improve readability, we will drop $M$, $\nu_1$, $\nu_2$  and the delimiters \# from the notation whenever there is no risk of ambiguity; furthermore we use some standard abbreviations in formulas, e.g., $\land$, $\forall$, $\oplus$ (the exclusive or), $\g{x} + 1$, $\g{x} -1$, $\g{x} = \g{y}$, $\g{x} \le \g{y}$.

The language of a formula $\varphi$ without free variables is 
$
L(\varphi) = \{ w \in L(M) \mid w \models \varphi \}.
$
\end{defi}

Whenever we will deal with logic definition of languages we will implicitly exclude from such languages the empty string, according with the traditional convention adopted in the literature\footnote{Such a convention is due to the fact that the semantics of monadic logic formulas is given by referring to string positions.} (see, e.g., \cite{McNaughtPap71}); thus, when talking about MSO or FO definable languages we will exclude empty rules from their grammars.

\begin{exa}\label{ex:MSO formula}
Consider the OP-alphabet with $\Sigma = \{a, b\}$ and $M$ any total OPM containing, among other precedence relations that are not relevant in this example, $a \lessdot a, a \doteq b, b \gtrdot b$. 
Thus, the universe $L(M)$ is the whole $\Sigma^*$. We want to build an MSO formula that defines the sublanguage consisting of an odd number of $a$ followed by the same number of $b$. We build such a formula as the conjunction of several clauses.

The first clause imposes that after a $b$ there are no more $a$: 
\[
\forall \g x (b(\g x) \Rightarrow \neg \exists \g y (\g x < \g y \land a(\g y))).
\]
Thus, the original $\Sigma^*$ is restricted to the nonempty strings of the language $\{a^*b^*\}$.
A second clause imposes that the first character be an $a$, paired with the last character, which is a $b$: 
\[ 
a(1) \land \exists \g y (1 \avv \g y \land  b(\g y) \land \#(\g y+1)).
\]
This further restricts the language to $\{a^nb^n \mid n > 0\}$ because the relations $a \lessdot a \doteq b \gtrdot b$
imply the reduction of $ab$, possibly with an $N$ in between. Hence, if the first $a$ and the last $b$ of the string are the context of such a reduction, the number of $a$ in 
the string must be equal to the number of $b$.

Finally, to impose that the number of $a$ ---and therefore of $b$ too--- is odd, we introduce two second-order variables $\g{O}$ ---which stands for odd--- and $\g{E}$ ---which stands for even--- and impose that i) all positions belong to either one of them, ii) the elements of $\g{O}$ and $\g{E}$ storing the $a$ alternate ---and therefore those storing the $b$ too---, iii) the position of the first and last $a$ belongs to $\g O$. Such conditions are formalized  below%
\footnote{Although it would be possible to use only one second-order variable, we chose this path to make more apparent the correspondence between the definition of this language through a logic formula and the one that will be given in Example \ref{ex:C vs NC} by using an OPG.}.

\[
\exists \g {O} \exists \g {E} 
\forall \g {x} 
\left(
a(\g{x}) \Rightarrow 
\begin{array}{l}
(\g {x} \in \g {O} \oplus \g {x} \in \g {E}) \land \\
( \g {x} \in \g {O} \land a(\g{x}) \land a(\g {x} + 1) \Rightarrow \g {x} + 1 \in \g{E}) \land \\
( \g {x} \in \g {E} \land a(\g{x}) \land a(\g {x} + 1) \Rightarrow \g {x} + 1 \in \g{O}) \land\\
1 \in \g{O} \land (a(\g{x}) \land b(\g{x}+1) \Rightarrow  \g{x} \in \g{O})
\end{array}
\right)
\]

\end{exa}


\noindent \emph{Remark}. 
The reader could verify that the same language can be defined by using a \emph{partial} OPM, precisely an OPM consisting \emph{exclusively} of the relations 
$\# \lessdot a, a \lessdot a, a \doteq b, b \gtrdot b, b \gtrdot \#$,
and restricting the MSO formula to the above clause referring only to second-order variables.
Using partial OPMs, however, does not increase the expressive power of our logic formalism ---and of the equivalent formalisms OPGs and OPAs---: we will show, in Section \ref{subsec:OPE}, that any ``hole'' in the OPM can be replaced by suitable (FO) subformulas.

We also anticipate that, as a consequence of our main result, defining languages such as the one of this example, necessarily requires a second-order formula.




In \cite{LonatiEtAl2015} it is proved that the above MSO logic describes exactly the OPL family.
As usual, we denote the restriction of the MSO logic to the first-order as FO\@. 

\subsection{The non-counting property for parenthesis and operator precedence languages}
\label{NCbasics}
In this section we resume the original definitions and properties of non-counting (NC) CF languages \cite{CreGuiMan78} based on parenthesis grammars \cite{McNaughton67} and show their relations with the OPL family.

In the following all Par-grammars will be assumed to be BDR, unless the opposite is explicitly stated.

\begin{defi}[Non-counting parenthesis language and grammar \textrm{\cite{CreGuiMan78}}]\label{def:NCLangAndGramm}
A parenthesis language $L$ is \emph{non-counting} (NC) or \emph{aperiodic} iff there exists an integer $n > 1$  such that, for all strings $x, u, z, v, y$ in $(\Sigma \cup \{ \llp, \rrp \})^*$ where $z$ and $uzv$ are well-parenthesized,  $xu^n zv^n y \in L$ iff $xu^{n+m} zv^{n+m} y \in L$, $\forall m \ge 0$.
\par\noindent
A  \emph{derivation} of a Par-grammar is \emph{counting} iff it has the form
$A \stackrel + \Longrightarrow u^m A v^m$, with $m > 1, |uv| > 1$, and there is not a derivation  $A \stackrel + \Longrightarrow u A v$.

A Par-grammar is \emph{non-counting} iff none of its derivations is counting. 
\end{defi}

\begin{thm}[NC language and  grammar (Th.\ 1 of \cite{CreGuiMan78})] A parenthesis language is NC iff its BDR grammar has no counting derivation.
\end{thm}

\begin{thm}[Decidability of the NC property (Th.\ 2 of \cite{CreGuiMan78})] \label{decidability NC}
It is decidable whether a parenthesis language is NC or not.
\end{thm}

\begin{defi}[NC OP languages and grammars]\label{def:NCOPL}
For a given OPL $L$ on an OP-alphabet $(\Sigma,M)$,
its corresponding parenthesized language $L_p$ is the language 
$\{M(x) \mid x \in L \}$. 
$L$ is NC iff $L_p$ is NC.

A derivation of an OPG $G$ is counting iff the corresponding derivation of the associated Par-grammar $G_p$ is counting.
\end{defi}

Thus, an OPL is NC iff its BDR OPG (unique up to an isomorphim of nonterminal alphabets) has no counting derivations.

\begin{exa}\label{ex:C vs NC}
Consider the following BDR OPG $G_{C}$, with $S = \{O\}$,
$
O \to 
a E b \mid
a b;
E \to 
a O b  
$.
Its parenthesized version generates the language 
$\{ (\llp a)^{2n+1} (b \rrp)^{2n+1}  \mid n \geq 0\}$ which is counting; thus so is $L(G_{C})$ which is the same language as that of Example \ref{ex:MSO formula}.

In contrast, the grammar $G_{NC}$, with $S = \{A\}$,
$
A \to 
a B b \mid
a b 
;
B \to 
a A c  
$
generates a NC language, despite the fact that the number of $a$ in $L(G_{NC})$'s sentences is odd, because substrings $aa$ are not paired with repeated substrings.\footnote{The above definition of NC parenthesized string languages is equivalent to the definition of NC tree languages~\cite{DBLP:conf/caap/Thomas84}.}
Notice however, that, if we parenthesize the grammar $G_{NoOP}$, with $S = \{A\}$,
$
A \to 
aa A bb \mid
a b
$
which is equivalent to $G_{C}$, we obtain a NC language according to Definition \ref{def:NCLangAndGramm}. This should be no surprise, since $G_{C}$ and $G_{NoOP}$ are not structurally equivalent and $G_{NoOP}$ is not an OPG, having a non-conflict-free OPM.
\end{exa}

The following important corollary immediately derives from Definition \ref{def:NCOPL} and Theorem~\ref{decidability NC}.

\begin{cor}[Decidability of the NC property for OPLs.] \label{decidability NC-OPL}
It is decidable whether an OPL is NC or not.
\end{cor}

In the following, unless parentheses are explicitly needed, we will refer to unparenthesized strings rather than to parenthesized ones, thanks to the one-to-one correspondence.

It is also worth recalling \cite{CreGuiMan81} the following peculiar property of
OPLs: whether such languages are aperiodic or not does not depend on their OPM; in other
words, although the NC property is defined for structured languages (parenthesis
or tree languages \cite{McNaughton67,Tha67}),  in the case of OPLs this property does not
depend on the structure given to the sentences by the OPM.
It is important to stress, however, that, despite the above peculiarity of OPLs, aperiodicity remains a property that makes sense only with reference to the structured version of languages. Consider, in fact, the following OPLs, with the same OPM consisting of $\{c \lessdot c, c \doteq a,c \doteq b, a \gtrdot b, b \gtrdot a  \}$ besides the implicit relations w.r.t.\ $\#$:

$L_1 = \{ c^{2n}(ab)^n \mid n \geq 1\}$, $L_2 = \{ (ab)^+\}$

They are both clearly NC and so is their concatenation $L_1 \cdot L_2$ according to Definition~\ref{def:NCOPL}, which in its parenthesized version is 
$ \{  \llp^{2(m-n)}(\llp c)^{2n}(a\rrp b \rrp)^m \mid m > n \geq 1\}$, (see also Theorem~\ref{concatenation-closure}); however, if we applied Definition~\ref{def:NCLangAndGramm} to $L_1 \cdot L_2$ without considering parentheses, we would obtain that, for every $n$, $c^{2n}(ab)^{2n} \in L_1 \cdot L_2$ but not so for $c^{2n+1}(ab)^{2n+1}$.

We mention that  the subfamily of OPLs which in \cite {DBLP:journals/ipl/Crespi-ReghizziM78} was proved NC and in \cite{LMPP15} was proved  FO logic definable,
  includes, as maximal elements, the maxlanguages of all OPMs.

\section{Expressions for operator precedence languages}\label{subsec:OPE}
Next we introduce \emph{Operator Precedence Expressions (OPE)} as another formalism to define OPLs, equivalent to OPGs and MSO logic.
An OPE uses the same operations on strings and languages as Kleene's REs, and just one additional operation, called \emph{fence}, that selects from a language the strings  that correspond to a well-parenthesized string.
In the past, regular expressions of different kinds have been proposed for string languages more general than the finite-state ones (e.g.\ the cap expressions for CF languages \cite{Yntema}) or for languages made of  structures instead of strings, e.g., the  tree languages or the picture languages. 
Our OPEs have little  in common with any of them and, unlike regular expressions for tree languages \cite{DBLP:conf/caap/Thomas84}, enjoy in the context of OPLs the same properties as regular expressions in the context of regular languages.

We recall that an OPM $M$ defines a function from unparenthesized strings to their parenthesized counterparts; such a function is exploited in the following definition.
For convenience, we define the homomorphism (projection) $\eta : \Sigma_\#  \to \Sigma $ as:  $\eta(a)= a$, for $a\in \Sigma$, and $\eta(\#)= \varepsilon$.

\begin{defi}[OPE]\label{def:ope}
  Given an OP-alphabet $(\Sigma, M)$ whose OPM is total, an OPE $E$ and its language $L(E)  \subseteq \Sigma^*$ are defined as follows.
  The meta-alphabet of OPE uses the same symbols as regular expressions, together with the two symbols `[', and `]'.
  Let $E_1$ and $E_2$ be OPE:
\begin{enumerate}
    \item $a \in \Sigma$ is an OPE with $L(a) = a$.
    \item $\neg E_1$ is an OPE with $L(\neg E_1) = \Sigma^* - L(E_1)$.

   \item $a[E_1]b$, called the {\em fence}\/ operation, i.e., we say $E_1$ in the
     fence $a,b$, is an OPE with:\\
  if  $a, b \in \Sigma$:  $L(a[E_1]b) = a \cdot \{x \in L(E_1) \mid M(a \cdot x \cdot b) = \llp a \cdot M(x) \cdot b \rrp \} \cdot b$\\
  if  $a = \#, b \in \Sigma$: $L(\#[E_1]b) = \{x \in L(E_1) \mid M( x \cdot b) = \llp  M(x) \cdot b \rrp \} \cdot b$\\
  if $a \in \Sigma, b = \#$: $L(a[E_1]\#) = a \cdot \{x \in L(E_1) \mid M( a \cdot x ) = \llp a \cdot M(x) \rrp \}$\\
    where $E_1$ must not contain \#.   \item $E_1 \cup E_2$ is an OPE with $L(E_1 \cup E_2) = L(E_1) \cup L(E_2)$.

    \item \label{concat-OPE} $E_1 \cdot E_2$ is an OPE with $L(E_1 \cdot E_2) = L(E_1) \cdot L(E_2)$, where
    $E_1$ does not contain $a[E_3]\#$ and $E_2$ does not contain $\#[E_3]a$, for some OPE $E_3$, and $a \in \Sigma$.

    \item $E_1^*$ is an OPE defined by $E_1^* := \bigcup^{\infty}_{n = 0} E_1^n$, where $E_1^0 := \{\varepsilon\}$, $E^1_1 = E_1$, $E_1^n := E_1^{n-1} \cdot E_1$;
    $E_1^+ := \bigcup^{\infty}_{n = 1} E_1^n$.
\end{enumerate}

Among the operations defining OPEs, concatenation has the maximum precedence; set-theoretic operations have the usual precedences, the fence operation is dealt with as a normal parenthesis pair.

Similarly to the case of regular expressions,
a {\em star-free (SF) OPE}\/ is one that does not use the $^*$ and $^+$ operators.
\end{defi}

The conditions on \# are due to the peculiarities of OPLs closure w.r.t.\ concatenation (see also Theorem~\ref{concatenation-closure}).
In point~\ref{concat-OPE}. the $\#$ is not permitted within, say, the left factor $E_1$ because delimiters are necessarily positioned at the two ends of a string.

Besides the usual abbreviations for set operations (e.g., $\cap$ and $-$), we will also use the following derived operators:
\begin{itemize}
  \item $a \Delta b := a [ \Sigma^+ ] b$.
  \item $a \nabla b := \neg( a \Delta b ) \cap a \cdot \Sigma^+ \cdot b$.
\end{itemize}

It is trivial to see that the identity $a[E]b = a \Delta b \cap a \cdot E \cdot b$ holds.

The fact that in Definition~\ref{def:ope} the matrix $M$ is total is without loss of generality:
to obtain the same effect as $M_{a,b} = \emptyset$
for two terminals $a$ and $b$,  (i.e.\ that there should be a ``hole'' in the OPM for them), we can use
the short notations
\begin{align*}
\hole(a,b) &:= \neg (\Sigma^* (a b \cup a \Delta b) \Sigma^*),\\
\hole(\#,b) &:= \neg (\# \Delta b \Sigma^*), \
              \hole(a,\#) := \neg (\Sigma^* a \Delta \#)
\end{align*}
and intersect them with the OPE.

The following examples illustrate the meaning of the fence operation, the expressiveness of OPLs w.r.t.\ less powerful classes of CF languages, and how OPEs naturally extend regular expressions to the OPL family.

\begin{exa}\label{ex:simple-OPE}
Let $\Sigma$ be $\{a, b \}$, $\{a \lessdot a, a \doteq b, b \gtrdot b\} \subseteq M$. The OPE $a[a^{*}b^{*} ]b$ defines the language $\{a^{n}b^n \mid n \geq 1 \}$.
 In fact the fence operation imposes that any string $x \in a^{*}b^{*} $ embedded within the context $a,b$ be well-parenthesized according to $M$.

The OPEs $a[a^{*}b^{*} ]\#$ and $a^{+}a[a^{*}b^{*} ]b \cup \{a^{+}\}$, instead, both define the language $\{ a^n b^m  \mid n > m \geq 0\}$ since the matrix $M$ allows for, e.g., the string $aaabb$ parenthesized as $\llp a \llp a \llp ab \rrp b\rrp \rrp$.

If instead $\Sigma = \{a, b, c\}$, with $\{a \lessdot a, a \doteq b, a \doteq c, b \gtrdot b, b \gtrdot c, c \gtrdot b\} \subseteq M$, then both $a[a^{*}(bc)^{*} ]b$ and $a[(aa)^{*}(bc)^{*} ]b$ define the language  $\{a(a^{2n}(bc)^{n})b \mid n \geq 0 \}$. 

It is also easy to define Dyck languages with OPEs, as their parenthesis structure is
naturally encoded by the OPM\@.
Consider $L_{\text{Dyck}}$ the Dyck language with two pairs of parentheses denoted by $a, a'$ and $b, b'$. This language can be
described simply through a partial OPM, shown in Figure~\ref{fig:dyck} (left). In other words it is  $L_{\text{Dyck}} = L(G_{max, M})$ where $M$ is the matrix of the figure.
\begin{figure*}
\begin{centering}
\begin{tabular}{m{0.3\textwidth}m{0.3\textwidth}}
$
\begin{array}{c|ccccc}
   & a        & a'      & b        & b'      & \#      \\
\hline
a  & \lessdot & \dot=   & \lessdot &         &         \\
a' & \lessdot & \gtrdot & \lessdot & \gtrdot & \gtrdot \\
b  & \lessdot &         & \lessdot & \dot=   &         \\
b' & \lessdot & \gtrdot & \lessdot & \gtrdot & \gtrdot \\
\# & \lessdot &         & \lessdot &         & \doteq  \\
\end{array}
$
&
$
\begin{array}{c|ccccc}
   & a        & a'       & b        & b'       & \#      \\
\hline
a  & \lessdot & \dot=    & \lessdot & \gtrdot  & \gtrdot \\
a' & \lessdot & \gtrdot  & \lessdot & \gtrdot  & \gtrdot \\
b  & \lessdot & \gtrdot  & \lessdot & \dot=    & \gtrdot \\
b' & \lessdot & \gtrdot  & \lessdot & \gtrdot  & \gtrdot \\
\# & \lessdot & \lessdot & \lessdot & \lessdot & \doteq  \\
\end{array}
$
\end{tabular}
\end{centering}
\caption{The partial OPM defining $L_{\text{Dyck}}$ (left) and a possible completion $M_\text{complete}$ (right).}
\label{fig:dyck}
\end{figure*}
Given that, for technical simplicity, we use only total OPMs, we must refer to the one in Figure~\ref{fig:dyck} (right),
and state in the OPE that some OP relations are not wanted, such as $a, b'$, where the open and closed parentheses are of the wrong kind, or
$a, \#$, i.e.\ an open $a$ must have a matching $a'$.

The following OPE defines $L_{\text{Dyck}}$ by suitably restricting the ``universe'' $L(G_{max, M_\text{complete}})$:
  \begin{align*}
\hole(a, b') \cap
\hole(b, a') \cap
\hole(\#, a') \cap
\hole(\#, b') \cap
\hole(a, \#) \cap
    \hole(b, \#)
  \end{align*}
\end{exa}

\begin{exa}\label{ex:interrupt}

\begin{figure*}
$
\begin{array}{c|ccccc}
       & call     & ret  & int      & \#      \\
\hline
call   & \lessdot & \dot=   & \gtrdot  &    \\
ret    & \gtrdot  & \gtrdot & \gtrdot  & \gtrdot  \\
int    & \gtrdot  &         & \gtrdot  & \gtrdot \\
\#     & \lessdot &         & \lessdot &
\end{array}
$
\caption{The partial OPM $M_\text{int}$ for the OPE describing an interrupt policy.}\label{fig:interrupt}
\end{figure*}

  For a more application-oriented case, consider the classical LIFO
  policy managing procedure calls and returns but assume also that
  interrupts may occur: in such a case the stack of pending calls is
  emptied and computation is resumed from scratch.

  This policy is already formalized by the partial OPM of Figure~\ref{fig:interrupt},
  with $\Sigma = \{call, ret, int \}$
  with the obvious meaning of symbols.
  For example, the string $call$ $call$ $ret$ $call$ $call$ $int$ represents a run where only the second call returns, while the other ones are interrupted.
  In contrast, $call$ $call$ $int$ $ret$ is forbidden, because a return is not allowed when the stack is empty.

  If we further want to say that there must be at least
  one  procedure terminating regularly, we can use the OPE: $\Sigma^* \cdot call \Delta ret \cdot \Sigma^*$.

  Another example is the following, where we state that the run must contain at least one sub-run where
  no procedures are interrupted: $\Sigma^* \cdot \hole(call, int) \cdot \Sigma^*$.

  Notice that the language defined by the above OPE is not a VPL since
  VPLs allow for unmatched returns and calls only at the beginning or
  at the end of a string, respectively.
\end{exa}

\begin{thm} \label{Th:OPE-OPG}
For every OPE $E$ on an OP-alphabet $(\Sigma,M)$, there is an OPG $G$, whose OPM is compatible with $M$, such that $L(E) = L(G)$. 
\end{thm}

\begin{proof}
By induction on $E$'s structure. The operations $\cup, \neg, \cdot$, and $^*$ come from the closure properties of OPLs.
The only new case is $a[E]b$, with $a, b \in \Sigma_\#$, which is given by the following grammar. 

If, by induction, $G$ defines the same language as $E$, then, 
for every axiom $S_E$ of $G$ we add to $G$ the following rules, 
where $S$ is a new axiom replacing $S_E$, and $S$, $S'$ are nonterminals not used in $G$:

\begin{itemize}
    \item $S \to \eta(a) S_E \eta(b)$, if $a \doteq b$ in $M$;
    \item $S \to \eta(a) S'$ and $S' \to S_E \eta(b)$, if $a \lessdot b$ in $M$;
    \item $S \to S' \eta(b)$ and $S' \to \eta(a) S_E$, if $a \gtrdot b$ in $M$.
\end{itemize}
Notice that in the first bullet $a, b \in \Sigma$, while in the second and third bullets $a$ or $b$ could be $\#$.
Let us call this new grammar $G'$.
The grammar for $a[E]b$ is then the one obtained by applying the construction for intersection between $G'$ and the maxgrammar for $M$.
This intersection is to check that $a \lessdot \mathcal{L}(S_E)$ and $\mathcal{R}(S_E) \gtrdot b$; if it is not the case, 
according to the semantics of $a[E]b$, the resulting language is empty.
\end{proof}

Next we show that OPEs can express any language that is definable through an
MSO formula as defined in Section~\ref{sec:logic}. Thanks to the fact that the
same MSO logic can express exactly OPLs \cite{LonatiEtAl2015} and to Theorem~\ref{Th:OPE-OPG} we will obtain our first main result, i.e., the equivalence
of MSO, OPG, OP automata (see e.g., \cite{DBLP:journals/csr/MandrioliP18}), and
OPE.

In order to construct an OPE from a given MSO formula we follow the traditional path adopted for regular languages (as explained, e.g., in \cite{Pin-LogicOnWords}) and augment it to deal with the new $\pmb{x}_i \avv \pmb{x}_j$ relation.
     For a MSO formula $\varphi$, 
let $\pmb{x}_1, \pmb{x}_2, \ldots, \pmb{x}_r$ be the set of first order variables occurring
in $\varphi$, and $\pmb{X}_1, \pmb{X}_2, \ldots, \pmb{X}_s$ be the set of second order
variables. We use the new alphabet
$B_{p,q} = \Sigma \times \{0,1\}^p \times \{0,1\}^q$,
where $p \ge r$ and
$q \ge s$. The main idea is that the $\{0,1\}^p$ part of the alphabet is used to encode 
the value of the first order variables (e.g.\ for $p=r=4$, $(1,0,1,0)$ stands for both the positions $\pmb{x}_1$ and $\pmb{x}_3$), 
while the $\{0,1\}^q$ part of the alphabet is 
used for the second order variables. Hence, we are interested in the language $K_{p,q}$ formed by all strings 
where the components encoding the first order variables contain
exactly one occurrence of 1. 
We also use this definition $C_k := \{ c \in B_{p,q} \mid \text{ the }(k+1)\text{-th component of }c = 1\}$.

\begin{thm}
For every MSO formula $\varphi$ on an OP-alphabet $(\Sigma, M)$
there is a OPE $E$ on the same alphabet such that $L(E) = L(\varphi)$. 
\end{thm}

\begin{proof}
By induction on $\varphi$'s structure; the construction is standard for 
regular operations, the only difference is $\pmb{x}_i \avv \pmb{x}_j$.

Following B\"uchi's theorem,
we use the alphabet $B_{p,q}$ to encode interpretations of free variables.
The set $K_{p,q}$ of strings
  where each component encoding a first-order variable is such that there exists
  only one 1 is given by the following regular expression:
\[
    K_{p,q} = \bigcap_{1 \le i \le p} (B_{p,q}^* C_i B_{p,q}^* - B_p^* C_i B_{p,q}^* C_i B_{p,q}^*).
\]
Disjunction and negation are naturally translated into $\cup$ and $\neg$.
Like in B\"uchi's theorem, the expression $E$ for $\exists \pmb{x}_i \psi$ (resp.\ $\exists \pmb{X}_j \psi$) is obtained from expression $E_\psi$ for $\psi$, on an alphabet $B_{p,q}$,
by erasing by projection the component $i$ (resp.\ $j$) from the alphabet $B_{p,q}$.
The order relation $\pmb{x}_i < \pmb{x}_j$ is represented by $K_{p,q} \cap B^*_p C_i B^*_p C_j B^*_p$.

Last, the OPE for $\pmb{x}_i \avv \pmb{x}_j$ is  
$B_{p,q}^* C_i [ B_{p,q}^+ ] C_j B_{p,q}^*$.
\end{proof}

\section{Star-free OPEs are equivalent to FO logic}\label{sec:SF-FO}

After having completed the characterization of OPLs in terms of OPEs, we now
enter the analysis of the critical subclass of aperiodic OPLs: in this section
 we show that the languages defined by star-free OPEs coincide
with the FO-definable OPLs; in Section~\ref{NC-closure} that NC OPLs are closed
w.r.t.\ Boolean operations and concatenation and therefore SF OPEs define NC OPLs; 
in Section~\ref{sec:contrMSO} we provide a new characterization of OPLs in terms of MSO formulas by exploiting a control graph associated with a BDR OPG;
finally, in Section~\ref{sec:FO} we show that such MSO formulas can be made FO when the OPL is NC.

\begin{lem}[Flat Normal Form]\label{th:flat-normal-form}
  Any star-free OPE can be written in the following form, called {\em flat normal form:}
  \[
    \bigcup_i \bigcap_j \ t_{i,j}
  \]
  where the elements $t_{i,j}$ have either the form
  $L_{i,j} a_{i,j} \Delta b_{i,j}  R_{i,j}$, or
  $L_{i,j} a_{i,j} \nabla b_{i,j}  R_{i,j}$, or $H_{i,j}$,
  for $a_{i,j}$, $b_{i,j} \in \Sigma$, and $L_{i,j}$, $R_{i,j}$, $H_{i,j}$ star-free regular expressions. 
\end{lem}

\begin{proof}
  The lemma is a consequence of the distributive and De Morgan properties, together with the following identities, where $\circ_1, \circ_2 \in \{\Delta, \nabla\}$, and
  $L_k$ are star-free regular expressions, $1 \le k \le 3$:
  \[
    a[E]b = a \Delta b \cap a E b
  \]
  \[
    L_1 a_1 \circ_1 a_2 L_2 a_3 \circ_2 a_4 L_3 =
    L_1 a_1 \circ_1 a_2 L_2 a_3 \Sigma^+ a_4 L_3 \cap
    L_1 a_1 \Sigma^+ a_2 L_2 a_3 \circ_2 a_4 L_3 
  \]
  \[
    \neg (L_1 a_1 \Delta a_2 L_2) = L_1 a_1 \nabla a_2 L_2 \cup \neg (L_1 a_1 \Sigma^+ a_2 L_2)
  \]
    \[
    \neg (L_1 a_1 \nabla a_2 L_2) = L_1 a_1 \Delta a_2 L_2 \cup \neg (L_1 a_1 \Sigma^+ a_2 L_2)
  \]
  The first two identities are immediate, while the last two are based
  on the idea that
  the only non-regular constraints of the left-hand negations are
  respectively $a_1 \nabla a_2$ or $a_1 \Delta a_2$, that represent
  strings that are not in the set only because of their structure.
\end{proof}

\begin{thm}
For every FO formula $\varphi$ on an OP-alphabet $(\Sigma, M)$
there is a star-free OPE $E$ on $(\Sigma, M)$ such that $L(E) = L(\varphi)$. 
\end{thm}

\begin{proof}
  Consider the $\varphi$ formula, and its set of first order
  variables: like in Section~\ref{subsec:OPE},
  $B_p = \Sigma \times \{0,1\}^p$ (the $q$ components are absent,
  being $\varphi$ a first order formula), and the set $K_p$ of strings
  where each component encoding a variable is such that there exists
  only one $1$.

  First, $K_p$ is star-free:
  \[
    K_p = \bigcap_{1 \le i \le p} (B_p^* C_i B_p^* - B_p^* C_i B_p^* C_i B_p^*).
  \]

  Disjunction and negation are naturally translated into $\cup$ and $\neg$; 
  $\pmb x_i < \pmb x_j$ is covered by the star-free OPE $K_p \cap B^*_p C_i B^*_p C_j B^*_p$.

  The $\pmb x_i \avv \pmb x_j$ formula is like in the second order case, i.e.\ 
  is translated into $B_{p}^* C_i [ B_{p}^+ ] C_j B_{p}^*$, which is
  star-free.

  For the existential quantification, the problem is that star-free
  (OP and regular) languages are not closed under projections. Like in
  the regular case, the idea is to leverage the encoding of the
  evaluation of first-order variables, because there is only one
  position in which the component is $1$ (see $K_p$). 
  Hence, we can use the two
  bijective renamings
  $\pi_0(a,v_1, v_2, \ldots, v_{p-1}, 0) = (a,v_1, v_2, \ldots, v_{p-1})$, and
  $\pi_1(a,v_1, v_2, \ldots, v_{p-1}, 1) = (a,v_1, v_2, \ldots, v_{p-1})$, 
  where the last component is the one encoding the quantified
  variable. Notice that the bijective renaming does not change the
  $\Sigma$ component of the symbol, thus maintaining all the OP
  precedence relations.

  Let $E_\varphi$ be the star-free OPE on the alphabet $B_p$ for the
  formula $\varphi$, with $\pmb x$ a free variable in it. Let us assume
  w.l.o.g.\ that the evaluation of $\pmb x$ is encoded by the last component
  of $B_p$; let $B = \Sigma \times \{0,1\}^{p-1} \times \{0\}$, and
  $A = \Sigma \times \{0,1\}^{p-1} \times \{1\}$.

  The OPE for $\exists \pmb  x \varphi$ is obtained from the OPE
  for $\varphi$ through the bijective renaming $\pi$, and considering
  all the cases in which the symbol from $A$ can occur. 

  First, let $E'$ be a OPE in flat normal form, equivalent to
  $E_\varphi$ (Lemma~\ref{th:flat-normal-form}).
  The FO semantics is such that
  $L(\varphi) = L(E') = L(E') \cap B^* A B^*$.

  By construction, $E'$ is a union of intersections of
  elements $L_{i,j} a_{i,j} \Delta b_{i,j} R_{i,j}$, or
  $L_{i,j} a_{i,j} \nabla$ $b_{i,j} R_{i,j}$, or $H_{i,j}$, where
  $a_{i,j}$, $b_{i,j} \in \Sigma$, and $L_{i,j}$, $R_{i,j}$, $H_{i,j}$
  are star-free regular languages.

  In the intersection between $E'$ and $B^* A B^*$, all the possible
  cases in which the symbol in $A$ can occur in $E'$'s terms must be
  considered: e.g.\ in $L_{i,j} a_{i,j} \Delta b_{i,j} R_{i,j}$ it
  could occur in the $L_{i,j}$ prefix, or in $a_{i,j} \Delta b_{i,j}$,
  or in $R_{i,j}$.  More precisely,
  $L_{i,j} a_{i,j} \Delta b_{i,j} R_{i,j} \cap B^* A B^* =$
  $(L_{i,j} \cap B^* A B^*) a_{i,j} \Delta b_{i,j} R_{i,j} \cup $
  $L_{i,j} (a_{i,j} \Delta b_{i,j} \cap B^* A B^*)$ $R_{i,j}$
  $\cup L_{i,j} a_{i,j} \Delta $ $b_{i,j} (R_{i,j} \cap B^* A B^*)$ (the
  $\nabla$ case is analogous, $H_{i,j}$ is immediate, being regular star-free).
  
  The cases in which the symbol from $A$ occurs in
  $L_{i,j}$ or $R_{i,j}$ are easy, because they are by construction
  regular star-free languages, hence we can use one of the standard regular
  approaches found in the literature (e.g.\ by using the {\em splitting lemma}\/ 
  in \cite{Diekert-Gastin-first-orderdefinable}).
  The only differences are in the factors $a_{i,j} \Delta b_{i,j}$, or
  $a_{i,j} \nabla b_{i,j}$.
  
  Let us consider the case $a_{i,j} \Delta b_{i,j} \cap B^* A
  B^*$. The cases $a_{i,j} \in A$ or $b_{i,j} \in A$ are like
  $(L_{i,j} \cap B^* A B^*)$ and $(R_{i,j} \cap B^* A B^*)$, respectively,
  because $L_{i,j} a_{i,j}$ and $b_{i,j} R_{i,j}$ are also regular star-free
  ($\nabla$ is analogous).
  
  The remaining cases are $a_{i,j} \Delta b_{i,j} \cap B^+ A B^+$ and
  $a_{i,j} \nabla b_{i,j} \cap B^+ A B^+$. \\
  By definition of $\Delta$,
  $a_{i,j} \Delta b_{i,j} \cap B^+ A  B^+ = a_{i,j} [B^* A B^*]  b_{i,j}$, and 
  its bijective renaming is\\
  $\pi_0( a_{i,j} ) [ \pi_0(B^*) \pi_1(A) \pi_0( B^* )] \pi_0(b_{i,j}) = a'_{i,j} [B_{p-1}^+]  b'_{i,j}$, where
  $\pi_0(a_{i,j}) = a'_{i,j}$, and
  $\pi_0(b_{i,j}) = b'_{i,j}$,
 which is a star-free OPE\@.
  By definition of $\nabla$,
  $a_{i,j} \nabla b_{i,j} \cap B^+ A B^+ =$
  $\neg( a_{i,j} [B_p^+] b_{i,j} ) \cap a_{i,j}  B_p^+  b_{i,j} \cap B^+ A B^+ = $
  $\neg( a_{i,j} [B_p^+] b_{i,j} ) \cap a_{i,j}  B^* A B^*  b_{i,j}$.\\
  Its renaming is
  $\neg( \pi_0(a_{i,j})[\pi_0(B_p^*) \pi_1(B_p)\pi_0(B_p^*) ] \pi_0(b_{i,j}) )$
  $ \cap\  \pi_0(a_{i,j}  B^*) \pi_1(A)$  $\pi_0(B^*  b_{i,j}) = $\\
  $\neg( a'_{i,j} [B_{p-1}^+] b'_{i,j} ) \cap a'_{i,j} B_{p-1}^+  b'_{i,j}$,
  a star-free OPE.
\end{proof}

\begin{thm}
  For every star-free OPE $E$ on an OP-alphabet $(\Sigma, M)$, there is a FO
  formula $\varphi$ on $(\Sigma, M)$ such that $L(E) = L(\varphi)$.
\end{thm}
\begin{proof}
The proof is by induction on $E$'s structure. 
Of course, singletons are easily first-order definable; for negation and union we use $\neg$ and $\lor$
as natural.

Like in the case of star-free regular languages, concatenation is less immediate, and it is 
based on formula {\em relativization}. Consider two FO formulae $\varphi$ and $\psi$,
and assume w.l.o.g.\ that their variables are disjunct, and 
let $\pmb x$ be a variable not used in neither of them.
To construct a relativized variant of $\varphi$, called $\varphi_{< \pmb x}$, proceed 
from the outermost quantifier, going inward, and replace every subformula $\exists \pmb y \lambda$ with
$\exists \pmb y ((\pmb y < \pmb x) \land \lambda)$. 
Variants $\varphi_{\ge \pmb x}$ and $\varphi_{> \pmb x}$ are analogous. We also call $\varphi(\pmb x, \pmb y)$ the relativization where quantifications $\exists \pmb z \lambda$ are replaced by
$\exists \pmb z ((\pmb x < \pmb z < \pmb y) \land \lambda)$.
The language $L(\varphi) \cdot L(\psi)$ is defined by the following formulas:
$\exists \pmb x (\varphi_{< \pmb x} \land \psi_{\ge \pmb x})$ if $\varepsilon \not\in L(\psi)$; 
otherwise 
$\exists \pmb x (\varphi_{< \pmb x} \land \psi_{\ge \pmb x}) \lor \varphi$.

The last part we need to consider is the fence operation, i.e. $a [E] b$.
Let $\varphi$ be a FO formula such that $L(\varphi) = L_M(E)$, for a star-free OPE $E$. 
Let $\pmb x$ and $\pmb y$ be two variables unused in $\varphi$. 
Then the language $L(a [E] b)$ is the one defined by 
$
\exists \pmb x \exists \pmb y (a(\pmb x) \land b(\pmb y) \land \pmb x \avv \pmb y \land \varphi(\pmb x, \pmb y))
$.
\end{proof}

\section{Closure properties of non-counting OPLs and star-free OPEs}\label{NC-closure}
Thanks to the fact that an OPM implicitly defines the structure of an OPL, i.e., its parenthesization,
aperiodic OPLs inherit from the general class the same closure properties w.r.t.\ the basic algebraic operations. 
Such closure properties are proved in this subsection under the same assumption as in the general case 
(see Proposition~\ref{PropOPL}),
i.e., that \emph{the involved languages share the same total OPM or have compatible OPMs}.

\begin{thm}
\label{complement-closure}
Counting and non-counting parenthesis languages are closed w.r.t.\ complement.
Thus, counting and non-counting OPLs are closed w.r.t.\ complement w.r.t.\ the max-language defined by any OPM.
\end{thm}

\begin{proof}
We give the proof for counting languages which also implies the closure of non-counting ones.

By definition of counting parenthesis language and from Theorem 1
of~\cite{CreGuiMan78}, if $L_p$ is counting there exist strings $x, u, v, z, y$
and integers $n, m$ with $n > 1, m > 1$ such that $xu^{n+r}zv^{n+r}y \in L$ for
all $r = km > 0$ but not for all $r>0$. Thus, the complement of $L_p$ contains
infinitely many strings $xu^{n+i}zv^{n+i}y \not\in L_p$ but not all of them since for
some $i$, $i = km$. Thus, for $\neg L_p$ too there is no $n$ such that
$xu^{n}zv^{n}y \in L$ iff $xu^{n+r}zv^{n+r}y \in L$ for all $r \geq 0$. 

The same holds for the unparenthesized version of $L_p$ if it is an OPL.
\end{proof}

\begin{thm}
\label{union-closure}
Non-counting parenthesis languages and non-counting OPLs are closed w.r.t.\ union and therefore w.r.t.\ intersection.
\end{thm}

\begin{proof}
  Let $L_{p1}, L_{p2}$ be two NC parenthesis languages/OPLs.
  Assume by contradiction
  that $L_p = L_{p1} \cup L_{p2}$ be counting. Thus, there exist strings $x, u, v, z, y$
  such that for infinitely many $n$, $xu^nzv^ny \in L_p$ but for no $n$ $xu^nzv^ny
  \in L_p$ iff $xu^{n+r}zuv^{n+r}y \in L_p$ for all $r \geq 0$. Hence, the same
  property must hold for at least one of $L_{p1}$ and $L_{p2}$ which therefore would
  be counting. 
\end{proof}

Notice that, unlike the case of complement, counting languages are not closed
w.r.t.\ union and intersection, whether they are regular or parenthesis or OP
languages.

\begin{thm}
\label{concatenation-closure}
Non-counting OPLs are closed w.r.t.\ concatenation.
\end{thm}

\begin{proof}
  Recall from \cite{Crespi-ReghizziM12} that OPLs with compatible OPM are closed w.r.t.\ concatenation.	Thus, let $L_1, L_2$ be NC OPLs,  and  $G_1 = (\Sigma, V_{N1}, P_1, S_1)$, 
 $G_2 = (\Sigma, V_{N2},  P_2, S_2)$ their respective BDR OPGs. 
	  Let also $L_{p1}$, $L_{p2}$, be their respective parenthesized languages  and $G_{p1}$, $G_{p2}$, their respective parenthesized grammars. 	We also recall 
 that in general the parenthesized version $L_p$ of $L = L_1 \cdot L_2$ is not the parenthesized concatenation of
  the parenthesized versions of $L_1$ and $L_2$, i.e., 
	 $L_p$ may differ from  $\llp L'_{p1} \cdot L'_{p2} \rrp$,
     where $\llp L'_{p1}\rrp =L_{p1}$ and  $\llp L'_{p2}\rrp =L_{p2}$, because
         the OP concatenation may cause the syntax trees of $L_1$ and $L_2$ to coalesce.
  
 The construction given in \cite{Crespi-ReghizziM12} builds a grammar $G$ whose nonterminal alphabet includes $V_{N1}$, $V_{N2}$ and a set of pairs $[A_1, A_2]$ with $A_1 \in V_{N1}$, $A_2 \in V_{N2}$; the axioms of $G$ are the pairs $[X_1, X_2]$ with $X_1 \in S_1$, $X_2 \in S_2$.\footnote{This is a minor deviation from the formulation given in \cite{Crespi-ReghizziM12} since in that paper it was assumed that grammars have only one axiom.} 
 In essence (Lemmas 18 through 21 of \cite{Crespi-ReghizziM12}) $G$'s derivations are such that $[X_1, X_2] \xLongrightarrow[G]{*} x[A_1,A_2]y $, $[A_1,A_2] \xLongrightarrow[G]{*} w$ implies
 $w = w_1 \cdot w_2$ for some $w_1, w_2$ and $X_1\xLongrightarrow[G_1]{*} xA_1$, $A_1 \xLongrightarrow[G_1]{*} w_1$, $X_2\xLongrightarrow[G_2]{*} A_2y$, $A_2 \xLongrightarrow[G_2]{*} w_2$. 
 Notice that some substrings of $x \cdot w_1 $, resp.\ $w_2 \cdot y $, may be derived from nonterminals belonging to $V_{N1}$, resp.\ $V_{N2}$, 
 as the consequence of rules of type $[A_1, A_2] \to \alpha_1 [B_1, B_2] \beta_2$ with $\alpha_1 \in V_1^*$, $\beta_2 \in V_2^*$, where $[B_1, B_2]$ could be missing;
 also, any string $\gamma$ derivable in $G$ contains at most one nonterminal of type $[A_1,A_2]$ (see Figure~\ref{fig:split-derivation}).
 
\begin{figure}[h]
   \centering
   \includegraphics[scale=0.4]{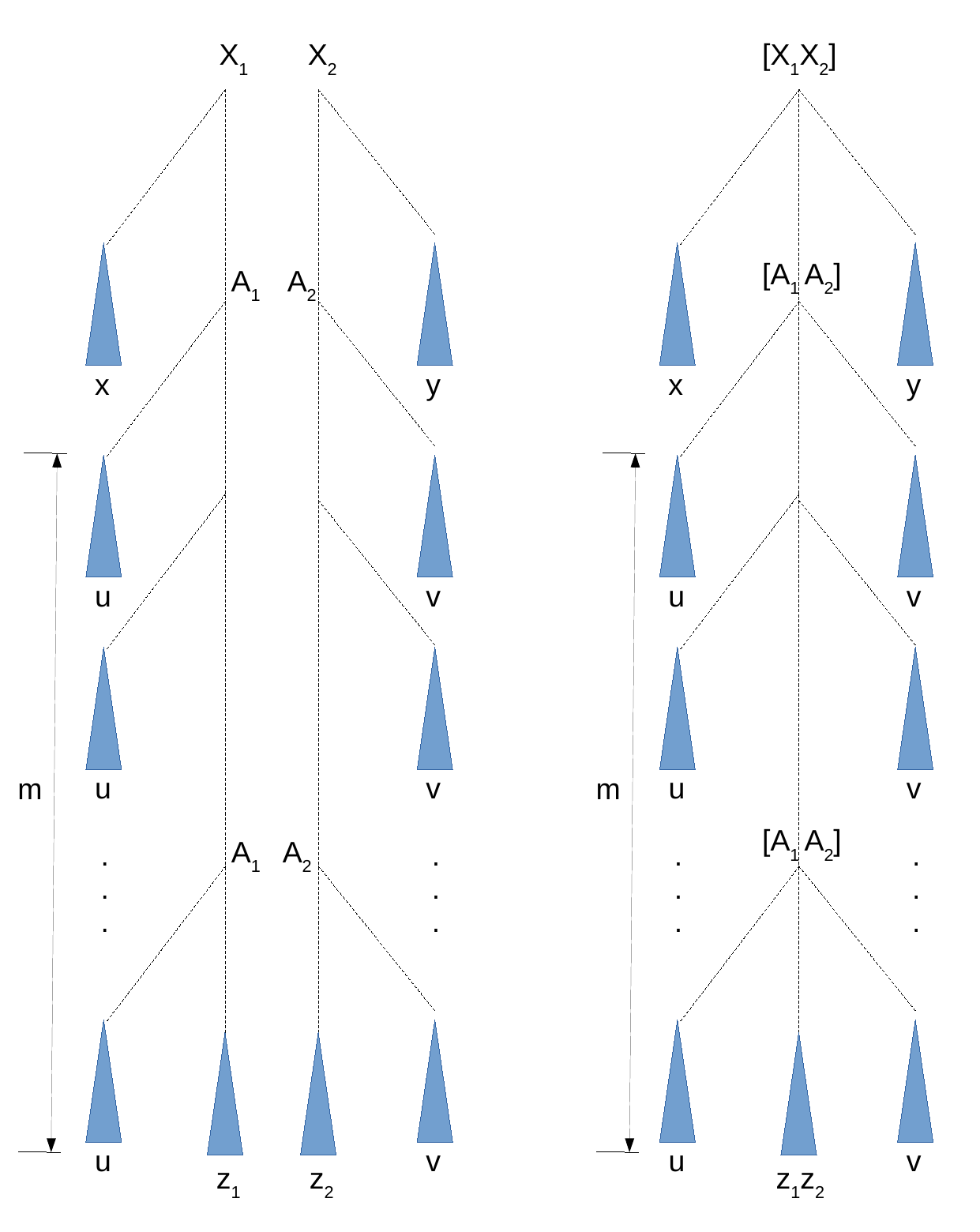}
 \caption{An example of paired derivations combined by the concatenation construction. In this case the last character of $u$ is in $\doteq $ relation with the first character of $v$.}
\label{fig:split-derivation}
\end{figure}
 
 Suppose, by contradiction, that $G$ has a counting derivation%
\footnote{Note that the $G$ produced by the construction is BD if so are $G_1$ and $G_2$, but it could be not necessarily BDR; however, if a BDR OPG has a counting derivation, any equivalent BD grammar too has a counting derivation.} 
 $[X_1, X_2] \!\xLongrightarrow[G]{*}$ $x[A_1,A_2]y \!\xLongrightarrow[G]{*}$ $xu^m[A_1,A_2]v^my \xLongrightarrow[G]{*}  xu^mzv^my$ (one of $u^m$, $v^m$ could be empty) whereas $[A_1, A_2]$ does not derive $u[A_1,A_2]v$: this would imply the derivations 
 $A_1 \xLongrightarrow[G_1]{*}  u^mA_1$, $A_2 \xLongrightarrow[G_2]{*}  A_2v^m$ which would be counting in $G_1$ and $G_2$ since they would involve the same nonterminals in the pairs $[A_i, A_j]$.
 Figure~\ref{fig:split-derivation} shows a counting derivation of $G$ derived by the concatenation of two counting derivations of $G_1$ and $G_2$; in this case neither $u^m$ nor $v^m$ are empty.
 
 If instead the counting derivation of $G$ were derived from nonterminals belonging to $V_{N1}$, (resp.\ $V_{N2}$) that derivation would exist identical for $G_1$ (resp.\ $G_2$).
\end{proof}

Thanks to the above closure properties we deduce the following important property of OPEs.

\begin{thm}\label{lem:SF-OPEs}
The OPLs defined through star-free OPEs are NC.
\end{thm}

\begin{proof}
  Thanks to Lemma~\ref{th:flat-normal-form} we only need to consider OPEs in
  flat normal form: they consist of star-free regular expressions combined
  through Boolean operations and concatenation with $a\Delta b$ and $a\nabla b$
  operators. $a\Delta b$ = $a[\Sigma^+]b$ is obviously NC; $a\nabla b$ is the
  intersection of the negation of $a\Delta b$ with the regular star-free
  expression $a\Sigma^+ b$. Thanks to the above closure properties of NC OPLs,
  star-free OPEs are NC. 
\end{proof}

\section{From grammar to logic through control graph}\label{sec:contrMSO}

In this cornerstone section we show how any OPL can be expressed as a combination of a ``skeleton
language" ---the max-language associated with the OPM---  with a ``regular control". 
Such a regular control, defined through a graph derived from the OPG, can be translated in the traditional way into MSO formulas, which become FO if the language defined by the graph is non-counting \cite{McNaughtPap71}.
These formulas, suitably complemented by the $\avv$ relation, express the language generated by the source OPG.

The following definition of \emph{control graph}
associates a regular language with every nonterminal symbol of the grammar.

\begin{defi}[control graph]
    \label{def:aut:G}
Let $G = (\Sigma, V_N, P, S)$ be an OPG\@. The {\em control graph of $G$}, 
denoted by $\mathcal{C}(G) = (Q, \Sigma, \pmb\delta)$, is the graph having 
vertices or states $Q$ and relation $\pmb\delta$ (see Section~\ref{subsec:FA}) defined as follows:
\begin{itemize}
    \item $Q = \dd V_N \cup \uu V_N$, where $\dd V_N$ (resp.\ $\uu V_N$) = $\{\dd A$ (resp.\ $ \uu
        A$) $\mid A \in V_N\}$. 
\item 
Let $W$ be the set:
\begin{equation}\label{eq:setW}
W = \left\{
w \in \Sigma^+ \mid
\begin{array}{l} \exists A \to \beta w \gamma \in P,
\\
\beta \in V^*\cdot V_N  \text{\ or\ } \beta=\varepsilon, 
\\
\gamma \in V_N \cdot V^* \text{\ or\ } \gamma =\varepsilon
\end{array}
\right\}.
\end{equation}
\noindent The \emph{macro-edges}  of $\pmb\delta$ are 
associated with the productions  according to the following table, where $w \in W$, $\alpha, \zeta \in V^*$:
\[
			\begin{array}{l|l}
				\text{rule} & \text{edge} 
				\\\hline
                A \to B \zeta & \dd A \boldtrans{\varepsilon}  \dd B 
				\\
                A \to w B \zeta & \dd A \boldtrans{w}  \dd B 
				\\			\hline	
                A \to \alpha B  & \uu B \boldtrans{\varepsilon} \uu A 
				\\
                A \to \alpha B w  & \uu B \boldtrans{w}  \uu A  
				\\	\hline
				A \to \alpha B w C \zeta  &	\uu B 	\boldtrans{w}  \dd C 
				\\\hline
				 A \to w & \dd A \boldtrans{w} \uu A 
				\\	\hline	
			\end{array}
				\]
\end{itemize}

For a given control graph, the regular languages consisting in the paths going from state to state are named \emph{control languages}; 
in particular, for any grammar nonterminal $A$, we will denote the set $\{ x \mid \dd A \boldtrans{x} \uu A\}$ as $R_A$,
where, with no risk of ambiguities, we use the same arrow to denote a single macro-edge and a  whole path of the graph. 
\end{defi}

The adoption of macro-steps to define a control graph
 allows us to state an  immediate correspondence between the terminal parts of grammar rules and graph macro-edges, without introducing useless intermediate steps. 

Intuitively, a state of type $\dd A$ denotes that a path of the control graph
visiting the syntax tree of a string generated by $G$ is touching the
nonterminal $A$ while following a top-down direction; conversely, it visits
$\uu A$ while following a bottom-up direction. We thus call those states, \emph{descending and ascending states} respectively.

We will see (Theorem~\ref{Th:GtoMSO}) that the frontier of a syntax tree rooted in nonterminal $A$ is
a path of the control graph, going from $\dd A$ to $\uu A$ (of course, 
such paths being regular languages, they also include strings that are not in
$L_G(A)$).

\begin{exa}
    \label{ex:contr-graph}
Consider the following OPG $G_{NL}$, with $S = \{A, B\}$.
\[
A \to 
a B c A \mid
a B c B \mid ac, \ 
B \to 
b A c A \mid 
b A c B \mid 
bc
\]
Its control graph $\mathcal{C}(G_{NL})$ is given in Figure~\ref{C-graphGNL}.

\begin{figure}
  \centering
  \includegraphics[scale=0.6]{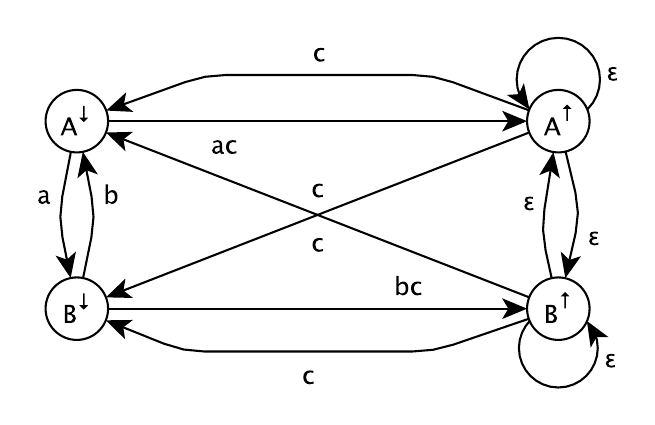}
  \caption{The control graph of $G_{NL}$}
    \label{C-graphGNL}
\end{figure}

\end{exa}


\subsection{Deriving MSO formulas from the control graph}\label{subsec:CGtoMSO}
We already know that the MSO logic defined in Section~\ref{sec:logic} as an extension of the traditional logic for regular languages defines exactly the family of OPLs. In this section we show a way to obtain an MSO formula equivalent to an OPG directly from its control graph: the final goal is to obtain from such a construction an FO formula instead of an MSO one in the case that the OPL is aperiodic.

Intuitively the $\avv$ relation, which is the only new element w.r.t.\ the traditional MSO logic for regular languages, ``embraces'' the string $x$ generated by some grammar nonterminal $A$,
thus it must be the case that $\dd A \boldtrans{x} \uu A$. Next we provide the details of the MSO construction.

First, we resume from previous papers about logic characterization of OPL \cite{LonatiEtAl2015,LMPP15} 
the following  $\treec$ formula which
states that the positions $ \g{x}_{1}, \ldots, \g{x}_{n}$, with $n \geq 1$, of a string 
are, in order, the positions of the terminal characters of a grammar rule rhs and 
$ \g{x}_{0}, \g{x}_{n+1}$ are the positions of the character immediately at the left and immediately at the right of the subtree generated by that rule:
\begin{equation}\label{eq:treec}
  \begin{array}{l}
\treec(\g{x}_{0}, \g{x}_{1}, \ldots, \g{x}_{n}, \g{x}_{n+1}) := \\
\g{x}_0 \avv \g{x}_{n+1} 
\land
 \bigwedge_{0 \leq i \leq n}  \left(
\begin{array}{c}
 \g{x}_{i}+1=\g{x}_{i+1} \\ 
 \lor\\ 
 \g{x}_{i} \avv \g{x}_{i+1}
\end{array}
\land
\bigwedge_{i+1 < j \leq n} \neg (\g{x}_{i} \avv \g{x}_{j})
    \right)
  \end{array}
\end{equation}

\begin{figure}
  \centering
\includegraphics[scale=0.35]{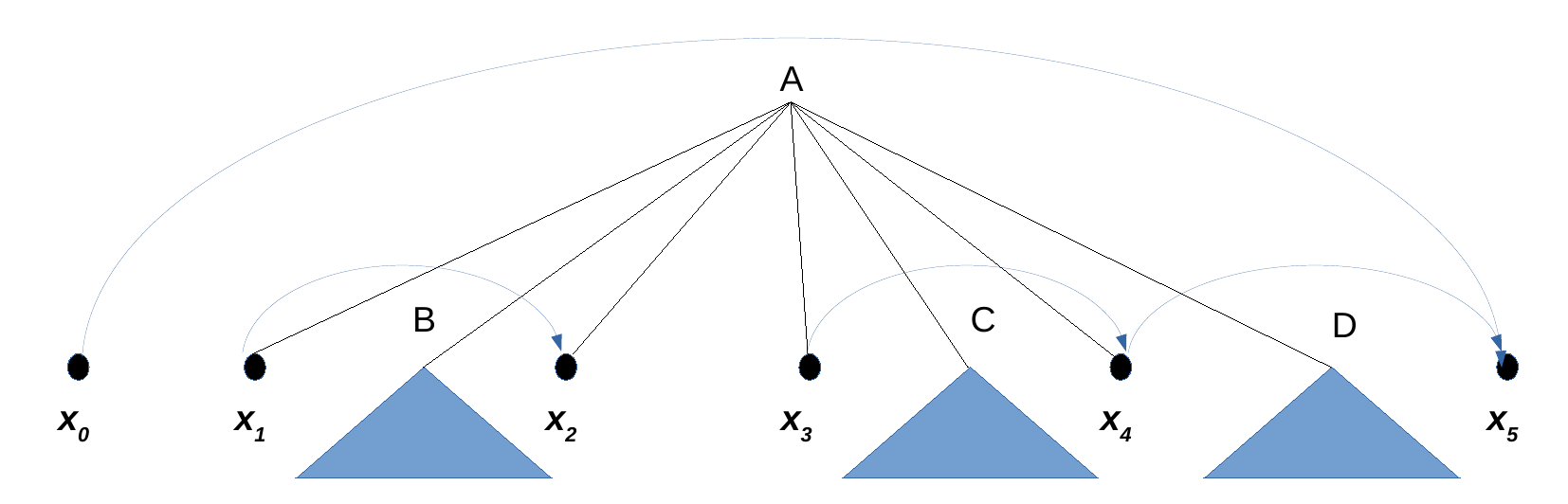}
  \caption{An example of the TreeC relation for a rule $A \to a B b c C d D$
    (with $a(\g{x}_1)$, $b(\g{x}_2)$, $c(\g{x}_3)$, $d(\g{x}_4)$).}
    \label{fig:TreeC}
\end{figure}

Figure~\ref{fig:TreeC} shows an example of the TreeC relation.

For any nonterminal $A$, let $\varphi_{A}$ be the MSO formula defining the regular language $R_A = \{ x \mid \dd A \boldtrans{x} \uu A\}$;
let $\varphi_A(\g{x},\g{y})$ be its relativization w.r.t.\ the new free variables $\g{x},\g{y}$, i.e.,
the formula obtained by replacing every subformula  $\exists \g z \lambda$ with
$\exists \g z ((\g x < \g z < \g y) \land \lambda)$. 

%
%

The following key formula $\psi_{A}$ states that for every pair of positions $\g{x} \avv \g{y}$, if $z$ 
is the string between the two positions, and 
$\dd A \boldtrans{z} \uu A$, then there must exist a rule of $G$ with $A$ as lhs,
and a rhs such that for all of its nonterminals $B_j$, if any, formula $\varphi_{B_j}$ holds.

\begin{equation}\label{eq:psi}
  \begin{array}{l}
\psi_A :=
\forall \g x, \g y  
\left(
 \begin{array}{c}
\varphi_A(\g{x}, \g{y})  \land \g{x} \avv \g{y}\\
\Rightarrow \\
   \displaystyle{\bigvee_{A \to B_{0} c_1 B_{1} c_2 
                             \ldots c_n B_{n}
   }} 
\! \exists \g{x}_1 \ldots \g{x}_{n} 
\left(  
 \begin{array}{c}
 \treec(\g{x}, \g{x}_{1}, \ldots, \g{x}_{n}, \g{y}) \ \land \\
  \displaystyle{\bigwedge_{1 \leq i\leq n} c_i(\g x_i) } \  \land \\
  \displaystyle{\bigwedge_{\substack{1 \leq j\leq n-1:\\B_{j} \neq \varepsilon}} \varphi_{B_{j}} (\g x_j, \g{x}_{j+1})} \ \land \\  
  \g x + 1 \neq \g x_1 \Rightarrow \varphi_{B_{0}} (\g x,\g x_1) \ \land \\
  \g x_n + 1 \neq \g y \Rightarrow \varphi_{B_{n}} (\g x_n,\g y)
 \end{array}
 \right)
 \end{array}
    \right)
  \end{array}
\end{equation}
where the disjunction is considered over the rules of $G$ and $B_{j}$ are either $\varepsilon$ or are the nonterminals occurring in the rhs of the production.

Finally, $\chi_G$ states that the strings included between \# must be derived by some axiom:
\begin{equation}\label{eq:chi}
  \chi_G :=
\bigwedge_{A \in V_N }\psi_A
 \land 
\exists \ee \left(  
\#(\ee+1) \land \neg \exists \g{y} (\ee + 1< \g{y}) 
\land
\displaystyle{\bigvee_{A \in S} \varphi_{A}(0,\ee+1) }
\right)
\end{equation}

\begin{exa}\label{ex:MSOformula}
Consider again the OPG $G_{NL}$ of Example~\ref{ex:contr-graph}.

Let $\varphi_A$ and $\varphi_B$ be the MSO formulas defining the regular languages $R_A$ and $R_B$, and $\varphi_A(\g{x}, \g{y})$ and $\varphi_B(\g{x}, \g{y})$ their respective relativized versions.
Then the $\psi_A$ formula for nonterminal $A$ of $G_{NL}$ is:

\begin{equation}\label{eq:psi-GNL}
  \begin{array}{l}

\forall \g x, \g y  
\left(
 \begin{array}{c}
\varphi_A(\g{x}, \g{y})  \land \g{x} \avv \g{y}
\Rightarrow \\
\! \exists \g{x}_1 , \g{x}_{2} 
\left(  
 \begin{array}{c}
 \treec(\g{x}, \g{x}_{1},  \g{x}_{2}, \g{y}) \ \land \\
   a(\g x_1) \land c(\g x_2) \  \land 
   \varphi_B (\g {x}_{1}, \g{x}_{2}) \ \land  \varphi_A (\g {x}_2, \g{y}) \land \\  
  \g x + 1 = \g x_1 \\
  
 \end{array}
 \right)
 \lor
 \\
 \! \exists \g{x}_1 , \g{x}_{2} 
\left(  
 \begin{array}{c}
 \treec(\g{x}, \g{x}_{1},  \g{x}_{2}, \g{y}) \ \land \\
   a(\g x_1) \land c(\g x_2) \  \land 
   \varphi_B (\g {x}_{1}, \g{x}_{2}) \ \land  \varphi_B (\g {x}_2, \g{y}) \land \\  
  \g x + 1 = \g x_1 \\
 \end{array}
 \right)
 \lor
 \\
 \! \exists \g{x}_1 , \g{x}_{2} 
\left(  
 \begin{array}{c}
 \treec(\g{x}, \g{x}_{1},  \g{x}_{2}, \g{y}) \ \land \\
   a(\g x_1) \land c(\g x_2) \  \land 
   
  \g x + 1 = \g x_1 \land  \g x_1 + 1 = \g x_2 \land  \g x_2 + 1 = \g y\\
 \end{array}
 \right)
 \end{array}
    \right)
  \end{array}
\end{equation}

We purposely avoided some obvious simplifications to emphasize the general structure of the $\psi$ formula.

\end{exa}

\begin{thm}[Regular Control]\label{Th:GtoMSO}
Let $G = (\Sigma, V_N, P, S)$ be a BDR $(\Sigma, M)$-compatible OPG,
$\mathcal{C}(G)$ its control graph, $\psi_A$ the formula \eqref{eq:psi} defined above for each $A\in V_N$.
Then, for any $A \in V_N$, $x \in L(A)$ if and only if $\# x \# \vDash \varphi_{A} (0,|x|+1)
\land \psi_A $.
\end{thm}

\begin{proof}
  First of all, we note that $\dd A \boldtrans{x} \uu A$ iff $\# x \#
  \vDash \varphi_A(0,|x|+1)$, i.e. $R_A = \{ x \mid \# x \# \vDash \varphi_A(0,|x|+1)\}$,
  by construction of $\mathcal{C}(G)$ and of $\varphi_A$.

  \smallskip 
  \noindent The proof is by induction on the height $m$ of the syntax trees rooted in $A$. 
  
  \noindent {\bf Base}: $m = 1$. If $A \xLongrightarrow[G]{} x$, with
  $x = c_1 \dots c_n$, i.e. $A \to x$ is a production of $G$, then
  $\# x \# \vDash \treec(0, 1 \ldots, n+1)$ and $\# x \# \vDash c_i(i)$ for
  every $i = 1 \ldots n$. Also, it is $\dd A \boldtrans{x} \uu A$,
  by construction of $\mathcal{C}(G)$. Hence, $\# x \# \vDash \varphi_{A} (0,|x|+1) \land \psi_A $.
  
  Conversely, we have $\# x \# \vDash \varphi_{A} (0,|x|+1) \land \psi_A $, with
  $x = \# \lessdot c_1 \doteq c_2 \doteq \ldots c_n \gtrdot \#$.
  Therefore: (i) $x \in R_A$, (ii) $\# x \# \vDash 0 \avv |x|+1$, and (iii) $\# x \# \vDash c_i(i)$ for
  every $i = 1 \ldots n$. (ii) and (iii) imply that there exists a production $B
  \to x$, but being $G$ BDR, $B$ must be $A$. Hence, $x \in L(A)$.

  \noindent {\bf Induction}: $ m > 1$. 
    Let us consider any $A \to B_0 c_1 B_1 \ldots c_n B_n \in P$,
    $c_i \in \Sigma$, where some $B_i$ could be absent --- we assume
    for simplicity that they are all present; the case where some of them are missing can be promptly adapted.

    \noindent
    {\bf Case $A \xLongrightarrow[G]{} B_0 c_1 B_1 \ldots c_n B_n
      \xLongrightarrow[G]{*} w_0 c_1 w_1 c_2 w_2 \ldots c_n w_n =  x$ implies
      $\# x \# \vDash \varphi_{A} (0,|x|+1) \land \psi_A $}.
   Induction hypothesis: for each $i = 0 \ldots n$, $B_i \xLongrightarrow[G]{*} w_i$
   implies $\# w_{i} \# \vDash \varphi_{B_i}(0, |w_i|+1) \land \psi_{B_i}$.

   Let $\g x_i$ be the position of $c_i$ in $\# x \#$ (i.e. $\# x \# \vDash
   c_i(\g x_i)$), $i = 1 \ldots n$.
   Being $A \xLongrightarrow[G]{} B_0 c_1 B_1 \ldots c_n B_n
   \xLongrightarrow[G]{*} w_0 c_1 w_1 c_2 w_2 \ldots c_n w_n =  x$, the
   structure of $x$ is such that
   $\# \lessdot w_0 \gtrdot c_1 \lessdot w_1 \gtrdot \ldots c_n \lessdot w_n
   \gtrdot\#$. Hence, $\# x \# \vDash \g x_{i-1} \avv \g x_i$, $i = 1 \ldots n$,
   and $0 \avv |x|+1$.
   By construction of $\mathcal C(G)$,
   $\dd A \boldtrans{\varepsilon} \dd B_0$,
   $\uu B_{i-1} \boldtrans{c_i} \dd B_i$, $i = 1 \ldots n$,
   $\uu B_n \boldtrans{\varepsilon} \uu A$, 
   so we have $\dd A \boldtrans{x} \uu A$. This means
   $\# x \# \vDash \varphi_{A}(0, |x|+1)$, and that the left-hand side of the implication in $\psi_A$ is true.  
   By induction hypothesis, $\# w_{i} \# \vDash \varphi_{B_i}(0, |w_i|+1)$ implies
   $\# x \# \vDash \varphi_{B_i}(\g x_i, \g x_{i+1})$; also,
   $\# x \# \vDash \varphi_{B_0}(0, \g x_{1})$ and 
   $\# x \# \vDash \varphi_{B_n}(\g x_n, |x|+1)$.
   Hence, 
   $\# x \# \vDash \treec(0, \g{x}_1 \dots  \g{x}_n, |x|+1)$. 
   Therefore, the right-hand side of the implication of $\psi_A$ is also true, where the big-$\lor$ is satisfied with
   the production $A \to B_0 c_1 B_1 \ldots c_n B_n$.
   Hence,  $\# x \# \vDash \varphi_{A} (0,|x|+1) \land \psi_A $.

   \noindent {\bf Case
     $\# x \# \vDash \varphi_{A} (0,|x|+1) \land \psi_A $
   implies
   $A \xLongrightarrow[G]{} B_0 c_1 B_1 \ldots $ $c_n B_n \xLongrightarrow[G]{*}$ $w_0 c_1 w_1 c_2 w_2$ $\ldots c_n w_n =  x$}.
 Induction hypothesis: for each $i = 0 \ldots n$,
 $\# w_{i} \# \vDash \varphi_{B_i}(0, |w_i|+1) \land \psi_{B_i}$
 implies
 $B_i \xLongrightarrow[G]{*} w_i$.

The hypothesis $\# x \# \vDash \varphi_{A} (0,|x|+1) \land \psi_A $ guarantees that for at least one rule of $G$, $A \to B_{0} c_1 B_{1} c_2 \ldots c_n B_{n}$ among $x$'s positions there exist $\g{x}_1 \ldots \g{x}_{n}$ such that 
$\# x \# \vDash \treec(0, \g{x}_1 \dots  \g{x}_n, \allowbreak |x|+1)$ and $c({\g{x}_i}) = c_i \mid i = 1 \dots n$. Thus $x= w_0c_1 \dots c_nw_n$ and, by the induction hypothesis, for each $i = 0 \ldots n$, there exist unique $B_i$ such that $B_i \xLongrightarrow[G]{*} w_i$. Since $G$ is BDR we conclude that $A$ is the unique nonterminal of $G$ such that $A \xLongrightarrow[G]{*} x$.
\end{proof}

From Theorem~\ref{Th:GtoMSO} we immediately derive the following main

\begin{cor} \label{cor:GtoMSO}
For any BDR $(\Sigma, M)$-compatible OPG $G$, $L(G)$ is the set of strings satisfying the corresponding formula $\chi_G$.
\end{cor}

In a sense, the above formula $\psi_{A}$ ``separates'' the formalization of the language structure defined by the OPM from that of the strings generated by the single nonterminals: 
the former part ---i.e., the $\avv$ relation and the $TreeC$ subformula--- are first-order. 
It is well-known from the classic literature \cite{McNaughtPap71} that NC regular languages can be defined by means of FO formulas. Thus, subformulas $\varphi_{A}$ of \eqref{eq:psi-GNL}, can be made FO if the regular control languages $R_A$ are NC\@. Thus,
we obtain a first important result:

\begin{cor}\label{cor:suff-cond-NC}
If the control graph of an OPG $G$ defines languages $R_A$, $A$ denoting any nonterminal character of $G$, that are all NC, then, $L(G)$ can be defined through an FO formula.
\end{cor}

The following example, besides illustrating the application of Theorem \ref{Th:GtoMSO} and its corollaries, presents an OPL version of a tree language that has been shown to be not definable through the FO restriction of the MSO logic for tree languages \cite{DBLP:journals/tcs/Potthoff94}. 
In contrast, formula \eqref{eq:chi} gives an FO-definition for the OPL version.

\begin{exa}\label{ex:Potthoff}
The OPG $G_\text{Logic}$, with terminal alphabet $\Sigma_{\llp \rrp} = \{\llp, \rrp, \land, \lor, 0, 1\}$ presented in Figure~\ref{fig:Glogic}, defines the language of fully parenthesized logical sentences making use of the $\land$ and $\lor$ operators only, that evaluate to $true$.
\begin{figure}
$
 	\begin{array}{ll}
&	S = \{ T \} \\
 	&	T \to \llp F \lor T \rrp \mid \llp T \lor  F\rrp \mid \llp T \lor  T \rrp \mid \llp T \land  T \rrp \mid 1  \\  	
 	&	F \to \llp T \land F\rrp \mid \llp F \land T \rrp  \mid \llp F \land F \rrp \mid \llp F \lor F \rrp \mid 0\\
 	\end{array}
  $
\hspace{0.5cm}
$
		\begin{array}{c|cccccccc}
		&\lor &\land & \llp & \rrp & 1 & 0 & \#  \\
		\hline
		\lor &  & &\lessdot &\doteq &\lessdot &\lessdot &  \\
		\land &  & &\lessdot &\doteq &\lessdot &\lessdot &   \\
		\llp & \doteq &\doteq &\lessdot & &\lessdot &\lessdot \\
		\rrp & \gtrdot &\gtrdot & &\gtrdot & & &\gtrdot\\
		1 & \gtrdot &\gtrdot & &\gtrdot & &  &\gtrdot\\
        0 & \gtrdot &\gtrdot & &\gtrdot & &  &\gtrdot\\
        \# & & & \lessdot & & \lessdot& \lessdot & \doteq
		\end{array}
		$
\caption{$G_\text{Logic}$ (left) and its OPM (right).}\label{fig:Glogic}
\end{figure}

Clearly the parenthesized sentences generated by the two nonterminals of $G_\text{Logic}$\footnote{Strictly speaking $G_\text{Logic}$ is not a parenthesis grammar since we omitted useless parentheses for the rhs $1$ and $0$.} are isomorphic to their STs (once the internal nodes are anonymized) and to the trees of the tree language defined on the alphabet
$\Sigma = \{\land, \lor, 0, 1\}$
partitioned into $\Sigma_0 = \{ 0, 1\}$ and $\Sigma_2 = \{\land, \lor\}$ where the indexes of the two subsets denote their arity. Furthermore, the sentences generated by the axiom $T$ are isomorphic to the set of trees that evaluate to $1$.

To give an intuition why this language is not FO definable using tree languages, we can refer to
\cite{DBLP:journals/ita/Heuter91}, where it is proved that ``a tree language is first-order definable
if and only if it is built up from finite set of special trees using the operations union, complement and concatenation,
all restricted to the class of special trees.'' 
{\em Special trees} are trees which can be labeled at the frontier with a single occurrence of a special symbol
(not in $\Sigma$) used for concatenation: two trees are concatenated by appending the second one to the first one in place of this special symbol. 
Intuitively, this kind of concatenation allows for a structure which is analogous to linear CF grammars, while
$G_\text{Logic}$ is clearly not linear.

\begin{figure}
\begin{tikzpicture}[
    scale=0.5,
    level 1/.style={sibling distance=5em},
    level 2/.style={sibling distance=3em},
    ]
\node{$T$}
    child{ node{$\llp$} }
    child{ node{$T$} child { node{$1$} } }
    child{ node{$\land$} }
    child{ node{$T$} 
        child{ node{$\llp$} }
        child{ node{$F$} child { node{$0$} }}
        child{ node{$\lor$} }
        child{ node{$T$} child { node{$1$} }}
        child{ node{$\rrp$} }
    }
    child{ node{$\rrp$} };
\end{tikzpicture}
\hspace{1cm}
\begin{tikzpicture}[
    scale=0.73,
    ]
\node{$\land$}
    child{ node{$1$} }
    child{ node{$\lor$} 
        child{ node{$0$} }
        child{ node{$1$} }
    };
\end{tikzpicture}
\hspace{1cm}
\begin{tikzpicture}[flush/.style={double, >=stealth, thin, rounded corners}]
\matrix (m) [matrix of nodes]
 {\# & $\llp$ & $1$ & $\land$ & $\llp$ & $0$ & $\lor$ & $1$ & $\rrp$ & $\rrp$ & \# \\
    0 & 1 & 2 & 3 & 4 & 5 & 6 & 7 & 8 & 9 & 10  \\
 };

\draw[->] (m-1-2)  to [out=60, in=120] (m-1-4);
\draw[->] (m-1-5)  to [out=60, in=120] (m-1-7);
\draw[->] (m-1-7)  to [out=60, in=120] (m-1-9);
\draw[->] (m-1-4)  to [out=60, in=120] (m-1-10);
\draw[->] (m-1-1)  to [out=60, in=120] (m-1-11);

\end{tikzpicture}
\caption{The ST of the $G_\text{Logic}$'s sentence  $\llp1 \land \llp0 \lor 1\rrp\rrp$ (left), the corresponding tree of the tree language (center),
and the $\avv$ relation for the string $\llp1 \land \llp0 \lor 1\rrp\rrp$ (right).}\label{fig:SynTree}
\end{figure}

Figure~\ref{fig:SynTree} (left) displays the ---only--- ST that the grammar associates to the string $\llp1 \land \llp0 \lor 1\rrp\rrp$, the corresponding tree in the tree language (center), and (right) the corresponding $\avv$ relation which illustrates the meaning of the $\treec$ formula. 
Figure~\ref{fig:CG} displays the control graph of the grammar.

\begin{figure}[h]
\centering\includegraphics[scale=0.6]{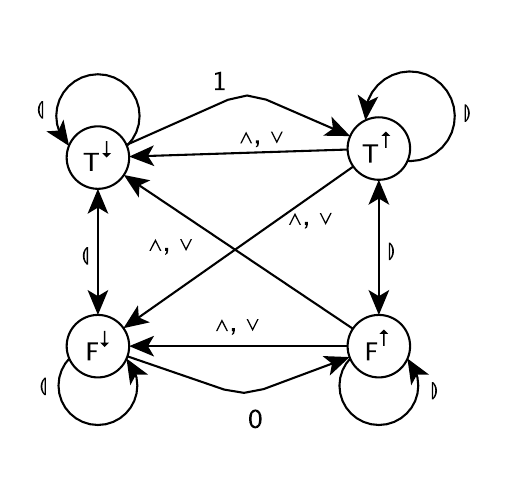}
\caption{The control graph of $G_\text{Logic}$.}\label{fig:CG}
\end{figure}

By following the left-to-right, bottom-up parsing of the string, we see that $1 ^\avv 3$ with $\llp(1)$ and $\land(3)$; 
the $1$ included in between belongs to $R_T$, since there exists one ---only--- rule with $1$ as rhs, i.e., $ T \to 1$, $1 \in L(T)$. 
The following parsing step leads to the relation $4 ^\avv 6$ with $\llp(4)$ and $\lor(6)$; the $0$ included in between belongs to $R_F$; since there exists one only rule $ F \to 0$, $0 \in L(F)$.
After a similar operation for positions $6$ through $8$, 
we have the string $\llp 0 \lor 1 \rrp \in R_T$ included within positions $3$ and $9$ for which relation $\avv$ holds; 
$\treec(3, 4, 6, 8, 9)$ holds too. 
There exists a rule $T \to \llp F \lor T\rrp$. 
By induction $0 \in L(F)$, $1 \in L(T)$; thus $\llp 0 \lor 1 \rrp \in L(T)$.
Completing the traversal of the syntax tree should now be a simple exercise leading to verify that formula $\psi_T$ holds for the string $\llp1 \land \llp0 \lor 1\rrp\rrp$.
Furthermore, by formula \eqref{eq:chi}, $\chi_{G_\text{Logic}}$ is satisfied, since $T$ is the only axiom of $G_\text{Logic}$.
A natural generalization leads to verify that a string in $\{\llp, \rrp, \land, \lor, 0, 1 \}^*$ belongs to $L(G_\text{Logic})$ iff it satisfies $\chi_{G_\text{Logic}}$.
The languages of the control graph are clearly NC, so that they can be defined through FO formulas $\varphi_T$, $\varphi_F$; the remaining part of $\psi$ is based on $\treec$, which is FO. Thus, we have obtained an FO definition of $L(G_\text{Logic})$.

\end{exa}

Corollary \ref{cor:suff-cond-NC} and Example \ref{ex:Potthoff} also hint at a much more attractive result: \emph{if a NC OPL is associated with NC control languages, then it can be defined through an FO formula.}
%
Unfortunately, we will soon see that there are NC OPLs such that the control graph of their (unique up to a nonterminal isomophism) BDR OPG defines counting regular languages $R_A$. 
Thus, the following ---rather technical--- section is devoted to transform the original BDR grammar of a NC OPL and its control graph into equivalent ones where the controlling regular languages involved in the above formulas are NC and therefore FO definable.

\section{NC regular languages to control NC OPLs}\label{sec:FO}

The previous section showed that, if an OPL is controlled by a control graph whose path labels from descending to corresponding ascending states
are NC regular languages, then the OPL can be defined through an FO formula; by adding the intuition 
that, if languages $R_A$, where $A$ denotes any nonterminal of the original grammar, are NC, then the original OPL is NC as well, we would obtain a sufficient condition for FO-expressibility of NC OPLs.

This is not our goal, however: we want to show that \emph{any} NC OPL can be expressed by means of an FO formula.
Unfortunately, it is immediate to realize that there are NC OPLs whose languages $R_A$ of the control graph of their BDR grammar are counting, as shown by the following simple example:
\begin{exa}\label{ex:Ncg_countcontr}
Consider the grammar $A \to a B c \mid d; \,
  B \to a A b$.
The regular control language  $R_A$ is $(aa)^* d (bc)^*$.
However, Theorem~\ref{Th:GtoMSO} still holds if we replace  $R_A$ by the NC language $a^*d(bc)^*$: intuitively, it is the OPM, and therefore the $\avv$ relation, which imposes that each $b$ and each $c$ are paired with a single $a$, so that for each sequence belonging to $(bc)^*$ we implicitly count an even number of $a$.
\end{exa}

Generalizing this natural intuition into a rigorous replacement of the original control graph of any OPG with a different NC one which preserves Theorem~\ref{Th:GtoMSO} is the target of this section. To achieve it, we need a rather articulated path which is outlined below:

\begin{enumerate}
\item First, in the same way as in \cite{CreGuiMan78} we build a linear grammar $G^L$
associated with the original OPG $G$ (which is always assumed to be BDR) such
that $L(G^L)$ is NC iff $L(G)$ is as well.

\item Then, we derive from the control graph of $G^L$  another control graph $\mathcal{\overline C}(G^L)$ whose regular languages are NC\@. 
This will require a rather sophisticated transformation of the original $\mathcal{C}(G^L)$.

\item The original grammar $G$ is transformed into an equivalent one $G'$, which is no longer BDR, whose nonterminals are
pairs of states of the transformed control graph $\mathcal{\overline C}(G^L)$
where one or more of them are homomorphically mapped into single nonterminals $A$ of $G$, and 
such that its control graph $\mathcal{ C}(G')$ exhibits only NC control languages.

\item Finally, the original Theorem~\ref{Th:GtoMSO} is extended to the case of the transformed grammar $G'$ and its new control graph. At this point, the MSO formalization of any OPL provided in Section~\ref{subsec:CGtoMSO} automatically becomes an FO one thanks to the fact that each subformula $\varphi_A$ defines a NC regular language.
\end{enumerate}

To obtain a first intuition of the final goal of the process outlined below consider the following grammar:
$
( \dd{AB}, \uu{A}) \to a ( \dd{AB}, \uu{B}) c \mid d; \,
 ( \dd{AB}, \uu{B}) \to a ( \dd{AB}, \uu{A})b.
$

 Apparently it is identical to the original grammar of Example \ref{ex:Ncg_countcontr} up to a simple renaming of its nonterminals.
 However, if we rebuild its control graph by using $\{\dd{AB}\}$ as $\dd {V_N}$ and $\{\uu{A},\uu{B} \}$ as $\uu {V_N}$ we obtain that  $R_{(\dd{AB}, \uu{A})}$ is $a^* d (bc)^*$, and $R_{(\dd{AB}, \uu{B})}$ is $a^+ d b(cb)^*$ which are both NC.

\subsection{Linearized OPG and its  control graph}

\begin{defi}[Bilateral linear grammar]\label{def:BilateralG}
A linear production of the form $A \to u B v$ such that $B\in V_N$, and $u, v \in \Sigma^+$ is called \emph{bilateral}. A linear grammar is bilateral if it contains only bilateral productions and terminal productions. 
\end{defi}
Thus, a bilateral grammar may not contain productions that are null, renaming, left-linear or right-linear.

The following definition slightly modifies  a similar one given in
\cite{CreGuiMan78}.

\begin{defi}[Linearized grammar]\label{defAssLinGr}
  Let $G = (\Sigma, V_N, P, S)$ be a BDR OPG\@. Its associated linearized grammar
  $G^L$ is $(\Sigma_L, V_N, P_L, S)$, where $\Sigma_L = \Sigma \cup \overline \Sigma
  \cup \{\overline\varepsilon_L, \overline\varepsilon_R \}$,
  $\overline\Sigma = \{ \overline C \mid C \in V_N \}$,
  $h$ is the homomorphism defined by $ h(a) = a$, $h(C) = \overline C$, and
  
  $
  P_L =
  \begin{array}{l}
   \{A \to h(\alpha) B h(\beta) \mid   A \to \alpha B \beta \in P, \alpha, \beta \neq \varepsilon\} \ \cup \\
   \{A \to \overline\varepsilon_L B h(\beta) \mid  A \to B \beta \in P \} \ \cup \\ 
   \{A \to h(\alpha) B \overline\varepsilon_R \mid  A \to \alpha B  \in P \}
  \ \cup \\ 
   \{A \to w \mid A \to w \in P, w \in \Sigma^+  \}.
  \end{array}
  $
\end{defi}

\begin{exa}\label{ex:linearization}
Consider the grammar $G_{NL}$ of Example~\ref{ex:contr-graph}. 
Its associated linearized grammar $G^L_{NL}$, with
$\Sigma_L = \{a,b,c, \overline A, \overline B, \overline \varepsilon_R\}$,%
\footnote{$\overline\varepsilon_L$ is useless in this case.}
and the same axioms as $G_{NL}$, has the following productions:

$
A \to 
a \overline B c A \overline \varepsilon_R \mid a B c \overline A \mid
a \overline B c B \overline \varepsilon_R  \mid  a B c \overline B \mid ac, $

$
B \to 
b \overline A c A \overline \varepsilon_R \mid b A c \overline A \mid 
b \overline A c B \overline \varepsilon_R \mid b A c \overline B \mid 
bc
$

Thus, the set $W$ of $G^L_{NL}$'s control graph is  $\{a,b,c, b \overline {A} c, a \overline {B} c, c \overline A, c \overline B, ac, bc, \overline \varepsilon_R \} $

\end{exa}

A linearized grammar is evidently bilateral and BDR (after some obvious clean-up). It has a different terminal alphabet ---and therefore OPM--- 
than the original grammar from which it is derived but it is still an OPG since its new OPM is clearly conflict-free 
(the two separate ``dummy $\varepsilon$'' have been introduced just to avoid the risk of conflicts). 
It is not guaranteed, however, that an OPG with $\dot=$-acyclic OPM  has an associated  linearized grammar enjoying the same property. 
Such a hypothesis, however, is not necessary to ensure the following results (indeed, it is only necessary to guarantee the existence of a maxgrammar generating the universal language $\Sigma^*$).
 
The following lemma is a trivial adaptation of the analogous Lemma 1 of \cite{CreGuiMan78} to Definition~\ref{defAssLinGr}.

\begin{lem}\label{L1JACM}
Let $G$ be a BDR OPG and $G^L$ its associated linearized grammar. $L(G^L)$ is NC iff $L(G)$ is as well.
\end{lem}

This simple but fundamental lemma formalizes the fact that the aperiodicity property can be checked by looking only at the paths traversing the syntax trees from the root to the leaves neglecting their ramifications. 

The next definition and property are taken from \cite{ChevalierDP07} with
a minor adaptation\footnote{The adaptation consists in allowing for the use
of macro-steps reading a \emph{nonempty} sequence of characters rather than
one single character per transition as in the traditional definition of FA
adopted in \cite{ChevalierDP07}. It is immediate to verify that
Proposition~\ref{prop:3stars} holds identically whether we consider FAs defined
in terms of macro-steps or the traditional ones.}.

\begin{defi}[Counter]\label{defCounter}
  For a given FA 
  (without $\varepsilon$-moves) 
  a {\em counter}\/ is a pair $(X, u)$,
  where $X$ is a sequence of different states $ q_1 q_2 \ldots q_k
  $, with $k > 1$ and
  $u$ is a
  nonempty string such that for $1 \le i \le k$,
  $q_i \boldtranss{\delta}{u} q_{(i+1) \bmod k}$; $k$ is called the {\em order}\/ of the counter. For a counter $C = (X, u)$, the sequence $X$
  is called the \emph{counter sequence} of $C$ and $u$ the \emph{string} of $C$.
\end{defi}

\begin{prop}\label{prop:3stars}
  If an FA $\mathcal A$ is counter-free, i.e., has no counters, then $L(\mathcal A)$ is non-counting.
\end{prop}

Notice that the converse of this statement only holds in the case of minimized deterministic FAs \cite{McNaughtPap71}.

Thus, for a linearized grammar $G^L$, every path of its control graph 
belonging to some $R_A$ is articulated into a sequence of macro-steps whose states belong to $\dd V_N$ followed by a sequence which traverses the corresponding
nodes of $\uu V_N$ in the reverse order 
---in between there is a single macro-step from some $\dd B$ to $\uu B$---. Accordingly, a counter sequence
may only contain nodes that either all belong to $\dd V_N$, or all belong to $\uu V_N$; thus, their corresponding counters will be said \emph{descending} or \emph{ascending}. 

Let $\mathcal{C}= (X, u)$ be a counter with $X =  A_1 A_2\ldots A_k $, $A_i \boldtrans{u} A_{(i+1)\bmod k}$, for $1 \le i \le k$. 
Let also $u = z_1 z_2 \ldots z_j$, $j \ge 1$ be the factorization into strings $z_i$ of the set $W$ corresponding to the macro-steps of the path $A_i \boldtrans{u} A_{(i+1)\bmod k}$: 
notice that such a factorization is the same for all $i$ since the OPM imposes the same parenthesization of $u$ in any path.

The following lemma allows us to reason about the NC property of linear OPLs without considering explicitly the parenthesis versions of their grammars.

\begin{lem}\label{lem:par-nopar}
Let $G^L$ be a bilateral linear OPG,  $\mathcal{C}(G^L)$ its control graph, $G^L_p$ the parenthesized version of $G^L$, and $\mathcal{C}(G^L_p)$ its control graph.
Then, for any nonterminal $A$ of $G^L$ the control language $R_{pA}$ is NC iff so is $R_{A}$.
\end{lem}

\begin{proof}
  If $R_{pA}$ is counting, then obviously so is $R_{A}$.
  
  Vice versa, suppose by contradiction that for all $k$
  $R_{A}$ contains
  a string $xy^kz$  but not $xy^{k+m}z$ for all $m \ge 0$.  Notice that for $k$ sufficiently
  large the parenthesized version $y_p^k$ of $y^k$ must contain either only open or
  only closed parentheses.  

  Let us assume w.l.o.g.\ that $y_p^k$ begins with an
  open (resp.\ ends with a closed) parenthesis; otherwise consider a suitable
  permutation thereof.  If all occurrences of $y_p$ itself begin with an open
  parenthesis (resp.\ end with a closed one), then $R_{pA}$ is counting too;
  otherwise for some $r$ with $1 < r \leq k$ there must exist an $u_p = y_p^r$ without a parenthesis
  between two consecutive occurrences of $y_p$; but this would imply a conflict
  in the OPM.  
\end{proof}

\begin{defi}[Counter table]\label{def:counterTable}
We use an array with the following scheme, called a \emph{counter table} $\mathcal T$,  
to completely represent, in an orderly fashion, the macro-transitions which may occur within a counter 
$\mathcal{C}= (X= T_1 T_2 \ldots T_k , u=z_1 z_2 \ldots z_j)$:
\begin{equation}
\label{eq:arrayCounterTransitions}
\setlength{\arraycolsep}{0.4cm}
\renewcommand{\arraystretch}{1.5}
\begin{array}{llll}
T_1^0 \boldtrans{z_1} T_1^1
&
\boldtrans{z_2} T_1^2
&
\ldots
&
T_1^{j-1}\boldtrans{z_j}   T_2^0
\\
T_2^0 \boldtrans{z_1} T_2^1
&
\boldtrans{z_2} T_2^2
&
\ldots
&
T_2^{j-1}\boldtrans{z_j}   T_3^0
\\
\multicolumn{4}{c}{\cdots}
\\
T_k^0 \boldtrans{z_1} T_k^1
&
\boldtrans{z_2} T_k^2
&
\ldots
&
T_k^{j-1}\boldtrans{z_j}   T_1^0
\end{array}
 \end{equation}
 
 \noindent where the $0$-th column is conventionally bound to the above counter $\mathcal{C}$.

 With reference to the above Table \eqref{eq:arrayCounterTransitions}
the sequence of macro-steps looping from $T_1^0$ to $T_1^0$ is called the \emph{path} of the counter table.
\end{defi}
Thus, a counter table defines a ``matrix of counters'' consisting of its columns:
in the case of Table \eqref{eq:arrayCounterTransitions} the first column $T_1^0, T_2^0, \ldots, T_k^0$ together with the string $u$ will be used as the \emph{reference counter} of the table. 
Each cyclic permutation of each column is another counter with the same string, whereas each column is the counter sequence of another counter whose string is a cyclic permutation of $u$, e.g. $( T_2^1 T_3^1 \ldots T_1^1$, $z_2 z_3 \dots z_j z_1)$.
For any counter of a counter table, its {\em associated path}\/ 
is the sequence of macro-steps looping from its first state to itself.
The above remarks lead to the following formal definition:
\begin{defi}\label{Def:counter-permutations}
Let $\mathcal T$ be a counter table expressed in the form of Table~(\ref{eq:arrayCounterTransitions});
the conventionally designated
counter $\mathcal{C}= ( T_1^0 T_2^0 \ldots T_k^0, z_1 z_2 \ldots z_j)$  is named its \emph{reference counter};
all columns $( T_{1}^m T_{2}^m \ldots T_{k}^m$, $z_{(m\bmod j)+1}z_{((m+1)\bmod j)+1 } \dots z_{((m+j-1)\bmod j)+1})$
 with $m = 1, 2 \dots j-1$ are named \emph{horizontal cyclic permutations} of the reference counter; 
all counters $( T_l^0 T_{(l \bmod k)+1}^0 \dots \allowbreak T_{l-1}^0, z_1 z_2 \ldots z_j)$, with $1 < l \leq k$, are named \emph{vertical cyclic permutations} of the reference counter;
horizontal-vertical and vertical-horizontal cyclic permutations, 
are the natural combination of the two permutations.
\end{defi}

If we apply cyclic permutations to the whole path producing a counter $\mathcal{C}= (X= T_1 T_2 \ldots T_k, u=z_1 z_2 \ldots z_j)$, and therefore a complete counter table, we obtain a family of counter tables associated with the original Table~\ref{eq:arrayCounterTransitions}. 
We decide, therefore, to choose arbitrarily an ``entry point'' of any path producing a counter. Such an entry point uniquely determines a counter table $\mathcal T$ and therefore a unique reference counter.
Furthermore, for convenience, if the same path $T_l \boldtrans{u} T_{(l+1)\bmod k}$, for $1 \le l \le k$ can also be read as $T_l \boldtrans{u'} T_{(l+1)\bmod k'}$, with $u= u'^r$, $k' = k \cdot r$ we represent the unique associated $\mathcal T$ by choosing the minimum of such $u$ (and the maximum of the $k$).
All elements of the table ---states, transitions, counter sequences--- will be referred through this unique $\mathcal T$, ignoring the other tables of its ``family''. Whenever needed, we will identify a counter table, its counter sequences, and any element thereof, through a unique index,  as $\mathcal T[i]$, $X[i]$, $T_{l}[i]$, respectively.

Notice that a counter table uniquely defines a collection of counters (among them the first column being chosen as its reference counter), but the same counter may be a counter, whether a reference counter or not, of different tables.
This case arises, for instance, when the linearized grammar contains two productions such as 
$A_1 \to z_1 B_1^1 v$ and $A_1 \to z_1 C_1^1 w$. Then the same counter $\mathcal{C}= (X= A_1 A_2 \ldots A_k, u=z_1 z_2 \ldots z_j)$ may occur  in two different counter tables that necessarily differ
in at least one of the intermediate states $B_h^i$.

Notice also that the various counters of a counter table are not necessarily disjoint.
Consider, for instance, the following sequence of transitions

$A \boldtrans{a} B$,
$B \boldtrans{b} C$,
$C \boldtrans{c} B$,
$B \boldtrans{a} D$,
$D \boldtrans{b} E$,
$E \boldtrans{c} A$

\noindent which constitute a counter table. In this counter table nonterminal $B$ occurs twice by using two different transitions; thus, we obtain 
the counters $(AB, abc), (BD, bca), (CE, cab)$. Furthermore, the same transition $B \boldtrans{b} C$,
can also be used to exit the counter table, after having executed the loop 
$B \boldtrans{b} C$,
$C \boldtrans{c} B$, instead of continuing the counter table with $B \boldtrans{a} D$.

\begin{defi}[Paired Paths]
   Let $\mathcal C(G^L)$ be the control graph of a linearized grammar $G^L$. 
   Let 
  $A_1 \Longrightarrow  u_1A_2v_1 \ldots  \allowbreak \Longrightarrow  u_1 \dots u_{n-1} A_n v_{n-1} \dots v_1$ with $u = u_1 u_2 \ldots
  u_{n-1}$,   $v = v_{n-1} \ldots v_1$ be a derivation for $G^L$. Then the  paths $\dd A_1 \boldtrans{u_1} \dd A_2, \ldots \dd A_{n-1} \boldtrans{u_{n-1}}
  \dd A_n$, and
  $\uu A_n \boldtrans{v_{n-1}} \uu A_{n-1}, \ldots \uu A_2  \boldtrans{v_1} \uu A_1$, called, respectively, \emph{descending} and
  \emph{ascending}, are   {\em paired (by such a derivation)}.
  
  Two counter tables are paired iff their paths, or cyclic permutations thereof, are paired; two counters are paired iff their associated paths $\dd T_1 \boldtrans{u^k} \uu T_1$, $\uu T_1 \boldtrans{v^h} \dd T_1$ are paired --- therefore so are the counter tables they belong to. 
\end{defi}

Notice that there could also be \emph{partially overlapping counter tables and counters}, which share one or more productions of $G^L$ but are not fully paired. 

\subsection{Transforming $G^L$ control graph}

If the control graph of a linearized grammar $G^L$ is counter free, then $L(G^L)$ is NC\@.
Notice, in fact, that
\begin{enumerate}
\item $\mathcal C(G^L)$ has no $\varepsilon$-moves, thus the Definition~\ref{defCounter}
of counter-free is well-posed for it;
\item  If, by contradiction, $G^L$, which is BDR, admitted a counting derivation, such a derivation would imply two paired counters of $\mathcal C(G^L)$.
\end{enumerate}

Unfortunately such a condition is only sufficient but not
necessary to guarantee that $L(G^L)$ is NC, as shown by Example~\ref{ex:Ncg_countcontr}.
Thus, according to the path outlined at the beginning of Section~\ref{sec:FO}, our next goal is to transform $\mathcal C(G^L)$ 
into a control graph, denoted as  $\overline{\mathcal C}(G^L)$, 
whose regular languages are NC and which will drive the construction of
a grammar $G'$, equivalent to the original $G$, such that its control graph
defines NC $R_A$ for its nonterminals. 
The construction of $\overline{\mathcal C}(G^L)$ will exploit
the following lemmas, which make use of the notion of paired counters:

\begin{lem}\label{lemma:gg}
  If $G^L$ is NC, then $\mathcal C(G^L)$ either has no paired counters or, for any two paired counters, the orders of the descending and ascending counter are coprime   numbers.
 
\end{lem}
\begin{proof}
  Assume, by contradiction, that the counters $\dd C_1 = (\dd X, u)$, $\uu C_2 =
  (\uu Y, v)$ are paired by the derivation   
  $A_1 \derives u^k A_1 v^h 
  $ and 
  that 
  for some $j, r, s > 1$, $k = j \cdot r$, $h = j \cdot s$.
  Let  $\dd X =  \dd A_1 \ldots \dd A_k $,
  $\uu Y =  \uu A_1 \ldots \uu A_h $.
  This means that for some $j$, $A_1 \derives u^j A_jv^j \derives
  u^{2j}A_{2j}v^{2j} \ldots \derives u^kA_1v^h$; thus $(\dd A_1 \dd A_j
  \dd A_{2j} \ldots \dd A_k, u^j)$
  and $( \uu A_1 \uu A_j  \uu A_{2j}  \ldots$  $\uu A_1, v^j)$, 
  where $\uu A_j$ and $\dd A_{(r-1)j}$, $\uu A_{2j}$ and $\dd A_{(r-2)j} \dots $ refer to the same nonterminal in the derivation  $A_1 \derives u^k A_1 v^h 
  $,  are two paired counters as well which correspond to a
  counting derivation of $G^L$. 
\end{proof}

\begin{exa}\label{ex:paired}
The productions $A \to aBb$ and $B \to aAb$ 
generate the two paired counters of order 2 of the control graph: $(\dd A \dd  B,a)$ paired with $( \uu B\uu  A, b)$.
Instead, the productions
$A_1 \to aA_2f$,
$A_2  \to bA_3g$,
$A_3  \to aA_4h$,
$A_4 \to bA_5f$,
$A_5  \to aA_6g$,
$A_6  \to bA_1h$
generate the following sequence of descending counters of order $3$ paired with ascending counters of order~$2$: 

$\begin{array}{l}
(\dd A_1\dd A_3 \dd A_5, ab), (\uu  A_1 \uu A_4, hgf)    \\
(\dd A_2\dd A_4 \dd A_6, ba), (\uu  A_2 \uu A_5, fhg)    \\
(\dd A_3\dd A_5 \dd A_1 , ab), (\uu  A_3 \uu A_6 , gfh)   \\
(\dd A_4\dd A_6 \dd A_2 , ba), (\uu  A_4 \uu A_1 , hgf)   \\
(\dd A_5\dd A_1 \dd A_3 , ab), (\uu  A_5 \uu A_2 , fhg)  \\
(\dd A_6\dd A_2 \dd A_4 , ba), (\uu  A_6 \uu A_3 , gfh)
\end{array}$
\end{exa}

By looking at the second case of Example~\ref{ex:paired} we notice that for each couple of paired counter sequences there is just one nonterminal that belongs to both of them. This remark is easily generalized to the following lemma:

\begin{lem}\label{lemma:qq}
    Let $L(G^L)$ be NC\@. If in $\mathcal {C}(G^L)$ there
    are two paired counters $\dd C_1 = (\dd X, u)$, $\uu C_2 =(\uu Y, v) $ there exists only one $A$ such that $\dd A \in \dd X$, $\uu A \in \uu Y$.
\end{lem}

\begin{proof} 
Let $|\dd X| = k$, and
$|\uu Y| = h$, with $h$ and
$k$ coprime, thanks to Lemma~\ref{lemma:gg}.
The two paired counters correspond to a NC derivation of $G^L$ 
$A_1 \derives x A_t y \derives u^k A_1 v^h$. 
The total length of
the derivation is $h \cdot k$ and each $A_t$ belongs to a set, marked $\dd{}$, of cardinality
$k$ in the table $\mathcal T[i]$ of $\dd C_1 $ and to a set, marked $\uu{}$, of cardinality $h$ in the
table $\mathcal T[f]$ of $\uu C_2$. Thus, for any couple $(\dd X, \uu Y)$ paired by the two counter tables,
there exists exactly one $A$, such that $\dd A \in \dd X$, $\uu A \in \uu Y$ by virtue of the Chinese
remainder theorem.
\end{proof}

On the basis of the above lemmas the construction of $\overline{\mathcal C}(G^L)$ aims at replacing any
ascending and descending counter with a loop
$X \boldtranss{\overline{\pmb\delta}}{u} X $
 where $X$ is a
suitable new state in $\overline{\mathcal C}(G^L)$ representing a whole counter sequence of $\mathcal {C}(G^L)$; thanks to
Lemma~\ref{lemma:gg}, the new loop will be paired
with a path that is not a counter or with another loop which in turn replaces a counter whose order is coprime
w.r.t.\ the order of the former one.
By virtue of Lemma~\ref{lemma:qq}, in turn, this will allow to disambiguate which element of the counter sequence
corresponds to the $G^L$'s nonterminal deriving the various instances of string $u$.

\begin{figure}[h]
\centering
  \includegraphics[scale=0.6]{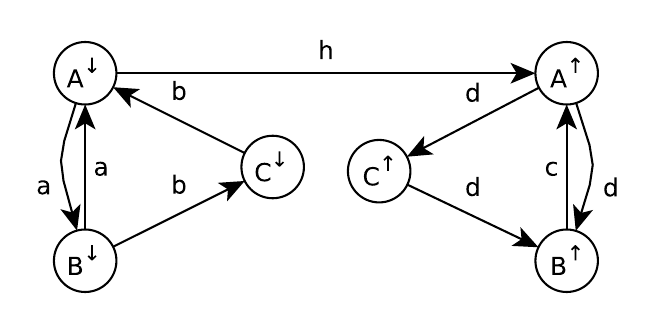}
  \caption{A control graph including a descending counter. }\label{fig:contres1}
\end{figure}

This basic idea, however, cannot be implemented in a trivial way such as replacing all states belonging to a counter sequence by a single state representing the whole sequence.
Consider, for instance, a grammar containing the following productions:
\[
\begin{array}{ll}
 	
 	& A \to aBc \mid h \\  	
 	& B \to aAd \mid bCd    \\
 	& C \to bAd   \\
 	\end{array}
\]
which produce the control graph depicted in Figure~\ref{fig:contres1}.

The control graph has a descending counter $(\dd A \dd B, a)$ paired with the
ascending path $\uu A \boldtranss{}{d} \uu B
\boldtranss{}{c} \uu A$.
If we simply replace the descending path $\dd A \boldtranss{}{a} \dd B
\boldtranss{}{a} \dd A$ with a self-loop $\dd {AB} \boldtranss{}{a}
\dd {AB} $ by coalescing the two states into one state denoted by $\dd {AB}$, we
obtain as a side effect a new counter $(\dd {AB} \dd C, b)$; if we further
collapse $\dd {AB} \dd C$ into $\dd {ABC}$ we reduce the descending part of the
control graph to a single state with two self-loops labeled $a,b$: at this
point, once a path reaches the state $\uu A$ and reads the symbol $d$ it is
impossible to decide whether such an ``ascending $d$'' should be paired with a
previous descending $b$ or $a$ since both are labeling a self-loop on the unique
state $\dd {ABC}$.

The construction we devised for such a $\overline{\mathcal C}(G^L)$ is therefore more complex: it is articulated into two steps:
first a $\hat{\mathcal C}(G^L)$ ``equivalent'' to $\mathcal C(G^L)$, in a sense that will be made precise in Lemma~\ref{lemma:zz},  is built.
$\hat{\mathcal C}(G^L)$ splits some states belonging to counters in such a way that each new instance thereof belongs to exactly one counter table; then the further construction $\overline{\mathcal C}(G^L)$ collapses all
counter sequences into single states that allow repeating the ``basic counter string $u$''
any number of times, instead of $k$ times. Thus, each path of the original control graph $\mathcal C(G^L)$ of type, say $\dd A \boldtrans{u^k} \dd A$ 
that realizes a counter $(\dd X, u)$ of order $k$ will be replaced by $k$ paths $\dd X \boldtrans{u} \dd X$ (apart from a transient that will be explained later). 
Thanks to Lemma~\ref{lemma:gg}, if $G^L$ is NC, it will not be paired with another counter $(\uu Y, v)$, or, if so happens, the order of the other counter will be an $h$ coprime of $k$;
thus, thanks to Lemma~\ref{lemma:qq}, it will be possible to associate each couple of paired counters of the control graph of $G^L$ with a unique derivation of the grammar.

\paragraph*{Construction of $\hat{\mathcal C}(G^L)$.}
Intuitively, the aim of $\hat{\mathcal C}(G^L)$ is to produce ``non-intersecting counter tables'', i.e., counter tables such that $\mathcal T[i] \neq \mathcal T[j]$ implies that the counter sequences of $\mathcal T[i]$ are all disjoint from those of $\mathcal T[j]$.
This is obtained by creating one instance of state $A$, say $A[i]$, for each counter table $\mathcal T[i]$ $A$ belongs to, where the index $i$ binds the state instance to the table. 

The construction below applies as well to states of type $\dd A$ and to states
of type $\uu A$, according to Definition~\ref{def:aut:G}. Notice that macro-transitions
 of the type $\dd A \boldtrans {z} \uu A$, which correspond to $G^L$'s productions $A \to z$, $z \in W$,
cannot belong to any counter table of $\mathcal C(G^L)$, but $\dd A$ and/or $\uu A$ can belong to some descending or ascending counter, respectively. 



The construction of $\hat{\mathcal C}(G^L) = (\hat Q, \Sigma, \hat{\pmb\delta})$ starts from $\mathcal C(G^L) = (Q, \Sigma, \pmb \delta)$, i.e., it is a process where $\hat Q$ and $\hat{\pmb\delta}$ are initialized as $Q$ and $\pmb \delta$, and modifies them in the following way. 
When the transformations below apply identically to descending and ascending paths we omit labeling the states of the control graph as $\dd{}$ or $\uu{}$:


First, we label all counter tables $\mathcal T$  with unique and different indexes $i$.

Then, all states
  belonging to $\mathcal T[i]$ are also labeled in the same way, so that if a state
  $A$ belongs to different counter tables, $\mathcal T[i]$ and  $\mathcal T[h]$, $i \ne h$,
  it will be
  split into different states $A[i]$ and $A[h]$; if instead it belongs to just one counter with only one associated table, for convenience it will be labeled with the same index $i$ identifying the table.
  If it does not belong to any counter table, it remains unlabeled.
  
  Then, $\hat{\mathcal C}(G^L)$'s transitions are defined as follows:
\begin{itemize}  
\item
  For every macro-transition $A \boldtranss{\pmb\delta}{f} B $ where $A$ and $B$ are both descending or both ascending, for all $m$ copies $A[1], A[2], \ldots A[m]$ of $A$ and $n$ copies $B[1], B[2], \ldots B[n]$ of $B$, 
  $A \boldtranss{\pmb\delta}{f} B $ is replaced by $m \cdot n$ macro-transitions $A[i] \boldtranss{\hat{\pmb\delta}}{f} B[h] $, 
  where $A[i]$ and/or $B[h]$ remain $A$ and/or $B$ if they do not belong to any counter table.
\item
For every transition $\dd A \boldtranss{\pmb\delta}{f} \uu A $, if $A$ belongs
to some descending and/or ascending counter ---thus it is labeled $ \dd A[i]$ and/or $ \uu A[h]$--- all possible $\dd A[i] \boldtranss{\hat{\pmb\delta}}{f} \uu A[h] $ replace the original macro-transition.
\end{itemize}

\begin{exa}\label{ex:beta}


Consider the fragment of a control graph  ${\mathcal C}(G^L)$ (which could be indifferently a descending or an ascending part thereof)
depicted in Figure~\ref{fig:nex1} (left). The corresponding fragment of $\hat {\mathcal C}(G^L)$ is given in Figure~\ref{fig:nex1-1} (right).

The example shows the case of two counter tables sharing some states. Notice that in
general the construction of $\hat {\mathcal C}(G^L)$ increases the number of
counters which are all isomorphic to the original one: for instance,
in the case of Figure~\ref{fig:nex1-1}, instead of the path
$A \boldtrans{a} H \boldtrans{b} L \boldtrans{a} B \boldtrans{b} A$, we have
$A[1] \boldtrans{a} H[1] \boldtrans{b} L[1] \boldtrans{a} B[1] \boldtrans{b} A[1]$,
but also $A[1] \boldtrans{a} H[2]
\boldtrans{b} L[1] \boldtrans{a} B[1] \boldtrans{b} A[1]$, $A[1] \boldtrans{a} H[1]
\boldtrans{b} L[2] \boldtrans{a} B[1] \boldtrans{b} A[1]$ \ldots. We will see, however, that, despite the increased number of paths, none of them will generate a counting path after the further transformation from $\hat {\mathcal C}(G^L)$ to  $\overline {\mathcal C}(G^L)$.

\begin{figure}[h]
\begin{tabular}{m{0.3\textwidth}m{0.5\textwidth}}
  \includegraphics[scale=0.5]{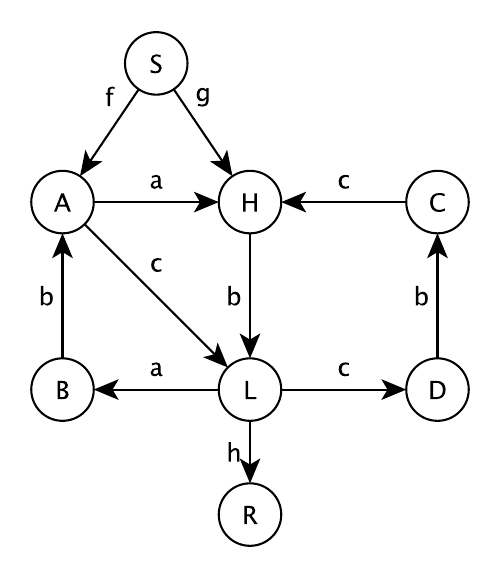}
  &
  \includegraphics[scale=0.5]{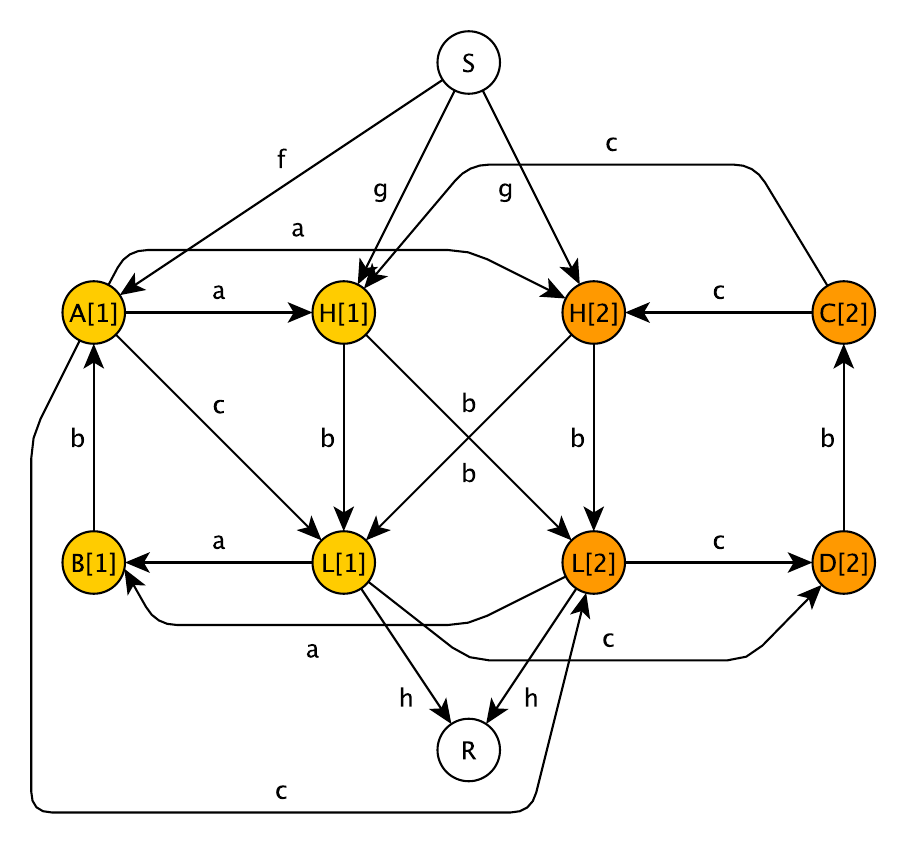}
\end{tabular}
  \caption{$\mathcal C(G^L)$ (left) and $\hat {\mathcal C}(G^L)$ (right); states belonging to different counter tables
    are depicted in different colors.}\label{fig:nex1}\label{fig:nex1-1}
\end{figure}

\end{exa}

\begin{lem}\label{lemma:zz}
  For each pair $(\dd A, \uu A)$ of $\mathcal C(G^L)$, and $z \in \Sigma^+$,
  $\dd A \boldtranss{\pmb\delta}{z} \uu A$ iff, either 
  $\dd A \boldtranss{\pmb{\hat\delta}}{z} \uu A$
  or, for all $\dd A[i]$, $\uu A[l]$,
  $\dd A \boldtranss{\pmb{\hat\delta}}{z} \uu A[l]$
  or
  $\dd A[i] \boldtranss{\pmb{\hat\delta}}{z} \uu A$ or
  $\dd A[i] \boldtranss{\pmb{\hat\delta}}{z}$  $\uu A[l]$.
By projecting the counters of $\hat{\mathcal C}(G^L)$ through the homomorphism $h(A[i]) = A$, $h(B) = B$ for all $B$ that do not belong to any counter, one obtains exactly the counter tables and the counters of $\mathcal C(G^L)$.
%
\end{lem}

\begin{proof} 
  Paths of $\mathcal C(G^L)$ that do not touch any state belonging to some counter table are found identically in $\hat{\mathcal C}(G^L)$.
  If the path of a counter table $\mathcal T[i]$ of $\mathcal C(G^L)$ touches a sequence of states $H, K, \dots L$, 
  $\hat{\mathcal C}(G^L)$ also has the path obtained
  by replacing $H$ by $H[i]$, $K$ by $K[i]$, etc., $i$ being the index of $\mathcal T[i]$. It is also always possible to ``jump'' from a table $\mathcal T[i]$ to another table $\mathcal T[l]$ by  using the transition target $B[l]$ instead of $B[i]$.
  
 Conversely, for each $A[i]$, $B[l]$, whether $i = l$ or not, if in $\hat{\mathcal C}(G^L)$ there is the macro-transition 
  $A[i] \boldtranss{\pmb{\hat\delta}}{f} B[l]$
 this means that in ${\mathcal C}(G^L)$ there was 
 $A \boldtranss{\pmb{\delta}}{f} B$.

  Furthermore, the construction of $\hat{\mathcal C}(G^L)$ does not produce counters
  that are not the image of $\mathcal C(G^L)$’s counters under
  $h^{-1}$, since all its transitions involving some $A[i]$ come from a corresponding ${\mathcal C}(G^L)$'s transition with $A$ in place of $A[i]$,
\end{proof}

\paragraph*{Construction of $\overline{\mathcal C}(G^L)$}
As anticipated, the core of $\overline{\mathcal C}(G^L)$'s construction moves from $\hat{\mathcal C}(G^L)$ and, roughly speaking, consists in collapsing all states labeled by the index of the same counter table and belonging to a counter sequence of a given counter  into a single new state named as the counter sequence itself and labeled by the index of the table it belongs to. 

The behavior of $\overline{\mathcal C}(G^L)$ is such that it is exactly like $\hat{\mathcal C}(G^L)$ (and as
$\mathcal C(G^L)$) until it reaches a state of some ---unique--- counter table, say state $T_1[i]$ of $\mathcal T[i]$ belonging to counter $C=(X[i],u)$ with $X[i] = T_1[i] \dots T_k[i]$. 
At that point it uses the single state $T_1[i]$
as an ``entry point'' to $\mathcal T[i]$; it follows the whole path $T_1[i] \boldtrans{u} T_2[i] \ldots T_k[i] \boldtrans{u} T_1[i]$ of the table up to the last macro-step that would ``close'' the counter; at this point 
its next transition, instead of going back to $T_1[i]$, enters a new state ---named \emph{counter sequence state}--- representing the whole counter sequence $X[i]$ that includes the state $T_1[i]$.

Then, $\overline{\mathcal C}(G^L)$ loops along the horizontal cyclic permutations of the counter ---a new counter sequence state is built for every column of the counter table---, 
therefore without counting the repetitions of the counter string $u$; in other words it ``forgets the vertical cyclic permutations'' of the counter table. 
When $\overline{\mathcal C}(G^L)$ exits from the loop
it nondeterministically reaches any node that can be reached by any state belonging to the counter sequence state it is leaving.
Notice that exit from the loop occurs only as a consequence of a transition that in ${\mathcal C}(G^L)$ was not part of the counter table; 
such a transition may lead either to a state that does not belong to the table, such as $L \boldtrans{h} R$ in Figure~\ref{fig:nex1}, 
or to a state that is still part of the table, such as $A \boldtrans{c} L$ in the same figure. 
In the latter case the same table can be re-entered, i.e., the original counting path may be resumed, but this must happen only by going into an entry point of the table, 
not directly into the counter sequence state containing it
(the reason of this choice will be clear later); 
for instance in the case of Figure~\ref{fig:nex1}, the transition that reads $c$ (from the counter sequence state containing $A$) leads to instances of $L$, not to the counter sequence state(s) containing it. 
Notice also that the transition $A \boldtrans{c} L$ may also occur in $\overline{\mathcal C}(G^L)$ during the ``transient'' before entering the counter sequence state: 
this means that the counting path is interrupted before being completed for the first time and possibly resumed from scratch (with a different entry point).

Obviously, $\overline{\mathcal C}(G^L)$ will exhibit all behaviors of $\mathcal C(G^L)$ plus more; 
we will see however, that pairing such, say, descending behaviors with the ascending ones 
will allow us to discard those that are not compatible with $G^L$'s derivations.

We now describe in detail the construction of $\overline{\mathcal C}(G^L)$.

Let $(X^m, u|m)$, where $m = 0, 1, \dots j-1$, denote any counter of a counter table $\mathcal T$ of $\mathcal C(G^L)$ with $X^m =  T^m_1 T^m_2 ... T^m_k$, $u|m = z_{(m \bmod j)+1} z_{((m+1) \bmod j)+1} ... z_{((m+j-1) \bmod j)+1}$,
$j \ge 1$, $z_i \in W$.
Thus, $(X^0, u)$ is the reference counter of $\mathcal T$ and $\{(X^m, u|m) \mid m = 1, 2, \dots j\}$ are its horizontal cyclic permutations (if any, i.e., if $j > 1$).
For every $m = 1, 2, \ldots j-1$,
$l = 1, 2, \ldots k$, $T^{m-1}_l \boldtranss{\pmb\delta}{z_m} T^{m}_l$,
$T^{j-1}_l\boldtranss{\pmb\delta}{z_j} T^0_{(l\bmod k)+1}$.

To simplify the notation we will avoid the index identifying the single tables whenever not necessary. 

Points 1 through 6 of
the construction below are identical whether they are applied to
states belonging to descending or ascending paths; thus we will not
mark those states with $\dd{}$ or $\uu{}$.

\begin{enumerate}

\item
For each counter sequence $X^m[i] = T^m_1[i] \dots T^m_k[i]$ of counter table $\mathcal T[i]$ 
we define the $l$-th \emph{pipeline} $PPL_l(X^m[i])$ as the sequence of all states
traversed by the whole path of the table starting from $ T^m_l[i]$ and ending in the state that precedes it in the counter table ---obviously traversed in cyclic way---.
It is followed by the new state $X^m[i]$, called a
\emph{counter sequence state}, which is therefore the same for all pipelines $PPL_l(X^m[i])$. The first state $ T^m_l[i]$ is called the \emph{entry point of the pipeline}.

For instance, with reference to
Figure~\ref{fig:nex1-1}, $PPL_1( A[1] L[1] )$ is 
$A[1] H[1] L[1] B[1]$ and\\ 
$PPL_2( A[1] L[1] )$ is $L[1] B[1] A[1] H[1]$ both followed by the state
$A L[1]$. \\
Similarly, $PPL_1(H[1] B[1])$ is
$H[1] L[1] B[1] A[1]$ and $PPL_2(H[1] B[1])$ 
is $B[1] A[1]$ $H[1] L[1]$, both followed by the state $HB[1]$.

For each counter table, all pipelines of its counters are
disjoint. Thus, for each table with counter sequences of order $k$ and string
$u$ consisting of $j$ elements in $W$ a collection of $(k \cdot j)^2$ different copies
of the original $k \cdot j$ states of the table plus $j$ counter sequence states are
in the state space $\overline Q$ besides all original states that do not participate
in any counter table.

\paragraph*{Notation} To distinguish the $k \cdot j$ replicas of the sequences that,
for each pipeline lead to the counter sequence states, we add a second index to
the one denoting the counter table, ranging from $0$ to $k \cdot j-1$; the $0$-th
copy, e.g., $H[2,0]$, will denote the \emph{entry point} of each
pipeline.

Let us now build $\overline{\mathcal C}(G^L)$'s (macro)transitions $\pmb{\overline\delta}$.
\item
All transitions that do not involve states belonging to counter tables are replicated identically from $\pmb{\hat\delta}$ and therefore from $\pmb{\delta}$.
\item 
  For all pipelines of all counters 
  $(X^m[i], u|m$) of all tables $\mathcal T[i]$,
 all original transitions of the table are replicated identically for each sequence by adding the further index $r$, 
 which is initialized to $0$ for the entry point, but the last one that would ``close the counter'';
  precisely:
  \begin{itemize}
  \item
  The entry point of pipeline $PPL_l(X^m[i])$ is $ T^m_l[i,0]$;
  
  the following transitions are added to to $\pmb{\overline\delta}$:
  
    \item
  for $1 \leq l \leq k$, $1 \leq m
  \leq j-1$, $0 \leq r <  k \cdot j-1$,
   $T^{m-1}_l[i,r]\boldtranss{\pmb{\overline\delta}}{z_{m }}  T^m_l[i,r+1]$, 
   \item
   if $r < k \cdot j -1$,
   $T^{j-1}_l[i,r]\boldtranss{\pmb{\overline\delta}}{z_j}  T^0_{(l \bmod k) + 1}[i,r+1]$,
  \item
 
  $T^m_l[i, k \cdot j-1]
  \boldtranss{\pmb{\overline\delta}}{z_{(m \bmod j)+1}} X^{(m+1)\bmod j}[i]$,
  which replaces the original\\ 
   $T^m_l[i]
  \boldtranss{\pmb{\hat\delta}}{z_{(m \bmod j)+1}} T^{(m+1) \bmod j}_p[i]$, where $p=(l \bmod k) + 1$ if $m = j-1$, $p = l$ otherwise, and $X^{(m+1)\bmod j}[i]$ is the counter sequence state containing $T^{(m+1) \bmod j}_p[i]$.

 
  \end{itemize}
  
  In other words, this first set of transitions allows to enter a counter sequence state from any state belonging to it, 
  only by starting from the entry point of the pipeline associated with that state,
  then to follow the whole path of the counter table and, at its last step, to enter the new state of type counter sequence, of which the entry point is a member.
  
 As a particular case, if $j=1$, there is only one counter sequence state $X[i]$, 
  all pipelines have length $k$, and consist of transitions
  $T_l[i,r] \boldtranss{\pmb{\overline\delta}}{u}  T_{(l \bmod k) + 1}[i, r+1]$,
  with $0 \leq r \leq k-1 $, but the last one which is $T_l[i, k] \boldtranss{\pmb{\overline\delta}}{u}   X[i]$, 
  where $X[i]$ is the counter sequence state containing $T_{(l \bmod k) + 1}[i]$ 
  which is also the entry point of the pipeline.
  
  Notice that in some cases the same transition could be used as part of a counter table path and as an exit way to it; since it leads to a state still belonging to the counter table, its target will be the entry point of a pipeline of the same counter table. Example~\ref{ex:double-transition} illustrates this case.
\item
For all counter sequence states $X^m[i] =T^m_1[i] \dots T^m_k[i] $, \\
$X^{(m+1) \bmod j}[i] = T^{(m+1) \bmod j}_1[i] \dots  T^{(m+1)\bmod j}_k[i] $ of a table $\mathcal T[i]$,\\
if for any $T^m_l[i]$, $T^{(m+1) \bmod j}_p[i]$, $z_{(m \bmod j)+1}$, $T^m_l[i]\boldtranss{{\pmb{\hat\delta}}}{z_{(m \bmod j)+1}} T^{(m+1) \bmod j}_p[i]$\\
(then it is also $T^m_{((l+o) \bmod k)+1}[i] \boldtranss{{\pmb{\hat\delta}}}{z_{(m \bmod j)+1}} T^{(m+1) \bmod j}_{((p+o)\bmod k)+1}[i]$ for all $o$; 
furthermore, either $p = l$ or $p = (l \bmod k)+1 $),\\
we set $X^m[i]\boldtranss{{\pmb{\overline\delta}}}{z_{(m \bmod j)+1}} X^{(m+1) \bmod j}[i]$.

Thus, once $\overline{\mathcal C}(G^L)$ entered a counter table with string $u$
it can accept any number of $u$, plus possibly a prefix and/or a suffix thereof, without counting them.
\item
\emph{Entering a counter.} Counters can be entered only through the entry points of their pipelines. 
This means that for each transition
$A \boldtranss{\pmb{\hat\delta}}{x} B $ 
that does not belong to the counter table $\mathcal T[i]$ but leads to a state $B = T^m_l[i]$
thereof (notice that $A$ could either belong or not to $\mathcal T[i]$) we add
---\emph{only}---
$A \boldtranss{\pmb{\overline\delta}}{x} B[i,0]$. 
All other elements of the pipelines  that are not entry point and the counter sequence states can be accessed only through the transitions built in points 3 and 4 above.
\item
\emph{Exiting a counter.} Counters can be exited in two ways: either in the transient before entering the counter sequence state, or exiting the loop that repeats the string $u$ any number of times without counting them.
In the former case this is obtained by adding, for each original transition of
${\mathcal C}(G^L)$ that departs from a state of the counter table $\mathcal T[i]$ and
does not belong to the table, say
$T^m_l \boldtranss{\pmb{\delta}}{x} B$,
an instance
thereof for all occurrences of $T^m_l[i,r]$ in the various pipelines of the
counters. Notice that the target state $B$ of such transitions could either
belong ---as in the case of transition $A \boldtrans{c} L$ of Figure~\ref{fig:nex1}--- or
not to the same table: in the positive case it should be ---\emph{only}--- the
entry point labeled $B[i,0]$ of the pipelines; in the negative case it could be
a single state not belonging to any counter table or the entry point of some
pipeline of a different table, say $B[p,0]$ (see Figure~\ref{fig:nex1-2} for the case of Figure~\ref{fig:nex1-1}).

Exiting the counter from the counter sequence state 
is obtained similarly by replicating the original transition
$T^m_l \boldtranss{\pmb{\delta}}{x} B $
for the target state $B$ in the same way as in the previous case and by replacing the source state $T^m_l$ with the counter sequence state $X^m[i]$ containing it.

\item
Finally, for each production $A \to x$ of $G^L$:

\begin{itemize}
    \item 
      If $A$ does not belong to any counter of $\mathcal C(G^L)$ only
      $\dd A \boldtrans{x} \uu A$
      is in $\overline{\pmb \delta}$
      (this is already implied by point 2 above).
    \item 
    If there is some $\dd A[i]$ in $\hat Q$ but no $\uu A[f]$, i.e., $A$ belongs
    to some descending counter sequence $\dd X[i]$ but to no ascending one, we set both
    $\dd A[i, r] \boldtrans{x} \uu A$
    for each $r$ and
    $\dd X[i] \boldtrans{x} \uu A$
    where $\dd A[i, r] $ may denote either an entry point of the pipeline ($r = 0$) or any other element thereof.
    \item 
    If instead $\dd A$ does not belong to any counter but there is some  $\uu
    A[f]$,
    we set only
    $\dd A \boldtranss{\pmb{\overline\delta}}{x} \uu A[f,0] $;
    no transition
    $\dd A \boldtrans{x} \uu X[f]$ or
    $\dd A \boldtranss{\pmb{\overline\delta}}{x} \uu A[f,r]$
    with $r \neq 0$ is set, however: this is due to our convention that counters can only be entered through the single states that are entry points of a pipeline, whereas, once they entered the counter sequence state they must be exited only therefrom.
    \item 
      If in $\hat{\pmb\delta}$ there are transitions
      $\dd A[i] \boldtrans{x} \uu A[f]$,
      i.e. $A$ belongs both to a descending counter
      $\dd X$ and to an ascending one $\uu X$ of $\mathcal C(G^L)$,
      then
      $\dd A[i,r] \boldtrans{x} \uu A[f,0]$,
      with $r \geq 0$, and
      $\dd X[i] \boldtrans{x} \uu A[f,0]$,
      are in $\pmb{\overline\delta}$ but neither
      $\dd A[i,r] \boldtrans{x} \uu X[f]$,
      nor
      $\dd X[i] \boldtrans{x} \uu X[f]$,
      nor
      $\dd A[i,r] \boldtrans{x} \uu A[f, s]$,  
      nor
      $\dd X[i] \boldtrans{x} \uu A[f,s]$,
      with $s \neq 0$ are included in $\pmb{\overline\delta}$ for the same reason as above.
    \end{itemize}
\end{enumerate}

\begin{figure}[h]
  \centering
  \includegraphics[width=\textwidth]
  {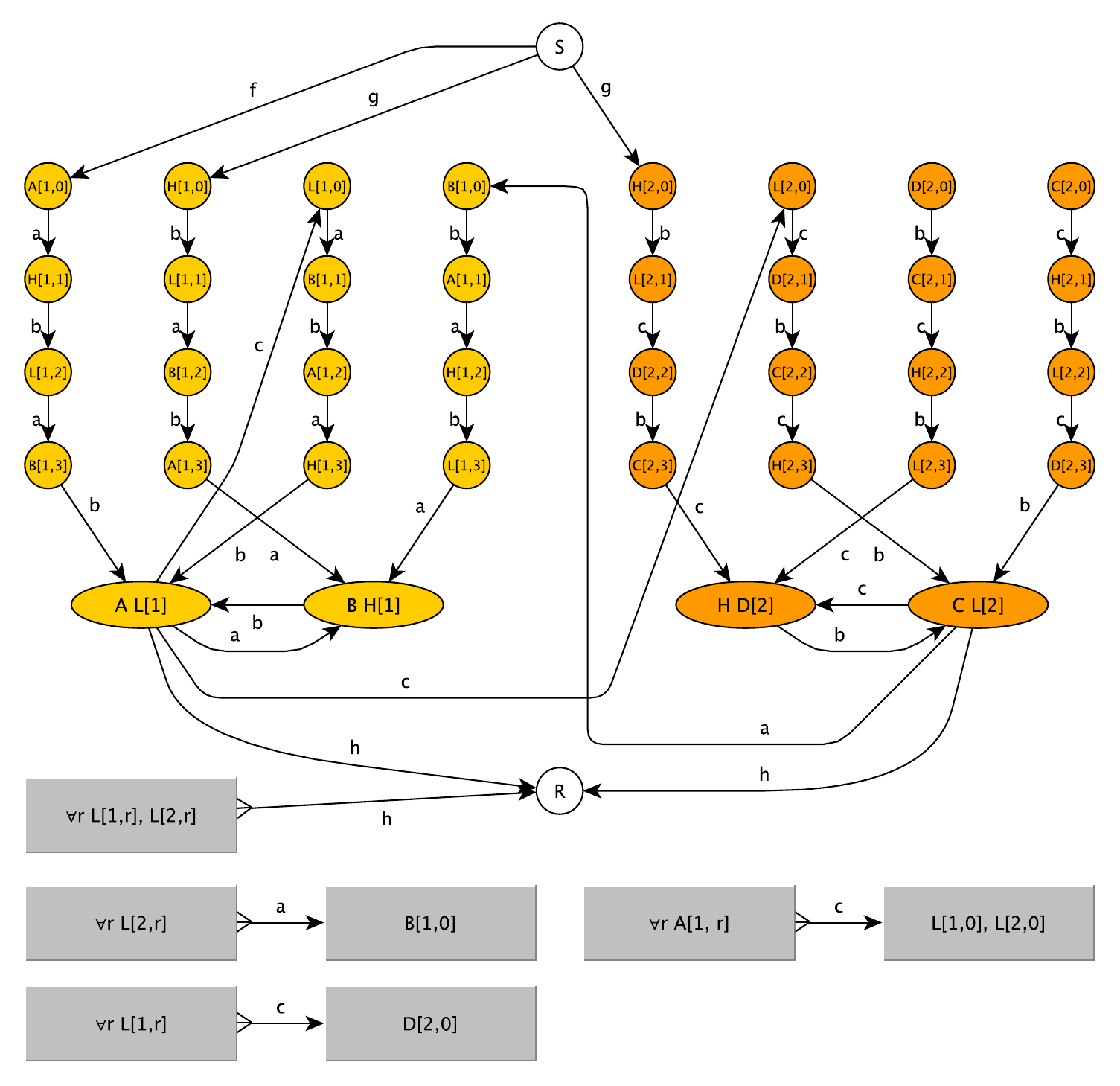}
  \caption{The $\overline{\mathcal C}(G^L)$ fragment derived from the $\mathcal C(G^L)$ and $\hat{\mathcal C}(G^L)$ of Example~\ref{ex:beta}. The gray boxes represent a collection of source or target states with the names indicated in the box.}
  \label{fig:nex1-2}
\end{figure}

To illustrate the main features of the above construction, as a first example, consider again the fragment of Example~\ref{ex:beta}: the corresponding fragment of $\overline{\mathcal C}(G^L)$ is depicted in Figure~\ref{fig:nex1-2}; see also the further Example~\ref{ex:double-transition}.

The following example, instead, explains why we introduced the pipelines as an input for counter sequence states.

\begin{exa}\label{ex:pipelines}
The control graph of Figure~\ref{fig:contres1} has shown that simply collapsing the states of a counter sequence into a single state produces undesired side effects, such as spurious counters. 
A first repair could consist in keeping the original states (of $\hat{\mathcal C}(G^L)$) and using them as an entry for the compound states, in some sense, a pipeline of length 1.

This solution too, however, is not enough. Consider, for instance, the fragment of control graph in Figure~\ref{fig:contr-pipe1} (left), no matter whether representing a descending or an ascending fraction of the whole graph; 
it contains just one counter table with counters $(AC, ab)$ and $(BD, ba)$; thus, the corresponding fraction of $\hat{\mathcal C}(G^L)$) is isomorphic to the original graph.
A possible version of $\overline{\mathcal C}(G^L)$) making use of single states to enter the counter sequence states is given in Figure~\ref{fig:contr-pipe2} (right)
which shows a new counter table with counters $(AP, ac)$ and $((BD)Q, ca)$ which do not correspond to the behavior of the original control graph.

\begin{figure}
\begin{tabular}{m{0.4\textwidth}m{0.5\textwidth}}
  \includegraphics[scale=0.5]{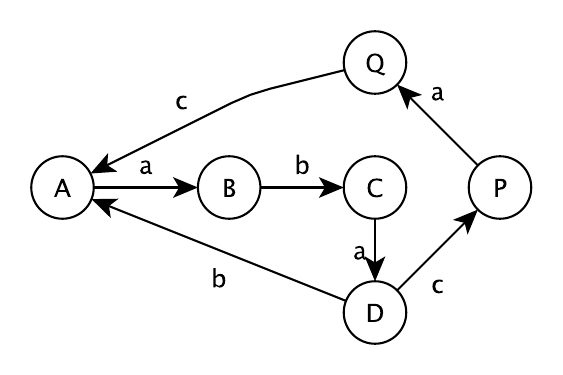}
  &
  \includegraphics[scale=0.5]{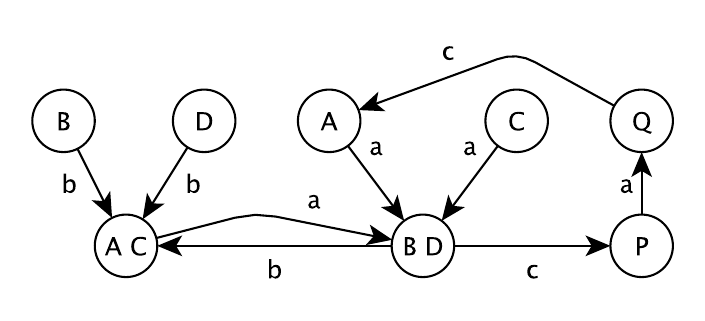}
\end{tabular}
    \caption{A fragment of control graph with one counter table (left), and
an erroneous attempt to build a $\overline{\mathcal C}(G^L)$ version of the control graph fragment (right).
    }
    \label{fig:contr-pipe1}
    \label{fig:contr-pipe2}
\end{figure}

The source of the problem abides in the fact that the path $cac$ reentering state $A$ after leaving $BD$ ``forgot'' that its source was $D$, not $B$; thus, it can go on in a way that does not separate the two cases.
The construction of $\overline{\mathcal C}(G^L)$) making use of the full pipelines, on
the contrary, ``compels'' to reenter the counter from scratch, i.e., from the
``real'' $A$, from which it would not be possible to bypass the path $aba$ to reach again the state $BD$.
This is why counters may be entered only through their entry points.

\end{exa}

Finally the example below points out that in some cases the same transition can be used to follow the path of a counter table, but also to exit it, depending on the context within which it occurs.

\begin{exa}\label{ex:double-transition}
Consider the counter table, say the $i$-th, consisting of the transition sequence
$A \boldtrans{a} B$,
$B \boldtrans{b} C$,
$C \boldtrans{c} B$,
$B \boldtrans{a} D$,
$D \boldtrans{b} E$,
$E \boldtrans{c} A$.
It produces pipelines with two occurrences of symbol $B$ with different indices as shown in Figure~\ref{fig:nex1-3}; 
this happens because the same transition, e.g., $B \boldtrans{b} C$ is used both to follow the path of the counter table, 
as in the above sequence, but could also exit it if applied after transition $C \boldtrans{c} B$. 
Notice that, as a consequence, the figure displays two different states with the same name, $ B[i, 5]$: we decided to tolerate this ``innocuous homonymy'' to avoid a further state renaming. 
\begin{figure}
  \centering
  \includegraphics[scale=0.6]{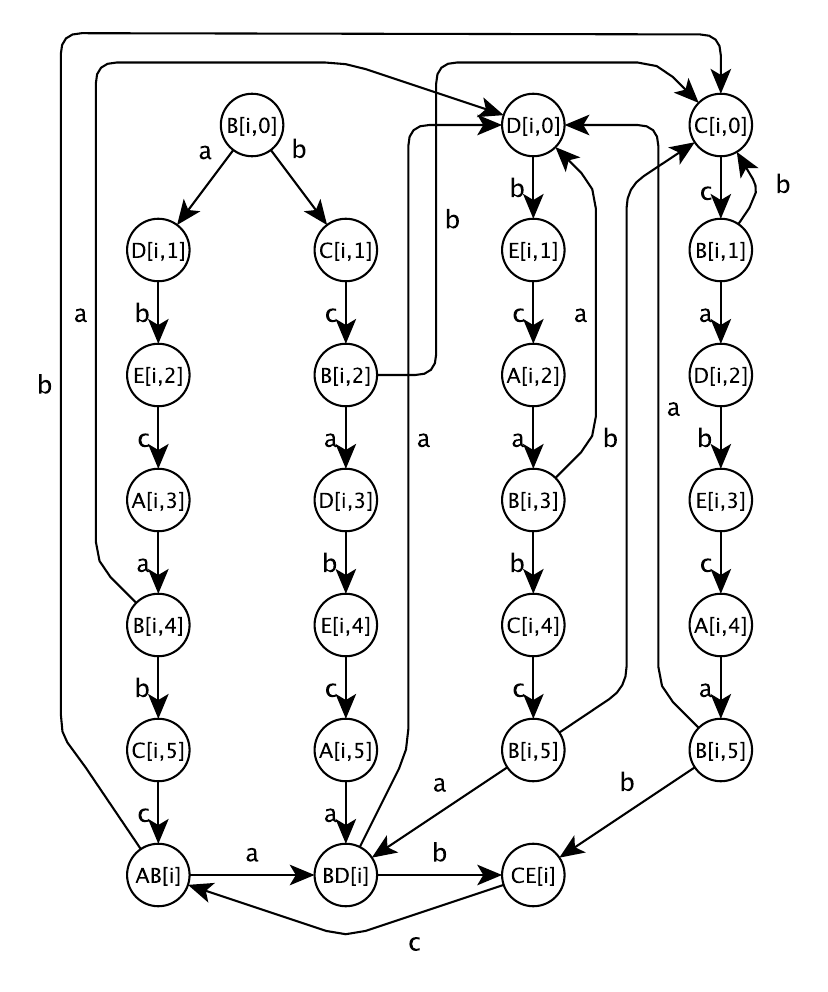}
  \caption{A significant fragment of the $\overline{\mathcal C}(G^L)$
  derived from the transition sequence 
$A \boldtrans{a} B$,
$B \boldtrans{b} C$,
$C \boldtrans{c} B$,
$B \boldtrans{a} D$,
$D \boldtrans{b} E$,
$E \boldtrans{c} A$. 
For simplicity other similar pipelines have been omitted.
}\label{fig:nex1-3}
\end{figure}
\end{exa}

\begin{lem}\label{lemma:rr}
For any nonterminal $A$ of $G^L$, the regular languages consisting of all paths of $\overline{\mathcal C}(G^L)$ 
going from anyone of $\dd A$, $\dd A[i,r]$, $\dd X[i]$, with $ A \in \dd X[i]$ 
to anyone of $\uu A$, $\uu A[f,r]$, $\uu X[f]$, with $ A \in \uu X[f]$ are NC.
\end{lem}
\begin{proof} 
    The original ``pure counters'' of $\hat{\mathcal C}(G^L)$ have been ``broken'' by
    replacing the arrows that would complete the string $u^k$ with transitions that enter a loop accepting $u^*$.
    Thus, any pipeline associated with a counter whose string is $u$ accepts sequences $u^m$, 
    with $m \geq k$. All paths of $\overline{\mathcal C}(G^L)$ 
    that do not touch counter sequence states existed in $\mathcal C(G^L)$ too up to the homomorphism that erases the indexes of the duplicated states.
   
   The only transitions that are not replicas of transitions existing in $\hat{\mathcal C}(G^L)$ (and in $\mathcal C(G^L)$) are those exiting the counter sequence states since they are derived from transitions originating by \emph{some} of the states belonging to the counter sequence, say $X$. 
   If such transitions originate paths that do not lead to any pipeline, i.e., that do not correspond to $\mathcal C(G^L)$'s paths leading to some counter table, 
   then such paths cannot contain any counter since they simply replicate $\mathcal C(G^L)$'s paths with no counters.
   Suppose, instead, that such a path, after reading a string $z$, reaches the entry point of a pipeline which, through a string $v^j$ leads to a new counter: 
   thus, the reading of $z$ is only a finite prefix of a path that leads from a counter sequence to another one 
   (if instead the path of the pipeline reading $v^j$ is abandoned before reaching the counter sequence state, it continues by replicating a path that existed already in $\mathcal C(G^L)$ without counters, up to a renaming of some states). 
   Notice that, as a particular case the new counter string $v$ could be $u$ again but referring to a different counter table, therefore with disjoint states.
   
   As a further special case, however, it could even happen that $z$ is $u^s$ 
   (it cannot be $u = z^s$ because by convention, $u$ is the minimal string that can be associated with the counter table --- see Definition~\ref{def:counterTable}) and, by reading $z$,  $\overline{\mathcal C}(G^L)$ re-enters a pipeline of the same table so that after going through the whole pipeline we reach again state $X$. 
   In this case we would have closed a loop from $X$ to $X$ by reading the string $u^{s+k}$, thus, $\overline{\mathcal C}(G^L)$  would not be counter free. Nevertheless, it is aperiodic since, together with $u^{s+k}$ we would also find all strings $u^{s+k+n}$ for any $n \geq 0$ because from $X$ we can read any string in $u^*$.
\end{proof}   
    
At this point it would be possible to prove again Theorem~\ref{Th:GtoMSO} and its Corollary~\ref{cor:GtoMSO} for any $G^L$ by suitably replacing formulas $\varphi_A$ with formulas referring to $\mathcal{\overline C}(G^L)$ instead of $\mathcal{C}(G^L)$. 
We would thus obtain FO definability of linear NC OPLs. This result however, has already been obtained with much less effort in~\cite{MPC20}. 
Here we want to achieve the general result for any NC OPL.

\subsection{NC control graph for general NC OPGs}
Let now $G$ be a BDR OPG, $G^L$ its associated linearized OPG, $\mathcal C(G^L)$ the original
control graph of $G^L$ and $\hat{\mathcal C}(G^L)$, $\overline{\mathcal C}(G^L)$ its respective transformations obtained
through their constructions (remember that $\overline{\mathcal C}(G^L)$ has been built starting from
$\hat{\mathcal C}(G^L)$). A new OPG 
$G'=(\Sigma, V_N',  P', S')$
structurally equivalent to $G$ is built according to the following procedure.

\paragraph*{Construction of $G'$}
\begin{itemize}
\item The nonterminal alphabet of $G'$, $V_N'$ consists of:

    \begin{itemize}
        \item All pairs $(\dd A, \uu A)$ where $\dd A$,
            $\uu A$ are \emph{singleton states} of $\overline Q$, i.e., 
            states of $\overline{\mathcal C}(G^L)$ other than counter sequence states.
           They include also singleton states belonging to pipelines, i.e., states of type $\dd {A}[i,r]$ or $\uu {A}[f,s]$ if $A$ belongs to some descending or ascending counter.
        

        \item All pairs $(\dd X_A,\uu A)$, $(\dd A, \uu X_A)$ where $\dd A$ and $\uu A$ are singleton states of $ \overline Q$ not belonging to any descending, resp.\ ascending, counter  and
            $\dd X_A$ and
            $\uu X_A$ are the counter sequence states containing $\dd A$ and $\uu A$, respectively.

        \item 
        The pairs $(\dd X_A, \uu X_A)$, $(\dd X_A, \uu A[f, s])$, $(\dd A[i,r ], \uu X_A)$ where $\dd X_A$ and $\uu X_A$ are the  counter sequence states belonging to two \emph{paired} counter tables $\mathcal T[i]$, $\mathcal T[f]$ and $(\dd A[i,r ]$, $\uu A[f, s])$ are elements of the corresponding pipelines.
        Thanks to Lemma~\ref{lemma:qq}, $(\dd X_A, \uu X_A)$ uniquely identifies a nonterminal $A$ of $G$.
        \item
        The same elements as in the point above where $\dd X_A$ and $\uu X_A$ are the  counter sequence states belonging to two \emph{non-paired} counter tables $\mathcal T[i]$, $\mathcal T[f]$, with the exclusion of the pair $(\dd X_A, \uu X_A)$.
        Notice that, if the counter tables are not paired, Lemmas~\ref{lemma:gg} and~\ref{lemma:qq} do not apply; thus, it might happen that $\dd X_A$ and $\uu X_A$ share more that one nonterminal of $G$.
  
      
      \end{itemize}         
\item For convenience, in the following construction we use the notation $\dd{[X_A]}$ (resp., $\uu{[X_A]}$) to denote either the singleton state $\dd A$ (resp.\ $\uu A$) or any counter sequence state
  $X_A$ containing $A$, or any element of the corresponding pipelines.\\

\item
    For every production $A \to x$ of G the following productions are in $P'$, for all $\dd{[X_A]}$:

   \begin{itemize}
   \item 
   if $A$ does not belong to any ascending counter, then $( \dd{[X_A]}, \uu A )\to x$;
   \item if $A$ belongs to an ascending counter, say $f$, then
       $( \dd{[X_A]}, \uu{A[f, 0]}) \to x$  (see point 7 of $\overline{\mathcal C}(G^L)$'s construction).
 \end{itemize}  

\item For every production $A \to B_0 x_1 B_1 \ldots x_n B_n$ of $G$ (with $x_i \in W$),
where, as usual, $B_0$ or $B_n$ may be missing, consider the following cases:
\begin{enumerate}
\item \label{G'-1}
$A$ does not belong to any counter, either descending or ascending.
Then the following productions are in $P'$:

$( \dd A, \uu A )\to  (\dd{[Y_{B_0}]}, \uu{[Y_{B_0}]} ) x_1 
              \ldots x_n
                ( \dd{[Y_{B_n}]}, \uu{[Y_{B_n}]} )$ 
 where, for each $h$, $\dd{[Y_{B_h}]}$ is $\dd B_h$ if $B_h$ does not belong to
 any descending counter, $\dd{B_h}[i,0]$ for any $i$ such that $B_h$ belongs to
 a counter table $\mathcal T[i]$. The $\uu{[Y_{B_h}]}$ components are all the ones defined in $V_N'$.
 \item \label{G'-2}
 $A$ belongs to a descending counter table $\mathcal T[i]$ but not to any ascending one.
Then the following productions are in $P'$:
\begin{itemize}
\item
if no $B_h$ belongs to $\mathcal T[i]$, then

$( \dd {[X_{A}]}, \uu A )\to  (\dd{[Y_{B_0}]}, \uu{[Y_{B_0}]} ) x_1 
              \ldots x_n
                ( \dd{[Y_{B_n}]}, \uu{[Y_{B_n}]} )$
where $\dd {[X_{A}]}$ stands for all $\dd A[i,r]$ plus $\dd {X_A}[i]$, and for each $h$, with $1 \leq h \leq n$,
$\dd{[Y_{B_h}]}$ is $\dd B_h$ if $B_h$ does not belong to any descending counter, $\dd{B_h}[l,0]$ for any $l$ such that $B_h$ belongs to a counter table $T[l]$, with $l \neq i$. 
The $\uu{[Y_{B_h}]}$ components are all the ones defined in $V_N'$.
\item
if there exists a $h$ such that $B_h$ belongs to $\mathcal T[i]$
---there can be at most one such $h$ 
because $\mathcal C(G^L)$ describes only paths through the STs of $G$ going from the root to a leaf and back and 
$\hat{\mathcal C}(G^L)$ ``separates'' possible intersecting counter tables from each other---
then

              $( \dd {[X_{A}]}, \uu A )\to  (\dd{[Y_{B_0}]}, \uu{[Y_{B_0}]} ) x_1 
              \ldots x_n
              ( \dd{[Y_{B_n}]}, \uu{[Y_{B_n}]} )$ 
 where if $\dd {[X_{A}]}$ is $\dd A[i,r]$, with $0 \leq r \leq p-1$, where $p$
 is the length of the pipeline,
 $\dd{[Y_{B_h}]}$ is $\dd B_h[i,r+1]$; if $\dd {[X_{A}]}$ is $\dd A[i,p]$
 or $\dd {X_{A}}[i]$ $\dd{[Y_{B_h}]}$ is $\dd {Y_{B_h}}[i]$; all remaining
 elements of the rhs, including $\uu{[Y_{B_h}]}$, are as in the previous item.
\end{itemize}
\item \label{G'-3}
$A$ belongs to an ascending counter table $\mathcal T[f]$ but not to any descending one. 
Then the following productions are in $P'$:
\begin{itemize}
\item 
If none of the $B_h$ belongs to $\mathcal T[f]$ then the lhs is $(\dd A, \uu A[f,0])$ and
the nonterminals $(\dd{[Y_{B_h}]},\uu{[Y_{B_h}]} )$ of the rhs are defined in the same way as in point (\ref{G'-1}) above.
\item
If there exists a \emph{unique}
$B_h$ belonging to $\mathcal T[f]$, then

 $( \dd A, \uu {[X_{A}]})\to  (\dd{[Y_{B_0}]}, \uu{[Y_{B_0}]} ) x_1 
              \ldots x_n
              ( \dd{[Y_{B_n}]}, \uu{[Y_{B_n}]} )$ 
              
where if $\uu {[Y_{B_h}]}$ is $\uu B_h[f,s]$, with $0 \leq s \leq p-1$,
$\uu {[X_{A}]}$ is $\uu A[f,s+1]$; if $\uu {[Y_{B_h}]}$ is $\uu B_h[f,p]$
or $\uu {Y_{B_h}}[f]$, $\uu {[X_{A}]}$ is $\uu {X_{A}}[f]$; all remaining
elements of the rhs, including $\dd{[Y_{B_h}]}$, are as in the previous bullet. 
\end{itemize}
\item
The case where $A$ belongs to a descending counter table $\mathcal T[i]$ and to a \emph{paired} ascending one $\mathcal T[f]$ can be treated as a natural combination of the previous  (\ref{G'-2}) and (\ref{G'-3}), keeping in mind Lemma~\ref{lemma:qq}.

Notice that, if the derivation involving the two paired counter tables is long enough 
---precisely, more than $2  \cdot k_i \cdot j_i = 2 \cdot k_f \cdot j_f$, where $k_i$, $j_i$, resp.\ $k_f$, $j_f$, are the order and the number of $W$'s elements of the descending, resp.\ ascending table--- 
then a number of consecutive nonterminals of $G'$ associated with $G$'s nonterminal $A$ will be of type $(\dd X_A, \uu X_A)$; 
the same will happen for the horizontal permutations of the counter which $A$ belongs to.

\item \label{G'-5}
$A$ belongs to a descending counter table $\mathcal T[i]$ and to an ascending one $\mathcal T[f]$ that are \emph{not paired}. In this case only one of the two tables can be followed by the derivation. 
In other words, a derivation $A \xLongrightarrow[] {*}  u^kAv $ is interrupted to move to another ``semicounting derivation'' $A \xLongrightarrow[] {*}  zAw^h $, possibly partially overlapping. 
In this case both possibilities are applied: 
all elements $\uu {[X_{A}]}$ of the ascending pipeline, including the counter sequence state, are paired with singleton elements of the descending pipeline \emph{excluding} the counter sequence state, and conversely, in all compatible ways.
The elements of the rhs are built in the same way as in points~\eqref{G'-2} and~\eqref{G'-3} above, respectively.

For instance, if $A$ belongs to a descending counter $(\dd A \dd B[1],a)$ 
and to an ascending one $(\uu A \uu C[2],b)$ a production $A \to aBb$
becomes the following $G'$'s productions
$( \dd {[X_{A}]}, \uu {[X'_{A}]} )\to  a (\dd{[Y_{B}]},\uu{[Y_{B}]} ) b$, 
$( \dd {[X'_{A}]}, \uu {[X_{A}]} )\to  a (\dd{[Y_{B}]},\uu{[Y_{B}]} ) b$
where $\dd {[X_{A}]}$ (resp.\ $\uu {[X_{A}]}$) stands for any element of the descending (resp.\ ascending) pipeline, \emph{including}
$\dd {AB}[1]$ (resp.\ $\uu {AC}[2]$) and 
$\dd {[X'_{A}]}$ (resp.\ $\uu {[X'_{A}]}$) stands for any element of the descending (resp.\ ascending) pipeline, 
\emph{excluding} $\dd {AB}[1]$ (resp.\ $\uu {AC}[2]$).\footnote{And, of course, the $(\dd{[Y_{B}]},\uu{[Y_{B}]})$ 
respect the rules stated in points~\eqref{G'-2} and~\eqref{G'-3} above.} 
See also Example~\ref{ex:crossing-counters}.
\end{enumerate}
\item
The axioms of $G'$ are:
\begin{itemize}
\item
the pairs $(\dd A, \uu A)$ where $A$ is an axiom of $G$ that does not occur in any counter table, whether descending or ascending;
\item
all pairs $(\dd A, \uu{[X_A]})$ where $A$ is an axiom of $G$ that does not occur in any descending counter table but occurs in some ascending ones;
\item
all pairs $(\dd A[i,0], \uu{[X_A]})$ where $A$ is an axiom of $G$ that belongs to the descending counter table $\mathcal T[i]$ and $\uu{[X_A]})$ denotes either $\uu A $ or any element of an ascending pipeline ---including the counter sequence set--- depending on whether or not $A$ belongs to some ascending counter table.
 \end{itemize}
\end{itemize}

Intuitively, $G'$ splits all of $G$'s nonterminals into pairs representing elements of  $\mathcal C(G^L)$'s descending and ascending paths involving the same nonterminal of $G$.
If one of $\mathcal C(G^L)$'s states belongs to a counter sequence,
this is recorded in the name of the new nonterminal symbol which can be an element of the corresponding pipeline of $\overline{\mathcal C}(G^L)$. If a derivation is following a descending or an ascending path of the syntax tree that is part of a counter table, say the $i$-th, then that part of the path must obey the constraints given by the $i$-th pipeline. Such constraints are given by $\overline{\mathcal C}(G^L)$ since all paths root-to-leaves and back of $G'$ are the same as those of $G^L$.
Notice that, whereas $G$ is BDR, $G'$ is not; it may also contain useless nonterminals.

The following examples illustrate the whole grammar transformation procedure.

\begin{exa}\label{ex:nnn}
Consider again the  grammar $G_{NL}$ of Examples~\ref{ex:contr-graph},~\ref{ex:MSOformula} 
and its linearized version $G^L_{NL}$ 
of Example~\ref{ex:linearization}.

The control graph of $G^L_{NL}$ is given in Figure~\ref{fig:C(GNLL)}: 
it exhibits three ascending counters $(\uu A \uu B, c\overline A )$, 
$(\uu A \uu B, c\overline B )$, $(\uu A \uu B, \overline \varepsilon_R )$; 
notice that the third one has no impact on the counting property since 
we also have the self loops 
$\uu A \boldtranss{\pmb{\overline\delta}}{\overline \varepsilon_R} \uu A$,
$\uu B \boldtranss{\pmb{\overline\delta}}{\overline \varepsilon_R} \uu B$.
The corresponding $\overline {\mathcal C}(G^L_{NL})$ is given in Figure~\ref{fig:C-bar(GNLL)}.

 \begin{figure}
  
 \centering
  \includegraphics[scale=0.6]{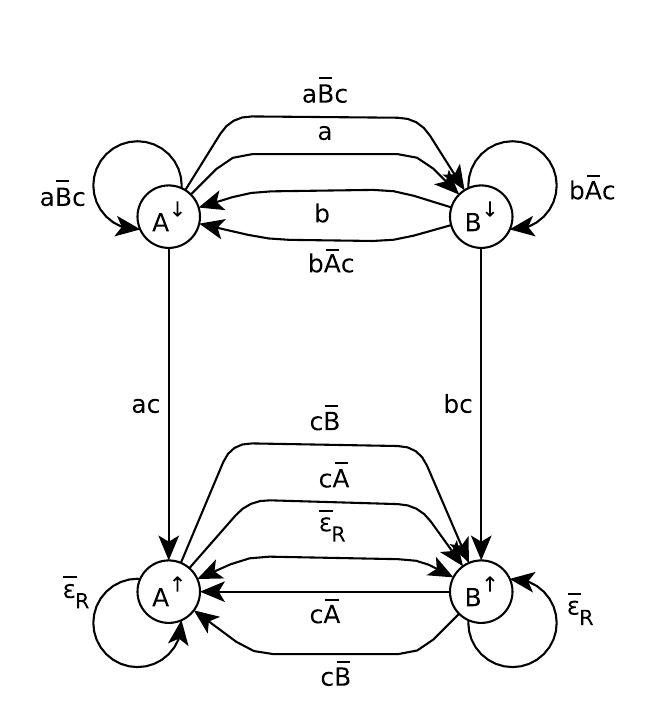}
\caption{The control graph $\mathcal C(G^L_{NL})$}
\label{fig:C(GNLL)}
\end{figure}

 \begin{figure}
  \centering
  \includegraphics[scale=0.6]{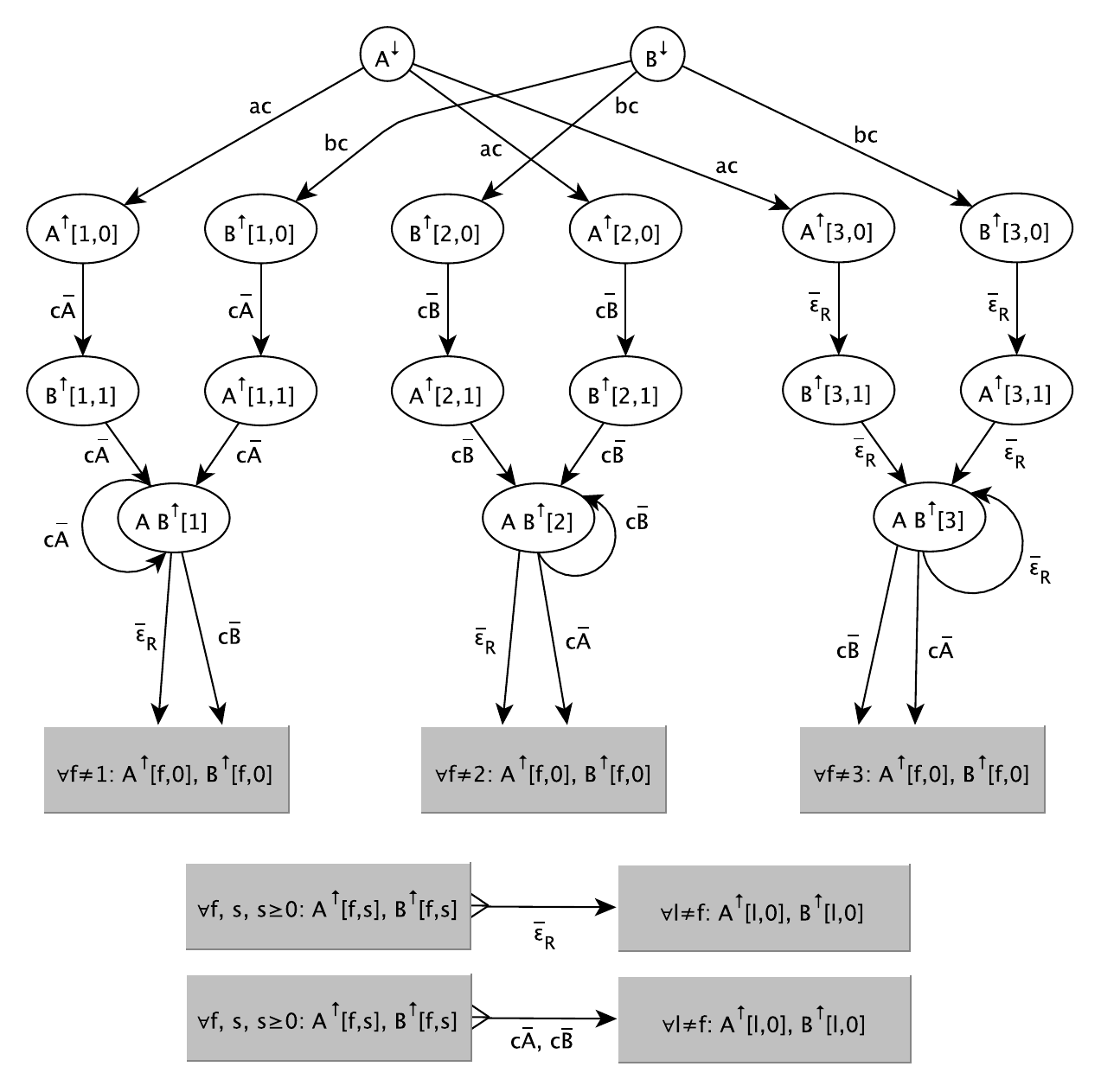}
 \caption{The control graph $\overline {\mathcal C}(G^L_{NL})$. The upper part of the graph concerning the descending paths is not reported being identical to the original one of $\mathcal C(G^L_{NL})$.}
\label{fig:C-bar(GNLL)}
\end{figure}

$G_{NL}'$'s nonterminal alphabet is the set:\\
$\{( \dd A, \uu {A}[f,s] ), ( \dd A, \uu  {AB}[f]  ), 
 ( \dd B, \uu {B}[f,s] ),  ( \dd B, \uu {AB}[f]  ) \mid f = 1, 2, 3, s = 0, 1\}$,

A significant sample of $G_{NL}'$'s rules is given below.

$( \dd A, \uu {A[f,0]} ) \to ac$

$( \dd B, \uu {B[f,0]} ) \to bc$

From the original $G$'s rule $A \to aBcB$ we obtain the following rules, where $[\uu{Y_{B}}[f]]$, resp.\ $[\uu{Y_{B}}[l]]$, stands for either $\uu {B}[f,1]$ or $\uu {B}[f,0]$ or $\uu {AB}[f]$, resp.\ $\uu {B}[l,1]$ or $\uu {B}[l,0]$ or $\uu {AB}[l]$, with $f, l = 1,2,3, f \neq l, h \neq f, l$:

$( \dd A, \uu {A}[h,0] ) \to a  ( \dd B, [\uu{Y_{B}}[f]] ) c  ( \dd B, [\uu{Y_{B}}[l]] )$, 

$( \dd A, \uu {A}[f,1] ) \to a  ( \dd B, \uu{B}[f,0] ) c  ( \dd B, [\uu{Y_{B}}[l]])$,

$( \dd A, \uu {AB}[f] ) \to a  ( \dd B, \uu{B}[f,1] ) c  ( \dd B, [\uu{Y_{B}}[l]])$,

$( \dd A, \uu {A}[l,1] ) \to a  ( \dd B, [\uu{Y_{B}}[f]] ) c  ( \dd B, \uu{B}[l,0])$,

$( \dd A, \uu {AB}[l] ) \to a  ( \dd B, [\uu{Y_{B}}[f]] ) c  ( \dd B, \uu{B}[l,1])$.

The rationale of the construction is that any (ascending, in this case) counter can be interrupted leading only to the entry point of a \emph{different} counter (or to a state not belonging to any counter, in the general case). 
If instead we are following a specific counter marked by its index $f$, the sequence of the states (in this case the ascending component of $G'$'s nonterminal) must follow the sequence imposed by the $f$-th pipeline,
whereas the other nonterminals, which correspond to the $\overline B$ terminals of $G^L_{NL}$, may be of any type. The remaining rules of $G'$ should now be easily inferred by analogy.

\end{exa}

The following example instead highlights the ambiguity of $G'$ as a consequence of introducing repeated rhs and the case of a grammar nonterminal belonging to both an ascending and a descending counter, but not paired.

\begin{exa}\label{ex:crossing-counters}
Consider the following grammar $G_\text{cross}$, with $S = \{A, B\}$:
$A \to a B c$, 
$B \to a A b  \mid a C b \mid h$,
$C \to dBb$.

It is easy to realize that $\mathcal C(G_\text{cross}) $ has a descending counter
$\dd C_1 = (\dd A \dd B, a) $ and an ascending one $\uu C_2 = (\uu B \uu C, b)$. 
Notice that nonterminal $B$ occurs in both counter tables but the two counters it belongs to are not paired.
Without providing explicitly the whole grammar $G_\text{cross}'$ we display
$\overline{\mathcal C}(G_\text{cross})$ in Figure~\ref{fig:C(Gcross)}.

\begin{figure}
  \centering
  \includegraphics[scale=0.6]{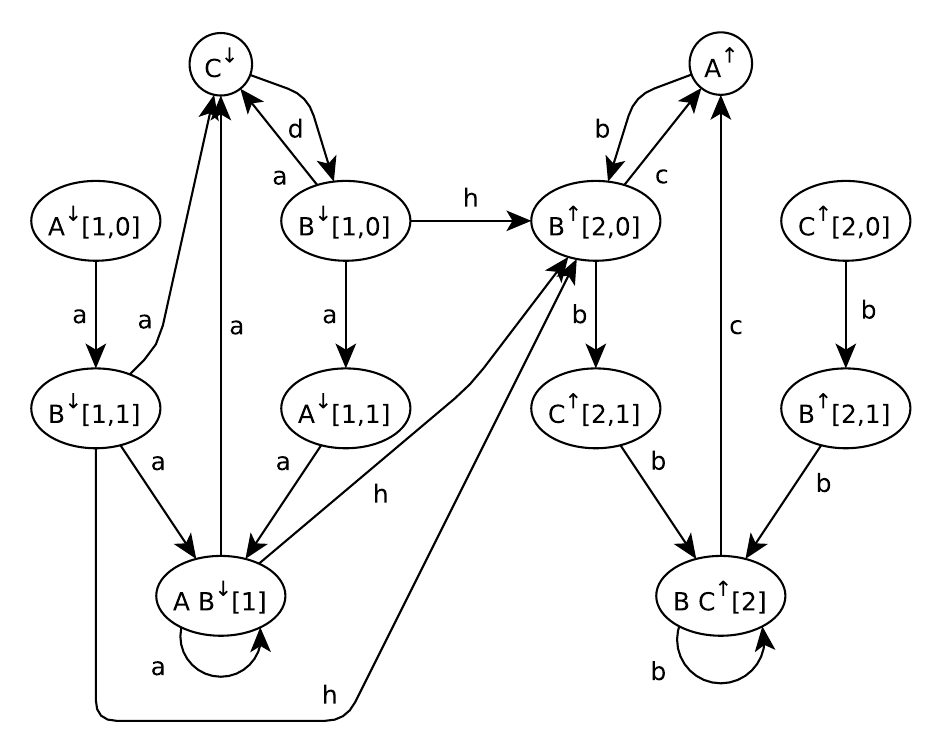}
\caption{The control graph $\overline{\mathcal C}(G_\text{cross})$. Notice that $\uu C[2,0]$ and $\uu B[2,1]$ are unreachable.}
\label{fig:C(Gcross)}
\end{figure}

A first derivation of $G_\text{cross}$ is 
$B \xLongrightarrow[G_\text{cross}]{} h$. Since $B$ is an axiom of $G_\text{cross}$, $h \in L(G)$. In $G_\text{cross}'$ $h$ can be derived ---in one step--- by the lhss
$( \dd {B}[1,0], \uu{B}[2,0] ) $, 
$( \dd {B}[1,1], \uu{B}[2,0] ) $, 
$( \dd {AB}[1], \uu{B}[2,0] ) $; 
however, since only $( \dd {B}[1,0], \uu{B}[2,0] ) $ is an axiom of $G'$, $h$ can be derived as a string of $L(G')$ only through that nonterminal; 
the derivation $( \dd {AB}[1], \uu{B}[2,0] )  \xLongrightarrow[G_\text{cross}']{} h$,
instead, could be used elsewhere as part of a longer $G_\text{cross}'$ derivation. 
The fact that in the lhs of $G_\text{cross}'$ rule occur the labels of two different counter tables denotes the possibility that it belongs to two different counters.

Imagine now that $h$ occurs in the context $d-b$. This means that $dhb$ has been derived in $G_\text{cross}$ by $C \xLongrightarrow[G_\text{cross}] {2} dhb$;
thus, no ambiguity remains and the only possible lhs for all rhs  
$d( \dd {B}[1,0], \uu{B}[2,0] ) b$, 
$d( \dd {B}[1,1], \uu{B}[2,0] ) b$, 
$d( \dd {AB}[1], \uu{B}[2,0] ) b$ is  $( \dd {C}, \uu{C}[2,1] ) $.

The next derivation step of $G_\text{cross}$ necessarily involves reducing the rhs $aCb$ to $B$. 
This step, however, could be a further step of the ascending counter $C_2$ or could interrupt the ascending counter and become an exit step from the descending counter $C_1$. Thus, we have two possible groups of lhs for
 $a( \dd {C}, \uu{C}[2,1] ) b$, namely 
  $\{( \dd {B} [1,1],\uu{BC}[2] ) $, 
  $( \dd B [1,0], \uu{BC}[2] ) \}$
  and
  $\{( \dd B [1,1] ,\uu{B}[2,0] ) $, 
  $( \dd B[1,0] , \uu{B}[2,0] ) $,
 $( \dd{AB} [1], \uu{B}[2,0] )  \}$.
 Notice, instead, that point~\ref{G'-5}. of $G'$ construction excludes the lhs $( \dd {AB[1]}, \uu{BC}[2] )$   which would be superfluous.
 
If the next reduction involves the context $a-c$ only $C_1$ will be followed by applying ambiguously one of the rules
 
 
 
 
 
 
 
$( \dd A [1,0], \uu{A} ) 
 \to a( \dd B [1,1], \uu{B}[2,0] ) c$,
 
 $( \dd A [1,1], \uu{A} ) 
 \to a( \dd{AB} [1], \uu{B}[2,0] ) c$,
 
 $( \dd {AB} [1], \uu{A} ) 
 \to a( \dd {AB} [1], \uu{B}[2,0] ) c$,
 
 $( \dd A[1,0], \uu{A} ) 
 \to a( \dd B[1,1], \uu{BC}[2] ) c$,
 
 

where the last production could be used in a derivation where the counter $\dd C_1$ is being followed but subsequently interrupted to conclude an instance of $\uu C_2$.
Symmetrically, if the next reduction involves the context $d-b$ only $C_2$ will be followed.

\noindent \emph{Remark}.
Notice that the construction of $G'$ produces in its control graph a transition $\uu C[2,1] \boldtrans{b} \uu B[2,0] $ ---and more--- that has no correspondent transition in $\overline {\mathcal C}(G^L_{cross})$. This is due to the fact that in this case the ascending pipeline could be interrupted but \emph{potentially } immediately resumed. Such new transitions could generate new counters which however would not make the control language counting as we already pointed out in Lemma \ref{lemma:rr}; see also the following Theorem \ref{theorem:L(CG)-non-counting}.
\end{exa}

\begin{lem}\label{lemma:gramm-equiv}
Let $G$ be a BDR OPG and $G'$ the grammar derived therefrom according to the above procedure.\\ 
For every $A \in V_N$, $A \xLongrightarrow[G] {*} x$ iff 
for some $( \dd{[X_A]}, \uu {[X_A]} ) $, $( \dd{[X_A]}, \uu {[X_A]} ) \xLongrightarrow[G'] {*} x$.
\end{lem}
\begin{proof}
    \noindent {\em Base of the induction. }\/ 
    By construction of $G'$, $A \xLongrightarrow[G] {} x$ iff for all $ \dd{[X_A]}$, either
    $( \dd{[X_A]}, \uu A ) \to x$, or $( \dd{[X_A]}, \uu A[f,0] ) \to x$, for any $f$ such that $A$ belongs to an ascending counter table $T[f]$.
    Moreover, by construction of $\overline{\mathcal C}(G^L)$,
    $\dd{[X_A]} \boldtranss{\overline{\pmb\delta}}{x} \uu A$
    or
    $\dd{[X_A]} \boldtranss{\overline{\pmb\delta}}{x} \uu A[f,0]$,
    for all $\dd{[X_A]}$, whether the ---possible--- corresponding counter table $\mathcal T[i]$ is paired with 
    $\mathcal T[f]$ or not.

    \noindent {\em Inductive step. }
    \begin{enumerate}
    \item From $G'$ to $G$.  Assume that for $m \le p$ and for each
     $A \in V_N$, $( \dd{[X_A]}, \uu{[X_A]} )   \xLongrightarrow[G'] {m} x$ for some $( \dd{[X_A]}, \uu{[X_A]} )$, implies $A \xLongrightarrow[G] {m}x$.
     Consider a derivation $( \dd{[X_A]}, \uu{[X_A]} ) \xLongrightarrow[G'] {*}$ 
      $x_1 ( \dd{[Y_{B_1}]}, \uu{[Y_{B_1}]} )$ $x_2 \ldots$ 
      $( \dd{[Y_{B_n}]}, \uu{[Y_{B_n}]} ) \xLongrightarrow[G'] {*}
       x_1 \ldots x_n w_n$,  with $( \dd{[Y_{B_h}]},$ $ \uu{[Y_{B_h}]} ) \xLongrightarrow[G'] {m_h} w_h$, $m_h \le p$, $x_h \in W$, $1 \leq h \leq n$ (notice that $W$ is the same for both $G$ and $G'$); 
      for simplicity, we treat only the case where $( \dd{[Y_{B_0}]}, \uu{[Y_{B_0}]} )$ is missing and  $( \dd{[Y_{B_n}]}, \uu{[Y_{B_n}]} )$ is present since the other cases are fully similar. 
       
       By the induction
      hypothesis $B_h\xLongrightarrow[G] {*}w_h$.
      By construction of $\overline{\mathcal C}(G^L)$, for some $\dd{[X_A]}$, $\uu{[X_A]}$, 
      $\dd{[Y_{B_h}]}$, $\uu{[Y_{B_h}]}$
      the following transitions are in $\overline {\pmb \delta}$:

      $\dd{[X_A]} \boldtrans{x_1}  \dd{[Y_{B_1}]}$,
      $\uu{[Y_{B_n}]} \boldtrans{ \overline\varepsilon_R} \uu{[X_A]}$; 
      
      $\uu{[Y_{B_1}]} \boldtrans{ x_{2} \overline B_{2} \ldots \overline B_{h-1} x_{h}} \dd{[Y_{B_{h}}]}$, 
             $2 \leq h \leq n$;
             
         $\uu{[Y_{B_h}]} \boldtrans{ x_{h+1} \overline B_{h+1} \ldots \overline B_n} \uu {[X_A]}$, 
           $1 \leq h \leq n-1$.
 
          By construction of $G'$, if $\uu{[X_A]}$ is an $\uu{X_A} [f]$ or $\uu{A}[f,s]$ for some $f,s$ with $s > 0$, then,
          for a unique $h$,
            $\uu{[Y_{B_h}]}$ is $\uu{B}[f,l]$, where $l$ is the length of the corresponding pipeline, or $\uu{B}[f,s-1]$, 
            respectively (see point (3) of $G'$'s construction). 
            Otherwise there are no constraints between the pipeline indexes of the nonterminals of the rhs and that of the lhs.
      This means that for some $\dd D$ in $\dd{[X_A]}$, $\dd H_h$ in $\dd{[Y_{B_h}]}$,
      $D$ was lhs of a production of $G$ such as $D \to x_1H_1 \ldots x_nH_n$. For each $h$, however,
      $\dd Y_{B_h}$
       is paired with a unique $\uu B_h$ or, by Lemma \ref{lemma:qq}, with an $\uu Y$ such that
      there is exactly one $B$ such that $\dd{B_h} \in \dd Y_{B_h}$ and  $\uu B_h \in \uu Y$ so that for a unique
      $B_h = H_h \xLongrightarrow[G] {*} w_h$. Thus, $ x_1B_1 \ldots x_n B_n$ is a unique rhs of $G$ with a
      unique lhs $D = A$, so that $A \xLongrightarrow[G] {*}  x$.
      
    \item From $G$ to $G'$.  Conversely, assume that for $m \le p$ and
      for each $A \in V_N$, $A \xLongrightarrow[G] {m} x$ implies that for some
      $( \dd{[X_A]}, \uu{[X_A]} )$,
      $( \dd{[X_A]}, \uu{[X_A]} ) \xLongrightarrow[G'] {m} x$
      (NB: there could be several  ones since $G'$ is not
      BDR).
      Consider a derivation $A \xLongrightarrow[G] {}  x_1B_1 \ldots B_n \xLongrightarrow[G] {*}   x_1w_n
      \ldots w_n$, with $B_h \xLongrightarrow[G] {m}  w_h$, $m \le p$. By the induction hypothesis there exists at least one
      derivation
      $( \dd{[X_{B_h}]}, \uu{[X_{B_h}]} ) \xLongrightarrow[G'] {m}  w_h$ for each $h$.

      The construction of $G'$ produces from $G$'s production $A \to x_0 B_1 \ldots B_n$
      all possible rules 
      $( \dd{[X_A]}, \uu{[X_A]} ) \to$
      $ x_1$ 
      $( \dd{[X_{B_1}]}, \uu{[X_{B_1}]} ) x_2 \ldots$ 
      $( \dd{[X_{B_n}]},$ $\uu{[X_{B_n}]} )$
      that are compatible with $\overline{\pmb\delta}$ according to the above
      construction. Thus, there exists at least one rule in $G'$
      $( \dd{[X_A]}, \uu{[X_A]} ) \to$
      $ x_1$ 
      $( \dd{[X_{B_1}]}, \uu{[X_{B_1}]} )$ $x_2 \ldots$ 
      $( \dd{[X_{B_n}]}, \uu{[X_{B_n}]} )$
      for each
      $( \dd{[X_{B_h}]}, \uu{[X_{B_h}]} ) \xLongrightarrow[G'] {*}  w_h$.
    \qedhere
    \end{enumerate}
\end{proof}

By taking into account how $G'$ axioms are derived from those of $G$ we immediately obtain the main theorem:

\begin{thm}\label{th:gramm-equiv}
The OPG $G$ and the OPG $G'$ built from it on the basis of the above construction are structurally equivalent.
\end{thm}
The structural equivalence is an obvious consequence of the fact that the two grammars share the same OPM.

The control graph of grammar $G'$, $\mathcal C(G')$, is defined
through a natural modification of the original
Definition~\ref{def:aut:G}: precisely, $\dd V_N$ is the set of the
left elements of $V_N'$, and $\uu V_N$ the set of right elements
thereof.

\begin{figure}
\centering \includegraphics[scale=0.6]{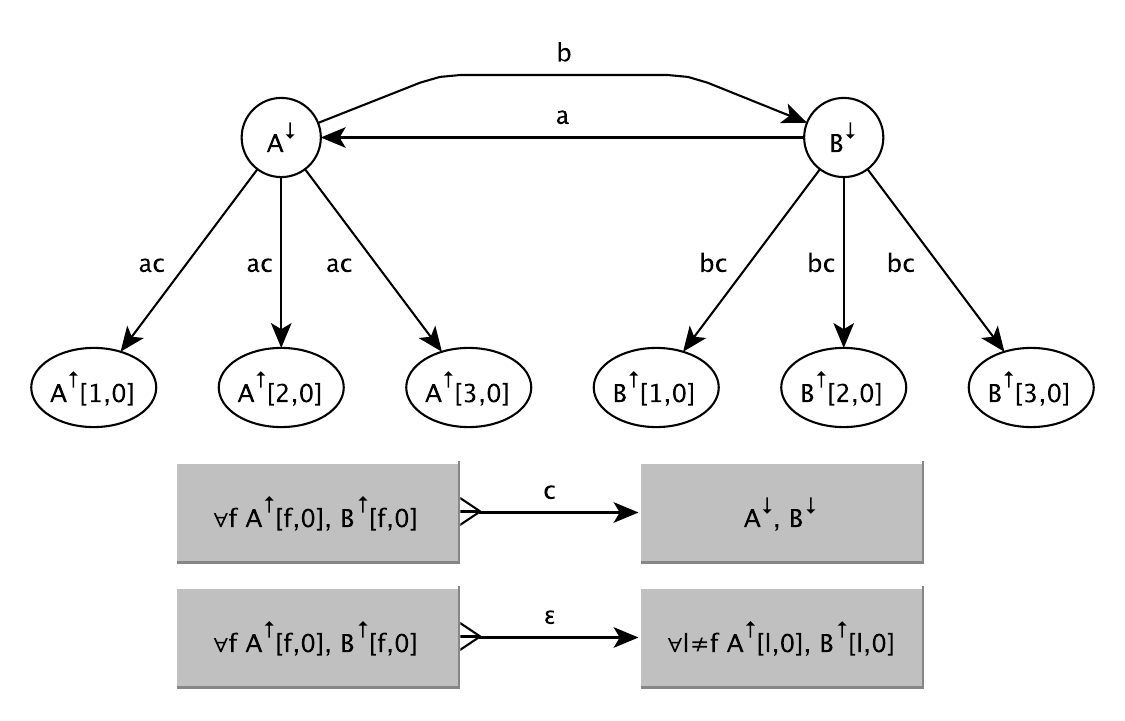}
 \caption{A fragment of the control graph $\mathcal {C}(G')$. The upper part of the graph depicts the descending (single) states; the lower part shows only the entry points of the ascending pipelines.
A significant sample of transitions involving other elements of the pipelines is:
$\forall f \ \uu B[f, 0] \boldtrans{\epsilon} \uu A[f, 1]$.}
\label{fig:C(G')}
\end{figure}

Figure~\ref{fig:C(G')} displays a fragment of $\mathcal {C}(G')$ for
the grammar of Example~\ref{ex:nnn}.  Whereas the transitions from
descending states are complete, for brevity only the entry points
of the ascending part of the graph are displayed.

The following theorem
extends Theorem~\ref{Th:GtoMSO} to the grammars such as $G'$ derived
from BDR OPGs.

\begin{thm}\label{theorem:G1-control-graph}
Consider formulas~(\ref{eq:psi}),~(\ref{eq:chi}) where the subscript $A$ is replaced by all pairs $( \dd {[X_{A}]}, \uu {[X_{A}]})$ as defined in the construction of $G'$.
Thus formula $\varphi_{( \dd {[X_{A}]}, \uu {[X_{A}]})}$ defines the set 
$\{ x \mid \dd{[X_{A}]} \boldtrans{x} \uu{[X_{A}]}\}$.
For any $( \dd {[X_{A}]}, \uu {[X_{A}]})\in V_N'$, $x \in L(( \dd {[X_{A}]}, \uu {[X_{A}]}))$ if and only if $\varphi_{( \dd {[X_{A}]}, \uu {[X_{A}]})} (0,|x|+1) \land \psi_{( \dd {[X_{A}]}, \uu {[X_{A}]})} $ hold.
\end{thm}

\begin{proof}
 The proof is almost identical to that of Theorem~\ref{Th:GtoMSO}, the only difference coming from the fact that $G'$ is not BDR\@. 
 Thus the $\bigvee$ of formula~(\ref{eq:psi}) must be extended to all $G'$ productions having any $( \dd {[X_{A}]}, \uu {[X_{A}]})$ as lhs. 
 E.g., in the base of the induction, instead of just one production $A \to x$ we may have several ones of type $( \dd {[X_{A}]}, \uu {[X_{A}]}) \to x$, each one of them satisfying  $\psi_{( \dd {[X_{A}]}, \uu {[X_{A}]})}$ with the corresponding lhs.
\end{proof}

The following theorem is the last step to achieve FO definability of aperiodic OPLs.

\begin{thm}\label{theorem:L(CG)-non-counting}
  Let $G'$ be the grammar built from any NC BDR OPG $G$ according to
  the procedure given above and let $\mathcal C(G')$ be its control
  graph. Then, for each $( \dd{[X_{A}]}, \uu{[X_{A}]} )$
  of $G'$ the set of paths $\dd{[X_{A}]} \boldtrans{w} \uu{[X_{A}]}$ is a
  NC regular language.
\end{thm}

\begin{proof}
  The fact that the set of paths is a regular language follows
  immediately from the definition of the automaton as in Definition~\ref{def:aut:G}.

  Consider a generic path $\dd{[X_{A}]} \boldtrans{w} \uu{[X_{A}]}$ of
  $\mathcal C(G')$  with $w = xv^ny$ with $n$ sufficiently large. 
  Thus, there must exist a subpath of
  $\dd{[X_{A}]} \boldtrans{w} \uu{[X_{A}]}$  such as $\dd{[X_B^1]} \boldtrans{v}
  \dd{[X_B^2]}$ $\boldtrans{v} \ldots \boldtrans{v} \dd{[X_B^n]}$, with $v = w_1 x_1 w_2 x_2 \ldots$ where $w_i$ are well
  parenthesized according to the OPM and $x_i\in W$, or similarly for an
  ascending path. Notice in fact that, being $v$'s parenthesization uniquely determined by the OPM, $[X_B^l], 1 \le l \le n$ are either all $\dd {[X_B^l]} $ or all $\uu {[X_B^l]} $.
  
  If for some $i$ $\dd{[X_B^i]} = \dd{[X_B^{i+1}]}$ then it is also $\dd{[X_{A}]} \boldtrans{xv^{n+r}y} \uu{[X_{A}]}$ for every $r \geq 0$. Suppose instead that for some $k > 1$  
  $\dd{[X_B^1]} \boldtrans{v} \dd{[X_B^2]}$ $\boldtrans{v} \ldots \dd{[X_B^k]} \boldtrans{v} \dd{[X_B^1]}$
  with  $\dd{[X_B^i]} \ne \dd{[X_B^j]}$
  for $i \ne j$.
 

  Since the original grammar $G$ is BDR, for each $w_i$ there exists a
  unique $C_i$ such that $C_i\xLongrightarrow[G]{*} w_i$. Thus,
  $\dd B_l \boldtranss{\pmb\delta}{\overline v} \dd B_{(l+1) \bmod k} $
  in
  $\mathcal C(G^L)$, where $\overline v$ 
 is obtained from $v$ by replacing each $w_i$ with $\overline C_i$;
  since $(\dd B_1 \ldots \dd B_k , \overline v)$ is a counter of $\mathcal C(G^L)$, 
  by construction of $\overline{\mathcal C}(G^L)$
  it is also
  $\dd X_B \boldtranss{ \overline{\pmb\delta}}{\overline C_1 x_1 \overline C_2 x_2 \ldots} \dd X_B$
  for $\dd X_B =  \dd {B_1 \ldots B_k}  $  and any path including $\overline{v}^k$ must also include the counter sequence state $\dd X_B$.
  By replacing back $\overline C_i$ with
  $w_i$ we obtain
  $\dd X_B \boldtrans{v} \dd X_B$
  as part of the path $\dd{[X_B^1]} \boldtrans{v} \dd{[X_B^2]}$ 
  $\boldtrans{v} \ldots \dd{[X_B^k]} \boldtrans{v} \dd{[X_B^1]}$; 
  thus $\dd{[X_{A}]} \boldtrans{w'} \uu{[X_{A}]}$ for  all $w' = xv^{n+r}y$ with $r \geq 0$.
\end{proof}

As a consequence of Theorem~\ref{theorem:L(CG)-non-counting} all formulas $\varphi_{( \dd {[X_{A}]}, \uu {[X_{A}]})}$ of Theorem~\ref{theorem:G1-control-graph} 
can be written in FO logic, so that the original MSO formulas~\ref{eq:psi},~\ref{eq:chi} become FO once applied to grammar $G'$. Finally we have obtained our main result:

\begin{thm}
Aperiodic operator precedence languages are FO definable.
\end{thm}

\section{Conclusion}
\begin{figure}
\begin{tabular}{@{\hspace{-8pt}}m{0.55\textwidth}m{0.42\textwidth}} 
  \centering{
   \includegraphics[scale=0.3]{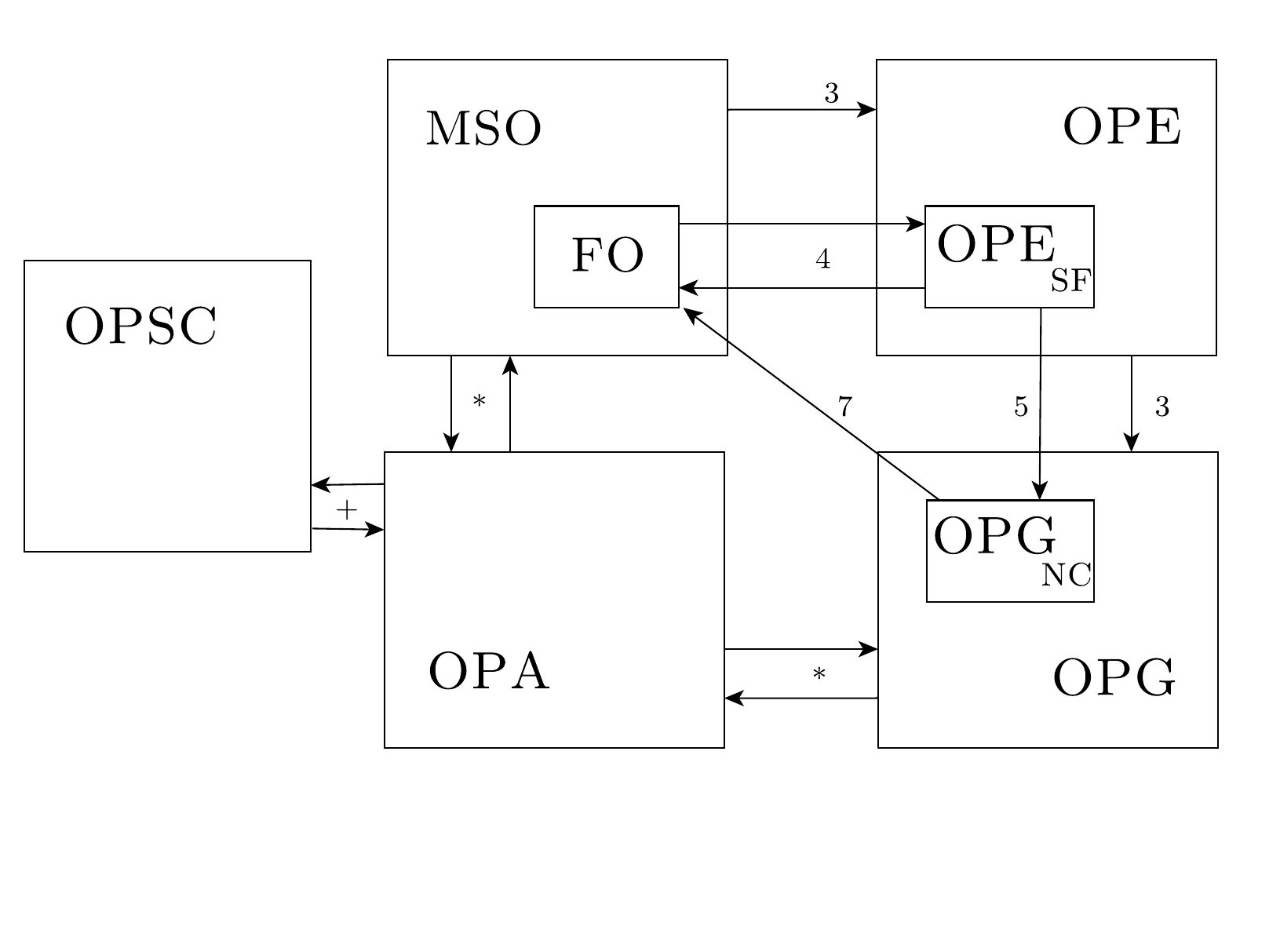}}
&
 \begin{flushleft}
   \begin{scriptsize}
 {\bf Legend}\\
  All boxes denote classes of OPLs with a common OPM:
  \begin{itemize}
      \item 
  MSO denotes OPLs defined through MSO formulas
      \item 
  FO denotes OPLs defined through FO formulas
      \item 
  OPA denotes OPLs defined through OP automata \cite{LonatiEtAl2015}
      \item 
  OPSC denotes OPLs defined 
  through  a Syntactic Congruence with a finite number of equivalence classes \cite{Henzinger23}
   \item 
  $\text{OPE}$ denotes OPLs defined through OPEs  
    \item 
  $\text{OPE}_{\text{SF}}$ denotes languages defined through star-free OPEs
      \item 
  $\text{OPG}_{\text{NC}}$ denotes aperiodic OPLs, i.e., languages defined through NC OPGs
  \end{itemize}
  Arrows between boxes denote language family inclusion; they are labeled by the reference pointing 
  to the section or paper where the property has been proved 
  (* is \cite{LonatiEtAl2015}, + is \cite{Henzinger23}).
\end{scriptsize}
\end{flushleft}
\end{tabular}
\caption{The relations among the various characterizations of OPLs and their aperiodic subclass. }
\label{fig:figurona}
\end{figure}
Figure~\ref{fig:figurona} summarizes the results presented in this paper together with previous related ones. 
The outer boxes represent equivalent ways to express general OPLs, whereas the inner ones represent equivalent ways to express aperiodic OPLs. 
Our results are in sharp contrast with the difficulties encountered in the literature with the same problems in the realm of tree languages.
 OPLs, being ``structured but not visible'' provide greater generality in terms of application fields 
 than ``structured \emph{and} visible'' languages such as tree languages and VPLs.
Consider, e.g., Example \ref{ex:Potthoff}: 
the FO-definition given there for fully parenthesized $\land$-$\lor$ expressions 
can be easily extended to expressions where parentheses may be omitted and replaced by the traditional \emph{precedence} of the  $\land$ operator over the $\lor$ one in the same way as it happens for arithmetic expressions. 
In that case it is the OPM that provides ``for free'' the implicit structure.

We believe that the process that we followed to obtain the characterization of aperiodic OPLs could be replicated in an analogous form to the case of VPLs. 
This path would start by defining an aperiodic subclass of VPLs, 
then pass through an extension of regular expressions and its star-free restriction, 
ending with the reduction of their logic characterization to first order under the hypothesis of aperiodicity.

Figure~\ref{fig:figurona} immediately suggests a further research step, i.e., making the inner triangle a pentagon, as well as the outer one.
We also hope that the articulated path that we used to prove that NC OPLs are FO definable could be made shorter and more direct, 
although we cannot forget that even in the case of regular languages such proof paths are rather complex, e.g.
\cite{McNaughtPap71}, or the more recent and shorter \cite{Wilke99,Diekert-Gastin-first-orderdefinable}.


Another natural extension of the results reported in this paper is their generalization to the case of $\omega$-OPLs \cite{LonatiEtAl2015}, complementing, once again, previous studies on aperiodic $\omega$-regular languages (see, e.g., \cite{DBLP:journals/ijfcs/Selivanov08,DBLP:journals/iandc/Wagner79}).

There are other subclasses of regular languages related to the aperiodic ones; among them we mention \emph{locally testable languages}:
intuitively, the distinguishing feature of these languages is that one can decide whether a string belongs to a given language or not by examining only substrings of bounded length. 
It is known \cite{McNaughtPap71} that aperiodic regular languages are the closure of the locally testable ones under concatenation. In \cite{CreGuiMan78} we proposed a first definition of locally testable parenthesis languages and showed that they are strictly included in the NC ones. 
More recently local testability has been investigated for tree languages and its potential application on data-managing systems has been advocated  \cite{DBLP:journals/corr/abs-1208-5129,DBLP:journals/corr/abs-1109-5851}.
It could be worth investigating the relation of this property too with aperiodicity and FO-definability for OPLs.


The most exciting goal that some researchers are pursuing, however, 
is the completion of the great historical path that, for regular languages, 
led from the first characterization in terms of MSO logic to the restricted case of FO characterization of NC regular languages, to the temporal logic one which in turn is FO-complete, 
thanks to Kamp's theorem \cite{DBLP:journals/corr/Rabinovich14}, and, ultimately, to the striking success of model checking techniques.

Some proposals of temporal logic extension of the classic linear or branching time ones to cope with the typical nesting structure of CF languages have been already offered in the literature. 
E.g., \cite{Marx2005} presents an FO-complete temporal logic to specify properties of paths in tree languages;  \cite{lmcs/AlurABEIL08,DBLP:journals/toplas/AlurCM11,DBLP:conf/cade/BozzelliS14} present different cases of temporal logics extended to deal with VPLs; 
they also prove FO-completeness of such logics. 

We too have designed a first example of temporal logic for OPLs~\cite{CMP20} which recently evolved into a new FO-complete and more user-friendly one \cite{CMP21arXiv}.
We also built an algorithm that derives an OPA from a formula of this logic of exponential size in the length of the formula and implemented a satisfiability and model checker which has been experimentally tested on a realistic benchmark \cite{CMP21}.
Thanks to the result of this paper, and to the fact that most, if not all, of the CF languages for practical applications are aperiodic, 
the final goal of building model checkers that cover a much wider application field than that of regular languages 
---and of various structured CF languages, such as VPLs, too--- with comparable computational complexity does not seem unreachable.

There are many jewels to extract from the old, but still rich, mine of OPLs.

\section*{Acknowledgments}
We are deeply thankful to the reviewers who dedicated time and effort to carefully read our paper and provided precious suggestions to improve its presentation.
We are also grateful to an anonymous reviewer of ICALP 2022 who, while rejecting this same result, claiming that it was wrong a priori just because it does not hold for tree languages, challenged us to produce a FO formula for the language of Example \ref{ex:Potthoff}.

\bibliographystyle{alphaurl}
\bibliography{opbib}
\end{document}